\documentclass[useAMS,usenatbib,usegraphicx,11pt]{article}
\pdfoutput=1
\usepackage{jcappub}
\usepackage[english]{babel}
\usepackage{xcolor}
\usepackage{multirow}
\usepackage{colortbl}
\usepackage{units}
\usepackage{mathrsfs}
\usepackage[super]{nth}
\usepackage{bm}
\usepackage[nameinlink,noabbrev]{cleveref}
\usepackage{subfigure}
\usepackage[hang, small, bf]{caption}
\newenvironment{alphafootnotes}
  {\par\edef\savedfootnotenumber{\number\value{footnote}}
   
   \setcounter{footnote}{0}}
  {\par\setcounter{footnote}{\savedfootnotenumber}}

\newcommand{\gr}{$\gamma$-ray}
\newcommand{\grs}{$\gamma$-rays} 
\newcommand{\jfactor}{{$J$-factor}}
\newcommand{\jfactors}{{$J$-factors}}

\newcommand{\beq}{\begin{equation}}
\newcommand{\eeq}{\end{equation}}
\newcommand{\balign}{\begin{align}}
\newcommand{\ealign}{\end{align}}

\newcommand{\lcdm}{{\ifmmode \Lambda{\rm CDM} \else $\Lambda{\rm CDM}$\fi}}
\newcommand{\mchi}{\ensuremath{m_{\chi}}}
\newcommand{\Msol}{\ensuremath{\rm M_\odot}}

\newcommand{\Msubtot}{\ensuremath{M_{\rm subs}}}
\newcommand{\Mtot}{\ensuremath{M_{\rm tot}}}

\newcommand{\fsub}{\ensuremath{f_{\rm subs}}}
\newcommand{\Rsol}{\ensuremath{R_\odot}}
\newcommand{\rhosol}{\ensuremath{\rho_\odot}}

\newcommand{\Nsub}{\ensuremath{N}}
\newcommand{\Nsubmean}{\ensuremath{\overline{N}}}
\newcommand{\alpham}{\ensuremath{\alpha_m}}
\newcommand{\Rvir}{\ensuremath{{R}_{\rm vir}}}
\newcommand{\Dvir}{\ensuremath{{\Delta}_{\rm vir}}}
\newcommand{\rvir}{\ensuremath{{r}_{\rm vir}}}
\newcommand{\Rtwohundred}{\ensuremath{{R}_{200}}}
\newcommand{\rscale}{\ensuremath{{r}_{\rm s}}}

\newcommand{\Mvir}{\ensuremath{{M}_{\rm vir}}}
\newcommand{\mvir}{\ensuremath{{m}_{\rm vir}}}

\newcommand{\clumpy}{{\tt CLUMPY}}

\newcommand{\healpix}{{\tt HEALPix}}

\newcommand{\degs}{{^{\circ}}}
\newcommand{\nside}{{N_{\rm side}}}

\newcommand{\jdrawn}{\ensuremath{{J_{\rm drawn}}}}
\newcommand{\jsubs}{\ensuremath{\langle J_{\rm subs}\rangle}}
\newcommand{\jcross}{\ensuremath{J_{\rm cross-prod}}}
\newcommand{\jsmooth}{\ensuremath{J_{\rm sm}}}

\newcommand{\junits}{\ensuremath{\rm{GeV^2\, cm^{-5}}}}
\newcommand{\thetaint}{\ensuremath{\theta_{\text{int}}}}

\newcommand{\thetahalf}{\ensuremath{\theta_{\text{h}}}}
\newcommand{\lum}{\ensuremath{\mathcal{L}}}
\newcommand{\sigmav}{\ensuremath{\langle \sigma v\rangle}}

\newcommand{\rhosub}{\ensuremath{\overline{\rho}_{\rm subs}}}
\newcommand{\rhotot}{\ensuremath{\overline{\rho}_{\rm tot}}}

\newcommand{\dd}{\ensuremath{\mathrm{d}}}
\newcommand{\fermi}{{\em Fermi}-LAT}
\newcommand{\TS}{\ensuremath{\mathrm{TS}}}

\definecolor{darkgreen}{rgb}{0,0.4,0}
\definecolor{modelref}{rgb}{0.247,0.3647,0.491}
\definecolor{firebrick}{rgb}{0.698,0.1333,0.1333}
\definecolor{seagreen}{rgb}{0.180,0.545,0.341}

\crefname{subsection}{\S}{\S}
\crefname{equation}{Eq.}{Eq.}

\title{\boldmath  Dark matter substructure modelling and sensitivity of the Cherenkov Telescope Array to Galactic dark halos}

\author[a,c]{M. H\"utten,}
\author[b]{C. Combet,}
\author[a]{G. Maier,}
\author[b]{D. Maurin.}
\affiliation[a]{DESY, Platanenallee 6, D-15738 Zeuthen,  Germany}
\affiliation[b]{LPSC, Universit\'e Grenoble-Alpes, CNRS/IN2P3,\\53 avenue des Martyrs, 38026 Grenoble, France}
\affiliation[c]{Humboldt-Universit\"{a}t zu Berlin, Newtonstra{\ss}e 15,\\ D-12489 Berlin, Germany}
\emailAdd{moritz.huetten@desy.de}
\emailAdd{celine.combet@lpsc.in2p3.fr}
\emailAdd{gernot.maier@desy.de}
\emailAdd{dmaurin@lpsc.in2p3.fr}

\abstract{Hierarchical structure formation leads to a clumpy distribution of dark matter in the Milky Way. These clumps are possible targets to search for dark matter annihilation with present and future \gr{} instruments. Many uncertainties exist on the clump distribution, leading to disputed conclusions about the expected number of detectable clumps and the ensuing limits that can be obtained from non-detection. In this paper, we use the \clumpy{} code to simulate thousands of skymaps for several clump distributions. This allows us to statistically assess the typical properties (mass, distance, angular size, luminosity) of the detectable clumps. Varying parameters of the clump distributions allows us to identify the key quantities to which the number of detectable clumps is the most sensitive. Focusing our analysis on two extreme clump configurations, yet consistent with results from numerical simulations, we revisit and compare various calculations made for the \fermi{} instrument, in terms of number of dark clumps expected and the angular power spectrum for the Galactic signal.  We then focus on the prospects of detecting dark clumps with the future CTA instrument, for which we make a detailed sensitivity analysis using open-source CTA software. Based on a realistic scenario for the foreseen CTA extragalactic survey, and accounting for a post-trial sensitivity in the survey, we show that we obtain competitive and complementary limits to those based on long observation of a single bright dwarf spheroidal galaxy.}


\begin{document}
\maketitle
\flushbottom


\section{Introduction}
\label{sec:intro}

Dark matter (DM) indirect detection in $\gamma$-rays was first discussed almost 40 years ago~\cite{1978ApJ...223.1015G,1978ApJ...223.1032S} in the context of the diffuse \gr{} emission. The most promising DM candidate is a weakly interacting massive particle (WIMP), which produces secondary {\grs} originating from the particles' self-annihilation \cite{2009NJPh...11j5006B}.  As the signal from self-annihilating DM roughly goes as the DM density squared divided by the distance squared, the most promising targets result from an interplay between being close and/or massive, highly concentrated DM reservoirs. This made the Galactic centre (GC) a prime target, as first considered in \cite{1987ApJ...313L..47S}. However, it was soon recognised that setting constraints on DM would be limited by astrophysical background at GeV energies \cite{1988PhLB..201..529S}. Higher energies face similar challenges, as illustrated by the first H.E.S.S. observations of the GC~\cite{2004A&A...425L..13A}, and following DM interpretations \cite{2005PhLB..607..225H}. Owing to their potentially high DM densities and small astrophysical backgrounds, dwarf spheroidal (dSph) galaxies orbiting the Milky Way \cite{1990ApJ...356L..43L} and dark clumps \cite{1990Natur.346...39L,1993ApJ...411..439S} were proposed next. In particular, the question of the DM clump population boost of the Galactic signal \cite{1998APh.....9..137B,2003PhRvD..68j3003B} and/or the detectability of individual clumps with future \gr{} satellites and ground-based instruments \cite{2000PhRvD..62l3005C,2002PhRvD..66h3006T,2003PhRvD..68j3003B} was raised.

Since these pioneering studies, a steady progress has been made in estimating the prospects of DM indirect detection (or in setting constraints from non-detection) with the \gr{} sky. The launch of the {\em Fermi} \gr{} Space Telescope in 2008 with its main instrument, the Large Area Telescope (LAT) \cite{2009ApJ...697.1071A}, was a huge step forward in exploring with unprecedented sensitivity the DM parameter space (see \cite{2015JETP..121.1104C} for a review). Thanks to the all-sky survey observing strategy, the \fermi{} collaboration has set constraints on \gr{} lines  \cite{2010PhRvL.104i1302A,2012PhRvD..86b2002A,2013PhRvD..88h2002A,2015PhRvD..91l2002A} or excesses in the \gr{} continuum spectrum of the diffuse Galactic and extragalactic emission \cite{2010JCAP...04..014A,2012ApJ...761...91A,2015JCAP...09..008T}, galaxy clusters \cite{2010JCAP...05..025A,2015ApJ...812..159A}, dSph galaxies \cite{2010ApJ...712..147A,2011PhRvL.107x1302A,2012ApJ...747..121A,2014PhRvD..89d2001A,2015PhRvL.115w1301A,2015ApJ...809L...4D}, and dark clumps \cite{2012ApJ...747..121A}. For dSph galaxies, whose known number is still growing from recent optical surveys, a stacking strategy allowed to exclude a thermal relic annihilation cross-section $\sigmav$ of a few times $10^{-26}~\unit{cm^3~s^{-1}}$ for WIMP masses $\mchi< 100$~GeV \cite{2015PhRvL.115w1301A,2015ApJ...809L...4D}.
Constraints from ground-based Cherenkov telescopes are competitive with \fermi{} limits to constrain DM particle masses above a few hundreds of GeV. These pointed instruments use a different observation strategy: the best constraints are obtained from DM searches in regions around the GC, leading to limits of $\sigmav \lesssim 3 \times 10^{-24}~\unit{cm^3~s^{-1}}$ for $\mchi{}\sim$~TeV \cite{2011PhRvL.106p1301A,2015PhRvL.114h1301A}; for dSph galaxies, the constraints either come from a stacking analysis (VERITAS \cite{2010ApJ...720.1174A,2015arXiv150901105Z} and H.E.S.S. \cite{2014PhRvD..90k2012A}) or from a deep observation of a single highly ranked object (MAGIC \cite{2014JCAP...02..008A,2016JCAP...02..039M}), leading to limits of $\sigmav \lesssim  1-3 \times 10^{-24}~\unit{cm^3~s^{-1}}$. 

To take full advantage of this wealth of data and to set the most reliable and effective limits on DM, the modelling of the DM signal received as much attention. Beside the complex GC region (see, e.g., the \fermi{} analysis \cite{2016ApJ...819...44A}), improvements were made in the DM modelling of dSph galaxies \cite{2013NewAR..57...52B,2013PhR...531....1S}, galaxy clusters \cite{2009PhRvD..80b3005J,2010PhRvD..82b3506Y,2011PhRvD..84l3509P,2012JCAP...07..017A,2012MNRAS.425..477N}, and diffuse emissions (see \cite{2015PhR...598....1F} for a recent review). To increase the sensitivity to DM indirect detection, multiwavelength analyses \cite{1992PhLB..294..221B,2015JCAP...02..032C} and observations of dSph galaxies \cite{2013ApJ...773...61S,2013PhRvD..88h3535N,2014JCAP...10..016R} and galaxy clusters \cite{2006A&A...455...21C,2013ApJ...768..106S} have been carried out. Enhancements of the signal-to-noise ratio arise from a better characterisation of \gr{} anisotropies and cross-correlation with other tracers (e.g., \cite{2014FrP.....2....6F}). Note that most of these calculations at both the Galactic and extragalactic level depend on the hierarchical structure formation and on the survival and distribution of subhalos in their host halos. In particular, the detectability of dark clumps observed as \gr{} point sources with no counterparts at other wavelengths or from their imprint in the angular power spectrum (APS) over the diffuse DM background (that can be boosted by these same micro-halos), depends crucially on the modelling of these clumps. Their properties are investigated by heavy numerical simulations \cite{2012PDU.....1...50K}. If state-of-the-art Milky Way like simulated halos, such as Via Lactea \cite{2007ApJ...657..262D}, Aquarius \cite{2008MNRAS.391.1685S}, and more recently COCO \cite{2016MNRAS.457.3492H} reach a mass resolution of a few $10^5$ solar masses, this must be compared with the very uncertain minimal mass of the subhalos (set by the DM candidate couplings) that could be ten orders of magnitude below (e.g., \cite{2006PhRvL..97c1301P}). Numerical simulations of the early universe have confirmed that such subhalos might survive until today \cite{2005Natur.433..389D}, a result also deduced from theoretical calculations (see \cite{2014PhyU...57....1B} for a review).

This paper revisits the question of the detectability of dark clumps for present and future \gr{} instruments. The prospects for the \fermi{} instrument were discussed by many authors in the light of high resolution numerical simulations \cite{2003MNRAS.345.1313S,2008ApJ...686..262K,2009Sci...325..970K,2010ApJ...718..899A,2011PhRvD..83b3518P}, or based on semi-analytical approaches which extrapolate the clump population down to the mass of the smallest micro-halos \cite{2004PhRvD..69d3501K,2004PhRvD..69d3512P,2005ApJ...633L..65O,2008MNRAS.384.1627P,2009JCAP...07..007L,2010PhRvD..82h3504K}. Whereas recent efforts are turned towards a better discrimination between astrophysical or DM signals in sources with no counterparts in the \fermi{} catalogue \cite{2012A&A...538A..93Z,2012JCAP...11..050Z,2012PhRvD..86d3504B,2012MNRAS.424L..64M,2013MNRAS.436.2461M,2014PhRvD..89a6014B}, the constraints that can be set depend ultimately on the number of expected DM clumps, which is still disputed \cite{2015JCAP...12..035B,2016JCAP...05..028S}. The sensitivity to dark clumps for the future Cherenkov Telescope Array (CTA) \cite{2013APh....43....3A,2013APh....43..171B} is discussed in \cite{2011PhRvD..83a5003B}, but a more up-to-date estimate is made here based on the foreseen CTA extragalactic survey \cite{2013APh....43..317D}. We use the \clumpy{} code \cite{2012CoPhC.183..656C,2016CoPhC.200..336B} to evaluate the impact of the clump distribution uncertainties on the ensuing \gr{} signal. Hundreds of Monte Carlo (MC) realisations are run per configuration to estimate the resulting uncertainties on the number of clumps, and to characterise the typical mass and distance of these detectable clumps. We also use realistic instrument responses in a plausible large-sky survey scenario to assess the sensitivity of CTA to these clumps. This provides a complementary view of CTA capabilities against pointed targets that will be part of the CTA DM programme \cite{2013APh....43..189D,2013arXiv1305.0302W,2015arXiv150806128C,2015JCAP...03..055S,2015PhRvD..91l2003L}. 

The paper is organised as follows: in \cref{sec:basics}, we present our modelling approach for the DM distribution in the Galaxy, focusing in particular on seven physical parameters important to the Galactic substructure distribution; \cref{sec:globalprop} presents the resulting \gr{} flux from the different models and describes several cross checks to demonstrate the consistency of our modelling; in \cref{sec:fermi}, we use our findings to revisit the possibility that DM subhalos might be present in the 3FGL catalogue of \fermi{}; \cref{sec:cta} presents the sensitivity of the future CTA to detect DM subhalos within its planned extragalactic sky survey and we finally discuss and summarise our findings in \cref{sec:discussion}.\footnote{The article is followed by a detailed appendix. In \cref{app:definitions}, we shortly review different definitions of cosmological matter overdensities used throughout this paper. In \cref{app:meanmedian}, we present useful formulae to analytically describe power-law source count distributions, and we use these results for a convergence study of the angular power spectra in \cref{app:APSconvergence}. In \cref{app:mvirJ}, we present the derivation and the mass and annihilation factor for the spectroscopically confirmed satellite galaxies in the Milky Way, including recently discovered objects. 
In \cref{app:CTAanalysis_details}, we provide additional details about the CTA analysis performed in \cref{sec:cta}}. Throughout the paper, we denote the mean of quantities $Q$ with a bar, $\overline{Q}$, and median values with a tilde, $\widetilde{Q}$. Global properties of the  Galactic host halo are denoted with capital letters (mass $M$, positions $R$ within the halo, distance $D$ to observer), and properties of individual subhalos with lowercase letters (mass $m$ at distance $d$, positions $r$ within the subhalo). When referring to the brightest subhalo, variables are indicated with an asterisk, $Q^{\star}$.


\section{Modelling the {\gr}-emission from Galactic DM subhalos}
\label{sec:basics}

To assess detection prospects of Galactic DM subhalos, we explore various parameter sets for the substructure density. The average total Galactic halo density is left unchanged.

\subsection{Modelling approach}
We use the \clumpy{} code\footnote{\url{http://lpsc.in2p3.fr/clumpy}} to model the \gr{} emission from Galactic DM subhalos. We refer the reader to \cite{2012CoPhC.183..656C,2016CoPhC.200..336B} for an extensive description of the \clumpy{} code features and validation. \clumpy{} has been used previously to study DM annihilation and/or decay in dSph galaxies \cite{2011ApJ...733L..46W,2011MNRAS.418.1526C,2015MNRAS.446.3002B,2015MNRAS.453..849B,2015ApJ...808L..36B,2015arXiv150608209B,2016ApJ...819...53W} and galaxy clusters \cite{2012PhRvD..85f3517C,2012MNRAS.425..477N,2012A&A...547A..16M}. For the purpose of this work, we mainly use \clumpy{} in the so-called `skymap mode' which allows the fast computation of full-sky maps of DM annihilation or decay signals. We focus on DM annihilation only, for which the expected DM differential \gr{} flux at energy $E$, in the direction $\vec{k}=(\psi, \vartheta)$ and per solid angle $\dd \Omega$ reads
\beq
\frac{\dd \Phi}{\dd E\,\dd\Omega}(E,\,\vec{k})=\frac{\dd\Phi^{\rm PP}}{\dd E}(E)\times \frac{\dd J}{\dd\Omega}(\vec{k}) \,,
\label{eq:flux-general}
\eeq
where 
\beq
\frac{\dd\Phi^{\rm PP}}{\dd E}(E)
      = \frac{1}{4\pi}\frac{\sigmav}{m_{\chi}^{2}\delta}\,\sum_{f}\frac{\dd N^{f}_{\gamma}}{\dd E}\, B_{f}\,,
       \label{eq:gammafluxpp}
\eeq
and where the \jfactor{} is generically written as
\begin{eqnarray}
J(\vec{k},\,\Delta\Omega) = \int\limits_{\Delta \Omega}\; \int\limits_{\rm{l.o.s}} \!\! \rho^2 \dd l\,\dd\Omega 
= \int\limits_{0}^{2\pi}\int\limits_0^{\thetaint}\int\limits_{\rm{l.o.s}} \!\! \rho^2(\vec{k};\;l,\,\theta,\,\phi)\, \dd l\,\sin\theta\,\dd\theta\,\dd\phi\,.
\label{eq:gammafluxastro}
\end{eqnarray}
In these equations, \mchi{} is the mass of the
DM particle $\chi$, \sigmav{} is the velocity-averaged annihilation cross-section, and $\dd N_{\gamma}^f/\dd E$ and $B_f$ correspond to the spectrum and branching ratio of annihilation channel $f$. The parameter $\delta$ is $\delta=2$ for a Majorana and $\delta=4$ for a Dirac particle, and we choose $\delta\equiv 2$ in the remainder of this paper. In \clumpy{}, \mchi{}, \sigmav{}, and $B_f$ are user-defined parameters, from which the \gr{} annihilation spectrum is calculated (based on the parametrisations of \citep{2011JCAP...03..051C}). The DM density $\rho$ is integrated along the line of sight (l.o.s.), and up to a maximum angular distance \thetaint{}. The overall DM density can be written $\rho_{\rm tot}=\rho_{\rm sm}+\rho_{\rm subs}$, where $\rho_{\rm sm}$ corresponds to the smooth component, and $\rho_{\rm subs}$ corresponds to the substructures of the Galactic DM halo.

Generating skymaps with \clumpy{} starts from setting DM properties: smooth DM profile, spatial and mass distribution of Galactic substructures, halo mass-concentration relation, DM particle mass, and annihilation/decay channels. The computation has been optimised as to draw only subhalos that outshine the mean DM signal (set by a user-defined precision), leading to a decomposition of the substructure signal  $J_{\rm subs}^{\rm tot} = \jdrawn + \jsubs$  into two components: \jdrawn{} is the signal from the substructures drawn in a realisation of the skymaps, and \jsubs{} is the average signal from all `unresolved' halos\footnote{For legibility purpose, we define \jsubs{} to be the sum of \jsubs{} and \jcross{} as defined in \cite{2016CoPhC.200..336B}.}, i.e., faint subhalos whose intrinsic \jfactors{} do not pass the threshold defined from the precision level required by the user. Additional levels of clustering within subhalos are also considered using this average description. We refer the reader to \cite{2012CoPhC.183..656C,2016CoPhC.200..336B} for details on the computation of these quantities. For the purpose of this work, suffice to say that the higher the precision requirement, the more halos are drawn and the smaller is the contribution of \jsubs{}. A convergence analysis with respect to the precision requirement is presented in \cref{app:APSconvergence}.

The flexibility of \clumpy{} allows the user to easily explore various models and configurations, calibrated (but not limited) to the results of various $\Lambda$CDM numerical simulations such as the Aquarius \cite{2008MNRAS.391.1685S}, Phoenix \cite{2012MNRAS.425.2169G} or Via Lactea~II (VL~II) \cite{2008ApJ...679.1260M} simulations, as well as their hydrodynamical updates \cite{2015MNRAS.447.1353M,2016MNRAS.457.1931S,2016ApJ...827L..23W}. 

\subsection{Milky Way dark matter halo}
The total DM density profile \rhotot{} of the Milky Way (MW) is modelled with an Einasto profile with a slope $\alpha_{\rm E}=0.17$ and scale radius $R_{\rm s} = 15.14\,\rm{kpc}$, as suggested by \cite{2010MNRAS.402...21N}. The normalisation of the profile is computed to satisfy $\rhosol\equiv \rho(\Rsol=8\,\text{kpc})= 0.4\,\mathrm{GeV\,cm^{-3}}$, as estimated by \cite{2014JPhG...41f3101R} (see also \cite{2012MNRAS.425.1445G,2015arXiv150101788F} for higher estimates). 
Although $\Rsol$ and $\rhosol$ suffer from large uncertainties \cite{2013ARep...57..128M,2015A&A...579A.123H,2015NatPh..11..245I}, their exact values are subdominant for the purpose of our study. All our calculations use a maximum radius $R_{\rm MW}=260\,\mathrm{kpc}$ of the MW  DM halo, yielding a total MW mass $M_{\rm MW}=1.1\times10^{12}\,\Msol$, in agreement with \cite{2013JCAP...07..016N}. As discussed in \cite{2012CoPhC.183..656C}, we define the average substructure density, $\rhosub{} = \fsub\times M_{\rm MW}\times\dd P/\dd V$, where $\dd P/\dd V$ is the spatial distribution of Galactic substructures, and $\fsub$ is the global fraction of the MW mass contained in subhalos. Given \rhotot{} and \rhosub{}, the   smooth MW profile then is defined as $\rho_{\rm sm} = \rhotot - \rhosub$. At sufficient angular distance from the GC ($\theta\gtrsim 10\degs$), this smooth component results in a diffuse background flux, \jsmooth,  that is found negligible by at least 3 orders of magnitude when compared to the flux of resolved subhalos  or to the residual background of CTA (see \cref{sec:CTAbackground}).

\subsection{Substructure properties and set of models}
\label{sec:sets_of_models}
Substructures in their host halo are characterised by their mass and spatial distribution, as well as the description of the DM distribution within each subhalo. In the $\Lambda$CDM hierarchical structure formation scenario, small structures collapse first and then fall and merge into larger structures. Subject to tidal forces, the `unevolved' initial substructure distribution turns into an `evolved' population whose properties differ from those of field halos \cite{2016MNRAS.457.1208H}. Semi-analytical models as well as numerical simulations have been used to characterise the properties of these substructures down (or extrapolated down) to the smallest mass scale. The spread in the various results yield significant uncertainties, e.g., on the cosmological annihilation signal \cite{2014MNRAS.439.2728M}.

To assess the detectability of dark clumps, we build sets of models by varying seven important properties of Galactic substructures, as described in \cref{tab:model-parameters}. For all these models, the threshold mass of the smallest and most massive subhalos are fixed to $m_{\rm min}=\unit[10^{-6}]{\Msol}$ \cite{2004MNRAS.353L..23G} and $m_{\rm max} = 0.01\,\Mtot$ \cite{2009MNRAS.399..983A} respectively: increasing $m_{\rm max}$ would slightly increase the median $J$-factor of the brightest subhalo, leaving unchanged our conclusions; decreasing $m_{\rm min}$ amounts to adding an extra-population of very low-mass halos that would contribute to the average \jsubs\ component, the effect being  dependent on the slope $\alpham$ of the mass distribution (see below). For the angular resolutions considered in our calculations, the brightest resolved objects always outshine the diffuse DM background emission from $\jsmooth + \jsubs$, moreover, for CTA, the diffuse background is dominated by the instrumental residual background and not by the \jsubs{} component.

The parameters and the choices made for their variation are briefly mentioned hereafter and their consequences on the flux are discussed in \cref{sec:globalprop}.
\begin{enumerate}

\item {\em Subhalo inner profile.}
As seen in \cref{eq:gammafluxastro}, the DM density profile is the fundamental ingredient to estimate the astrophysical part of the DM annihilation flux. When ignoring baryonic effects, DM substructures are characterised by cuspy profiles, with two standard parametrisations being the Navarro-Frenk-White (NFW) and Einasto  descriptions \cite{1996ApJ...462..563N,1989A&A...223...89E}. For a given subhalo, we calculate the normalisation and scale radius of the DM inner profile by providing the subhalo mass, \mvir{}, and its concentration $c_{\rm vir}(\mvir,R)\equiv\rvir/\rscale$, where $R$ is the distance from the GC, and \rvir{}, \rscale{} are the virial and scale radius of the subhalo, respectively. The meaning of $c_{\rm vir}$ is further discussed in the next paragraph. For the Einasto profile, the shape parameter $\alpha_{\rm E}$ introduces an additional degree of freedom, which we fix to $\alpha_{\rm E}\equiv 0.17$.\footnote{We also investigated the impact of a scattering in $\alpha_{\rm E}$. The authors of \cite{2008MNRAS.391.1685S} find the Aq-A subhalos equally well described by Einasto profiles with $0.16\leq \alpha_{\rm E} \leq 0.20$. For $\alpha_E=0.16$, the \jfactors{} increase by $\lesssim 20\%$ compared to $\alpha_E=0.17$, and decrease  by $\lesssim 60\%$ when choosing $\alpha_E=0.20$. The DM constraints set from the brightest subhalo change by the same amount, whereas the effect on \jsubs{} is subdominant.}  An Einasto profile is the default configuration, and the effect of switching to NFW is performed in the VAR0 model (see \cref{tab:model-parameters}). Note that recent simulations have shown that halos close to the free streaming scale \cite{2010ApJ...723L.195I,2013JCAP...04..009A,2014ApJ...788...27I,2016arXiv160403131A} could be cuspier than the NFW profile. Including in \clumpy{} this mass-dependent inner slope---Eq.~(2) of \cite{2014ApJ...788...27I}---leads to a 13\% to 43\% increase of \jsubs{} (comparable to the 12\% to 67\% increase found in \cite{2014ApJ...788...27I}), but more importantly here, we checked that these boosted microhalos provide no detectable dark clumps.

\item {\em Mass-concentration parametrisation $c(m)$.} 
Once the parametrisation of a subhalo inner profile is chosen, its structural parameters (normalisation and scale radius) are fully determined from the mass $m_\Delta$ and concentration-mass $c_\Delta-m_\Delta$ relation (see \cref{app:definitions} for a definition of $\Delta={\rm vir},\,200\,,500\,\dots$). The latter depends on the subhalo evolution in its host halo, i.e. its location and `evolved' mass. Several parametrisations, based on the results of numerical simulations have been proposed in the last few years, the most recent suggesting a flattening of the relation at low masses \cite{2014MNRAS.442.2271S} and a higher concentration of subhalos compared to field halos. The latter effect was shown to yield an extra $\sim 5$ boost factor \cite{2015PhRvD..92l3508B,2016MNRAS.457..986Z,2016arXiv160304057M} on \jsubs{} compared to previous calculations.
\begin{figure}[t!]
\captionsetup{format=plain}
\centering
    \includegraphics[width=0.8\textwidth]{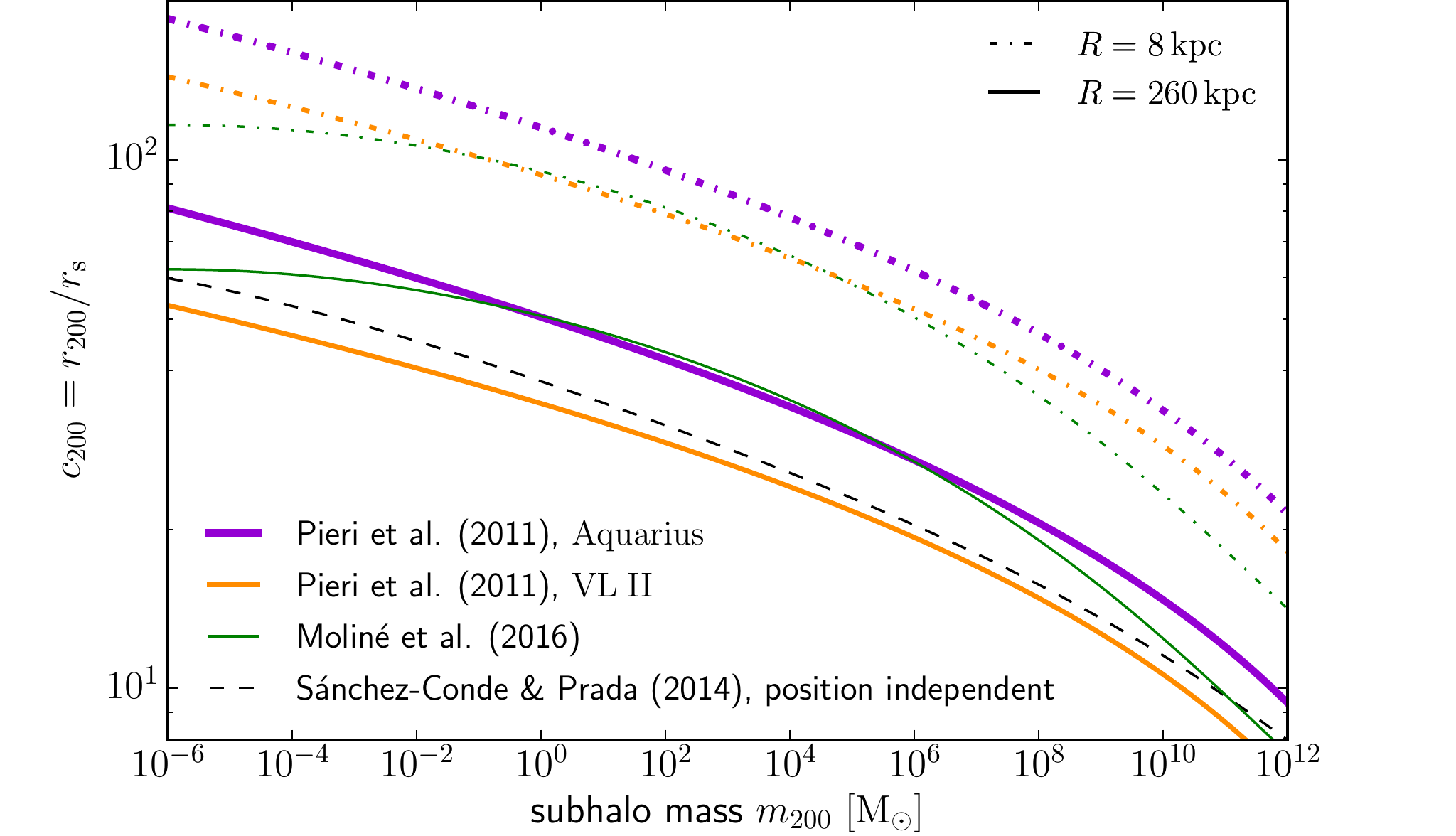}
  \caption{
Models for the concentration $c_{200}$ of Galactic subhalos at redshift $z=0$. Beside the distance-independent parametrisation SP \cite{2014MNRAS.442.2271S}, shown are the values at the galactocentric distances $R=\unit[8]{kpc}$ (dashed-dotted lines) and $R=\unit[260]{kpc}$ (solid lines) from  \cite{2011PhRvD..83b3518P} (violet and orange) and \cite{2016arXiv160304057M} (green).}
  \label{fig:cdelta-mdelta}
\end{figure} 
Here, we consider the field halo S{\'a}nchez-Conde~\& Prada \cite{2014MNRAS.442.2271S} parametrisation (SP), the distance-dependent Pieri et al.\!\! \cite{2011PhRvD..83b3518P} parametrisations (P-VLII based on the Via Lactea~II simulation and P-Aq based on Aquarius) and the Molin\'e et al.\!\! \cite{2016arXiv160304057M} parametrisation.\footnote{As for several other refinements discussed in this section, the Pieri et al.\!\!  and Molin\'e et al.\!\!  parametrisations have been implemented in \clumpy{} for the purpose of this work, but are not yet available in the public version.} These mass-concentration prescriptions are displayed in \cref{fig:cdelta-mdelta}. 
The Pieri et al.\!\! and Molin\'e et al.\!\! approaches account for the fact that the closer a subhalo of a given mass is to the GC, the more concentrated it is. The various parametrisations are compared at a distance of 8~kpc from the GC (dotted-dashed lines), where the P-VLII and Molin\'e et al.\!\! parametrisations produce similar concentrations, while P-Aq appears systematically higher. At large distance from the GC (solid lines), all parametrisations yield lower concentrations that become more compatible with the SP field halo distance-independent parametrisation (dashed line). In the following, we use the SP description as our conservative baseline and investigate the distance-dependent effects of  the P-VLII and Molin\'e parametrisations in both the {\sc VAR6} and {\sc HIGH} models. We will discard the P-Aq prescription in the following, but remind that using this parametrisation would result into even larger \jfactors{} than used in our optimistic model {\sc HIGH}.\footnote{We also investigated the P-Aq model,  and found a $\sim 60\%$ increase for the flux from the brightest subhalos compared to the P-VLII description. This increase would also improve the sensitivities presented in \cref{fig:IndividualSourceSensitivity-CTA} by the corresponding factor.}

\item {\em Width of the mass-concentration distribution, $\sigma_c$.}
Rather than assuming a single concentration for a given halo mass from the mean parametrisations above, the concentration is drawn from a log-normal distribution of width $\sigma_c$ around the values given by these parametrisations. This is incorporated to account for the intrinsic scatter of the $c(m)$ relation found in numerical simulations. We consider $\sigma_c = 0.14$ as our default value \cite{2014MNRAS.442.2271S, 2002ApJ...568...52W}, and study the impact of a larger scatter in the VAR2 model where $\sigma_c = 0.24$ \cite{2001MNRAS.321..559B} is used.

\item {\em Number of halos $N_{\rm calib}$ between $10^8$ and $\unit[10^{10}]{\Msol}$.}
This number is used as a calibration for the total number of subhalos. $\Lambda$CDM simulations of MW size halos predict an overabundance of high-mass subhalos compared to the currently known satellite galaxies; this is the so-called `missing satellites' problem \cite{2010arXiv1009.4505B}, which is linked to the `too-big-to-fail' problem \cite{2011MNRAS.415L..40B}. Baryonic feedback onto the cusps of DM subhalos could possibly solve this tension \cite{2014JCAP...04..021D,2015ApJ...806..229M}. Indeed, hydrodynamical simulations roughly show half as many  high-mass subhalos as DM only simulations. About $100-150$ objects are obtained above $\unit[10^8]{\Msol}$ \cite{2015MNRAS.447.1353M,2016MNRAS.457.1931S}, and we choose $N_{\rm calib}=150$ as our default value. A more subhalo-rich configuration, used in models VAR4 and HIGH (see \cref{tab:model-parameters}), is defined by $N_{\rm calib}=300$ as motivated by the results of DM-only simulations \cite{2008MNRAS.391.1685S}.

\item {\em Index of the clump mass distribution, \alpham{}.} 
Numerical simulations show that the DM halo mass distribution is well described as $\dd N/\dd m \propto m^{-\alpham}$. For the reference model we take $\alpham=1.9$ as suggested by numerical simulations of MW-like halos \cite{2008MNRAS.391.1685S,2008ApJ...679.1260M} and investigate a slightly steeper mass function in the VAR1 model, where $\alpham=2.0$. Together with $N_{\rm calib}$, $m_{\rm min}$, and $m_{\rm max}$, the choice of \alpham{} determines the total number of clumps $\Nsub_{\rm tot}$ and their total mass \Msubtot{} (see \cite{2012CoPhC.183..656C} for details). In \cref{tab:model-parameters}, we provide the resulting $\Nsub_{\rm tot}$ and, with the knowledge of the total mass of the Galaxy, $M_{\rm MW}$, the global portion of DM bound into subhalos, $\fsub = \Msubtot{}/M_{\rm MW}$.

\item {\em Spatial distribution of Galactic substructures, $\dd P/\dd V=\rhosub{}/\Msubtot{}$.} 
The fraction of mass bound into substructures is expected to decrease towards the GC, as subhalos are tidally disrupted by the strong gradient of the gravitational potential.  This is discussed in detail in \cite{2016MNRAS.457.1208H}, where the authors argue that this effect is the result of a selection effect of the `evolved' subhalos (suffering from tidal stripping), the `unevolved' distribution following the host smooth distribution.
\begin{figure}[t!]
\centering
    \includegraphics[width=0.8\textwidth]{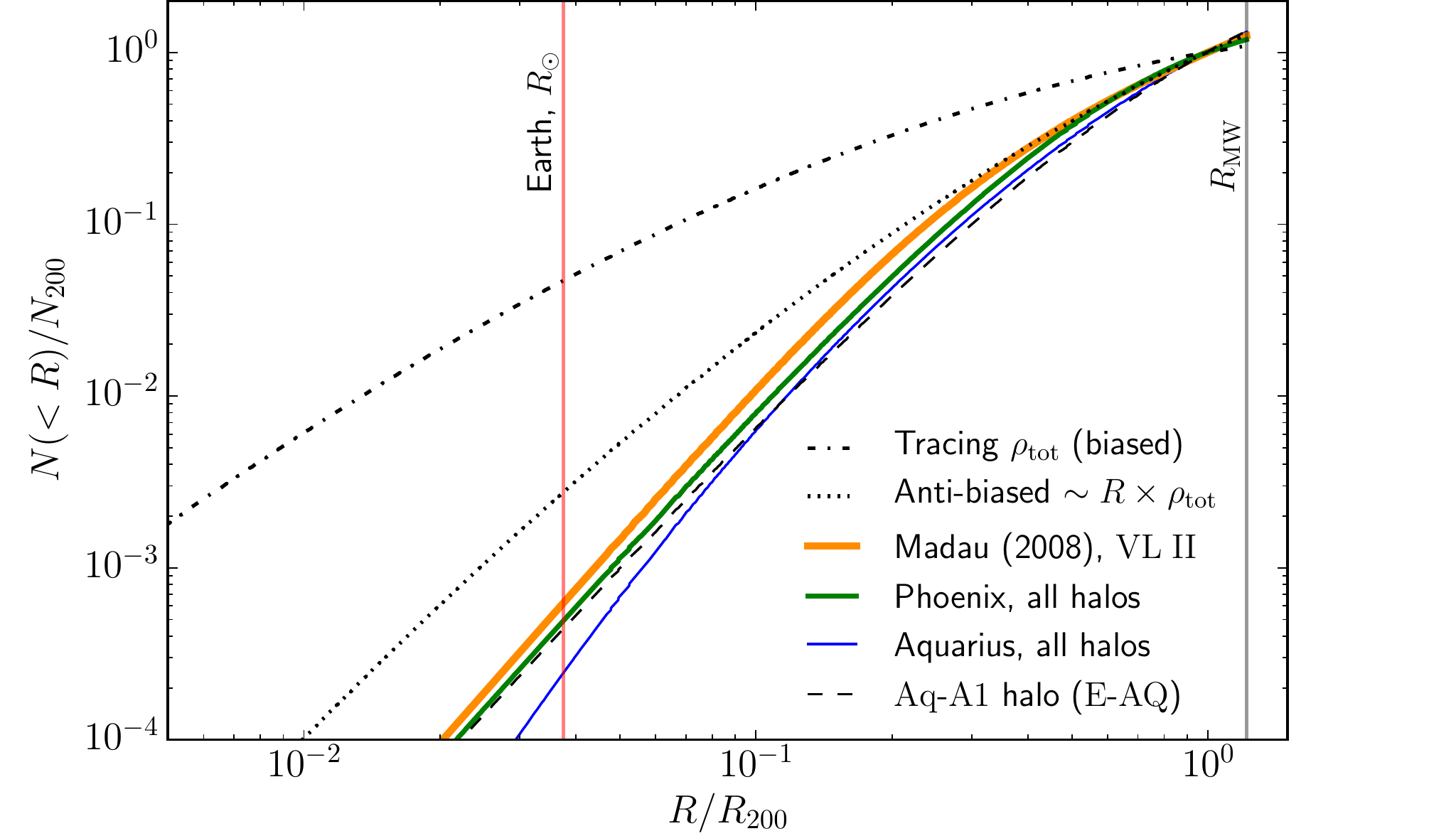}
  \caption{
Number of subhalos within $R$, normalized to $N_{200}$, the number of subhalos within \Rtwohundred{},  up to $R_{\rm MW}/\Rtwohundred=1.22$ ($\Rtwohundred=\unit[213.5]{kpc}$ for our Galactic halo). We also show the position of the observer at $\Rsol/\Rtwohundred=0.037$. The number of subhalos, $\Nsub(R)$, results from the subhalo number density $\dd \Nsub/\dd V = \Nsub_{\rm tot}\cdot\dd P/\dd V$.
}
\label{fig:fsub_comp}
\end{figure} 
\Cref{fig:fsub_comp} displays the cumulative number of halos as a function of the distance to the GC, where $\rhosub{}\propto\rhotot$ is shown as dash-dotted line. To correctly describe \rhosub{}, a generic `anti-biased' parametrisation $\rhosub/\rhotot\propto R$ was proposed in \cite{2007ApJ...671.1135K,2008JCAP...10..040S}  (dotted line). This is to be compared to the result obtained from numerical simulations, namely, a fit to Aq-A1 (well described by an Einasto profile `E-AQ', long-dashed line) or to all Aquarius halos (curved power-law, cyan) halos \cite{2008MNRAS.391.1685S}, or subhalos of the Via Lactea~II simulation \cite{2004MNRAS.355..819G,2008ApJ...679.1260M} (orange). We also show the profile for the Phoenix simulation of galaxy clusters (rescaled to $\Rtwohundred$, green line) described by an Einasto profile \cite{2012MNRAS.425.2169G}, which leads to a comparable dependence. Simulations produce slopes steeper than $\rhosub/\rhotot\propto R$, with the Aquarius and Phoenix results found to be well described by $\rhosub/\rhotot\propto R^{1.3}$ in \cite{2016MNRAS.457.1208H}'s model. In \cref{tab:model-parameters}, we choose the Einasto profile fitted to the Aquarius A-1 halo as our default setup and use the Via Lactea~II parametrisation in the VAR3 and HIGH models. \Cref{tab:model-parameters} also gives the local DM mass fraction under the form of subhalos in the solar neighbourhood, $\fsub(\Rsol) = \rhosub(\Rsol)/\rhotot(\Rsol)$, which is dependent on the chosen \rhosub{} profile.

\item {\em Flag for sub-substructures.}
The default calculation assumes no further substructures within Galactic subhalos. However, several levels of substructures (clumps within clumps) can be taken into account by \clumpy{} for the flux calculation, which would boost the signal further. This is tested in model VAR5, where we assume self-similarity, with respect to the model LOW, between subhalos and sub-subhalos (following what has been shown by both the Aquarius and Via Lactea~II simulations \cite{2008Natur.456...73S,2008ApJ...686..262K}), i.e. $f_{\rm subsub}=0.19$, E-AQ for the spatial distribution of sub-subhalos within their host, and SP for the $c(m)$ parametrisation. 

\end{enumerate}

\begin{table}[t!]
\resizebox{\textwidth}{!}{%
\centering
  \begin{tabular}{| c c | c | c | c | c | c | c | c | c |c |c |c |}
    \hline
	&& \multirow{2}{*}{Model}& \multirow{2}{*}{VAR0} & \cellcolor{modelref!60} & \multirow{2}{*}{VAR1} & \multirow{2}{*}{VAR2} & \multirow{2}{*}{VAR3}& \multirow{2}{*}{VAR4}& \multirow{2}{*}{VAR5} & \multirow{2}{*}{VAR6a}&  \multirow{2}{*}{VAR6b}&   \cellcolor{firebrick!60}  \\
	&& 	  	 					& & \multirow{-2}{*}{\cellcolor{modelref!60}LOW}		& &  	 	& 	 		& 	 		& 	 & 	&	&  \multirow{-2}{*}{\cellcolor{firebrick!60}{HIGH}}	 \\\hline
 \parbox[c]{0mm}{\multirow{7}{*}{\rotatebox[origin=c]{90}{Varied }}} &	\parbox[c]{4mm}{\multirow{7}{*}{\rotatebox[origin=c]{90}{parameters }}} 
     & inner profile			 & \cellcolor[gray]{0.8}NFW  & E	& E	&  E 		& E  		& E 		& E 	& E	& E  & E 	  \\
    && $\alpham$ 				 & 1.9  & 1.9 		& \cellcolor[gray]{0.8} 2.0		&  1.9 		& 1.9  		& 1.9 		& 1.9 	& 1.9	& 1.9  &1.9 	  \\
 	&& $\sigma_c$ 			     & 0.14 	& 0.14 	& 0.14		& \cellcolor[gray]{0.8}0.24 		& 0.14 		& 0.14 		& 0.14 	& 0.14	& 0.14 & 0.14   \\
 	&& $\overline{\varrho}_{subs}$& {\rm E-AQ} 	& {\rm E-AQ} 	&  {\rm E-AQ} 	& {\rm E-AQ}	&  \cellcolor[gray]{0.8} {\rm M-VLII}		& {\rm E-AQ} 	& {\rm E-AQ} &{\rm E-AQ} & {\rm E-AQ} 	& {\rm M-VLII} \\
 	&& $N_{\text{calib}}$ 		 & 150	 	& 150	 	& 150	 	& 150	 	& 150	 	& \cellcolor[gray]{0.8}300	 	&  150 & 150 & 150	 	&300   \\
 	&& sub-subhalos? 			 & no	 	& no	 	& no	 	& no	 	& no	 	& 	no 	& \cellcolor[gray]{0.8}yes & no  & no	 	& no   \\
 	&& $c(m)$ 				 & {\rm SP}  & {\rm SP}  & {\rm SP} 	& {\rm SP} 	& {\rm SP} 	& {\rm SP} 	& {\rm SP} & \cellcolor[gray]{0.8} {\rm Molin\'e}& \cellcolor[gray]{0.8} {\rm P-VLII} 	& {\rm P-VLII}  \\
     \hline\hline 
      \parbox[c]{0mm}{\multirow{8}{*}{\rotatebox[origin=c]{90}{Derived }}} &	\parbox[c]{4mm}{\multirow{8}{*}{\rotatebox[origin=c]{90}{parameters }}} 
        & $N_{\rm tot}\; (\times 10^{14})$		& 6.1	& 6.1	& 150 & 6.1		& 6.1		& 12 & 6.1& 6.1 & 6.1 & 12  \\
    \multicolumn{2}{|c|}{}   & $\fsub \; [\%]$ 	& 19 	& 19 	& 49		& 19		& 19			& 38	 & 19& 19	& 19	& 38  \\
     \multicolumn{2}{|c|}{}  & $\fsub(\Rsol)\;[\%]$		& $0.30$	& $0.30$	& $0.77$		& 	$0.30$	& $0.47$		& 	$0.59$ & $0.30$& 0.30  & $0.30$		& $0.93$  \\[0.15cm]
	\multicolumn{2}{|c|}{} & $\widetilde{D}_{\rm obs}^{\star}\; [\rm kpc]$			&  $22^{+32}_{-16}$	& $19^{+27}_{-14}$	& $13^{+27}_{-10}$	& $21^{+33}_{-15}$ &  $20^{+22}_{-15}$	& $17^{+26}_{-13}$ 	& $21^{+30}_{-14}$&  $8^{+18}_{-6}$	& $9^{+14}_{-6}$		&  $8^{+11}_{-6}$  \\[0.25cm]
	\multicolumn{2}{|c|}{} & $\log_{10}(\widetilde{m}_{\rm vir}^{\star}/\Msol)$			&  $9.0^{+0.8}_{-1.4}$	& $8.8^{+0.8}_{-1.4}$	& $8.5^{+0.9}_{-1.5}$	& $8.9^{+0.8}_{-1.4}$ &  $9.0^{+0.7}_{-1.3}$	& $8.9^{+0.9}_{-1.4}$ 	& $9.0^{+0.7}_{-1.4}$& $7.9^{+1.4}_{-1.6}$	& $7.9^{+1.4}_{-1.5}$		& $8.2^{+1.2}_{-1.5}$  \\[0.25cm]

	\multicolumn{2}{|c|}{} & $\log_{10}\left(\frac{\widetilde{J}^{\star}}{\junits}\right)$			& $19.9^{+0.4}_{-0.3}$ & $20.0^{+0.5}_{-0.3}$	& $20.0^{+0.4}_{-0.3}$	& $20.0^{+0.4}_{-0.3}$	& $20.1^{+0.4}_{-0.3}$	& $20.2^{+0.4}_{-0.3}$ 	& $20.3^{+0.5}_{-0.3}$& $20.3^{+0.5}_{-0.4}$	& $20.4^{+0.5}_{-0.3}$	& $20.8^{+0.5}_{-0.4}$  \\ [0.15cm]

\hline
  \end{tabular}
} 
\caption{Parameters for the different models investigated in this study. The first seven lines correspond (from top to bottom) to: the subhalo density profile, the slope of the subhalo mass distribution, the width of the concentration distribution, the subhalo spatial distribution, the number of objects between $10^{8}$ and $\unit[10^{10}]{\Msol}$, the flag for sub-subhalos, and the mass-concentration relation. The columns are ordered by increasing flux of the brightest object. `NFW' stands for a Navarro-Frenk-White profile and `E' for an Einasto profile with $\alpha_{\rm E}=0.17$. `E-AQ' is the Einasto parametrisation fitted to the substructure distribution in Aquarius simulation \cite{2008MNRAS.391.1685S}, while `M-VLII' corresponds to the Via Lactea~II parametrisation of Madau et al.\!\! \cite{2008ApJ...679.1260M}. The mass concentration relation is `SP' for the S{\'a}nchez-Conde \& Prada parametrisation \cite{2014MNRAS.442.2271S}, or the distance-dependent description by Molin\'e et al.\!\! \cite{2016arXiv160304057M}, respectively by  Pieri et al.\!\! \cite{2011PhRvD..83b3518P}, `P-VLII'. Derived parameters in the six bottom rows are the following: $N_{\rm tot}$ is the total number of subhalos in the MW; $\fsub$ is the global mass fraction contained in subhalos; $\fsub(\Rsol)$ is the mass fraction contained in subhalos at the solar distance from the GC; $\widetilde{D}_{\rm obs}^{\star}$, $\widetilde{m}_{\rm vir}^{\star}$, and $\widetilde{J}^{\star}$ are the median distance from the observer, mass, and {\jfactor} of the brightest subhalo from the 500 realisations of each model.
 }
\label{tab:model-parameters} 
\end{table}


\section{Global properties of the models}
\label{sec:globalprop}
Before performing analyses dedicated to \fermi{} and CTA in \cref{sec:fermi} and \cref{sec:cta}, we describe here the overall behaviour of the models with respect to the substructure properties described above. This is done by the inter-comparison of the various models (\cref{subsec:global}), by comparison to the known MW satellites (\cref{subsec:compare_dSph}), and by comparison of the angular power spectrum to previous calculations (\cref{subsec:powspec}).

\subsection{Impact of the substructure description}
\label{subsec:global}

Each substructure property that is varied according to \cref{tab:model-parameters} essentially impacts the number of substructures and/or their associated $J-$factors. Five hundred skymap realisations for each modelling (LOW, HIGH, VAR0 to VAR6) have been simulated. The bottom half of \cref{tab:model-parameters} gives the global properties of each modelling, averaged over the 500 realisations. Compared to the LOW model, only the calibration number $N_{\rm calib}$ and the slope of the mass distribution affect the {\it total} number of subhalos. However, the number of halos within a given $J$-factor range will depend on all substructure-related properties and one may therefore use histograms of the subhalos $J$-factors to assess the importance of each property.

\begin{figure}[t!]
  \begin{center}
    \includegraphics[width=0.8\textwidth]{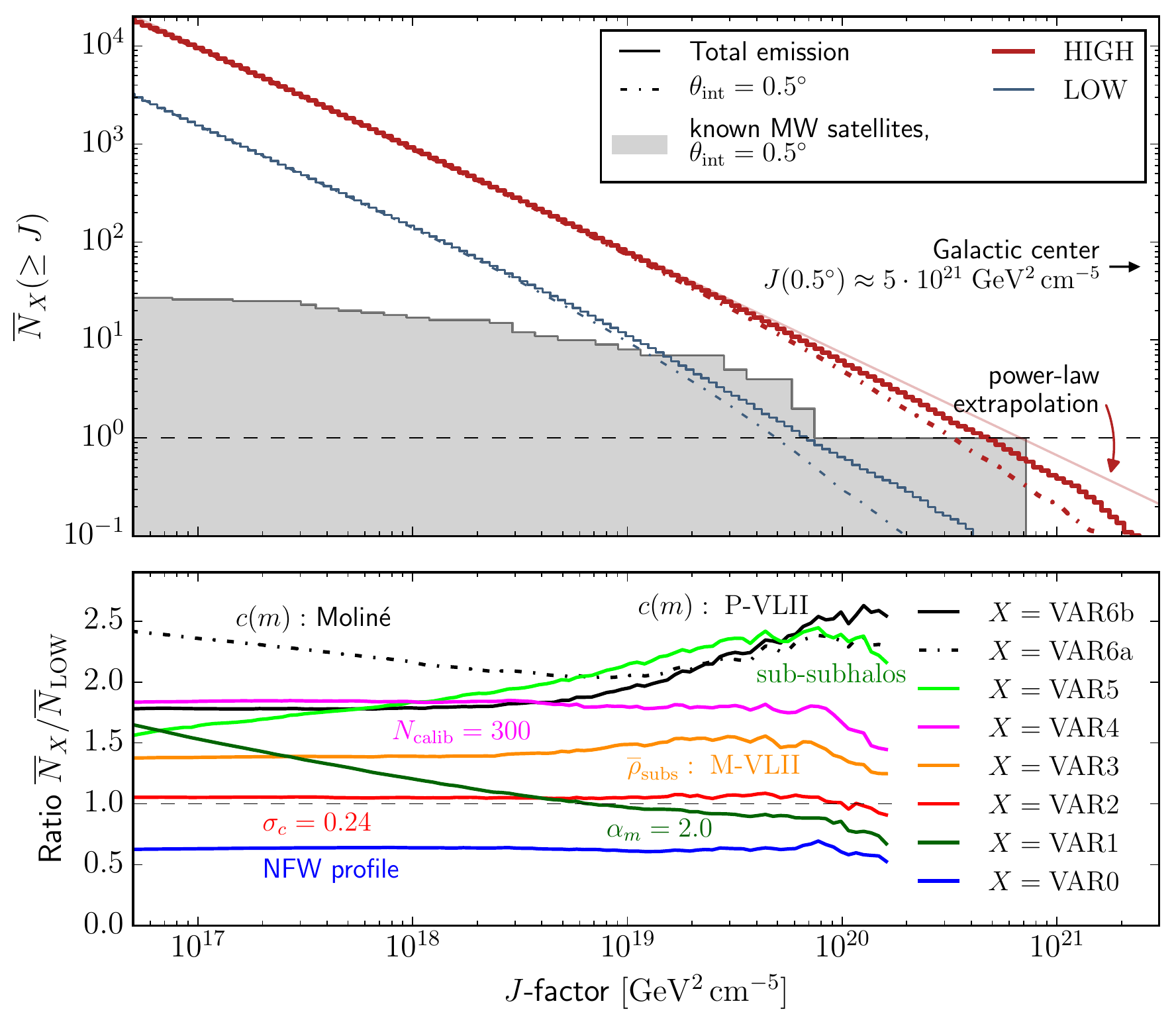}
  \captionof{figure}{
{\em Top:} cumulative source count distribution of Galactic subhalos (full sky, averaged over 500 simulations) for model LOW and HIGH. The solid lines show the distribution of the total {\jfactors}, $J(\theta_{\rm vir})$, the dashed-dotted lines the distribution when only taking into account the emission from the central $0.5\degs$ of the subhalos, $J(0.5\degs)$. The grey-shaded histogram shows the cumulative distribution of all known dSph galaxy objects, including the Large and Small Magellanic cloud, with the values listed in \cref{tab:M_and_J}. {\em Bottom:} ratio of all VAR$i$ models to LOW.}
  \label{fig:Histogram1D_Jfactor-Nsub-fullsky}
  \end{center}
\end{figure}
For a given model $X$, we define $\Nsubmean_{X}(\,>J)$ as the mean number of halos with a \jfactor{} above a certain threshold $J$, averaged over the 500 realisations. This quantity is shown in the top panel of \cref{fig:Histogram1D_Jfactor-Nsub-fullsky} for model LOW (pale blue) and HIGH (red). The behaviour of the histograms is well approximated by a power-law $\propto J^{1-\alpha}$, with $\alpha \sim 2$ over a large range of $J$ values (thin solid red line). We also define the median $J$-factor of the brightest halo in a given model, $\widetilde{J}^{\star}$, and report these values in \cref{tab:model-parameters}, along with the median mass, $\widetilde{m}_{\rm vir}^\star$, and distance from the observer, $\widetilde{D}_{\rm obs}^\star$, of this brightest halo. We refer the reader to \cref{app:meanmedian} for a detailed discussion of the $J^{\star}$ distribution and how $\widetilde{J}^{\star}$ and $\Nsubmean(\,>J)$ are connected.

The ratio $\Nsub_{X}(\,>J)/\Nsub_{\rm LOW}(\,>J)$ is plotted in the bottom panel of \cref{fig:Histogram1D_Jfactor-Nsub-fullsky}, where $X={\rm VAR}i,\;\; i\in[0,6]$:
\begin{itemize}
   \item Changing the subhalo inner profile (VAR0, blue), the substructure spatial distribution (VAR3, orange), the normalization of the mass distribution (VAR4, magenta), or the width of the mass-concentration description (VAR2, red) yields an increase or decrease of the number of clumps, uniformly over the entire $J-$factor range.
   
   \item Changing the subhalo inner profile or changing the subhalo spatial distribution yields a $\sim 40\%$ change compared to the LOW model, while the width of the concentration distribution only affects the result by a few percents. The other substructure properties do not affect the number of subhalos in the same way for low and high $J$-factor values.
   
   \item Including a boost from sub-subhalos (VAR5, light green) or having a distance-dependent concentration prescription (VAR6, black) both produce a similar effect: a larger number of halos in each $J$-factor bin, the effect increasing with increasing $J$ (from $\sim 50-70\%$ to more than a factor 2).
   
   \item  For the highest $J$-decade, both distant-dependent concentration prescriptions VAR6a and  VAR6b result in compatible \jfactors{}.  This is well understood given that the high-$J$ end is populated by subhalos close to us, i.e. also close to the GC with $R\approx \unit[10]{kpc}$. At these distances, P-VLII and Molin\'e concentrations are in agreement on a wide mass range (\cref{fig:cdelta-mdelta}). The prescriptions then differ for less luminous subhalos, which represent subhalos at larger galactocentric distances. For those objects, Molin\'e et al. predict larger concentrations, which results into brighter objects and a steepening of the source count distribution.
   
   \item Steepening the slope of the mass function (VAR1, dark green) increases (decreases) the number of faint (bright) halos, the amplitude of the effect remaining $\lesssim 50\%$ for the $\sim10^3$ most luminous subhalos.
\end{itemize}
\begin{figure}[t!]
  \centering
    \includegraphics[width=0.62\textwidth]{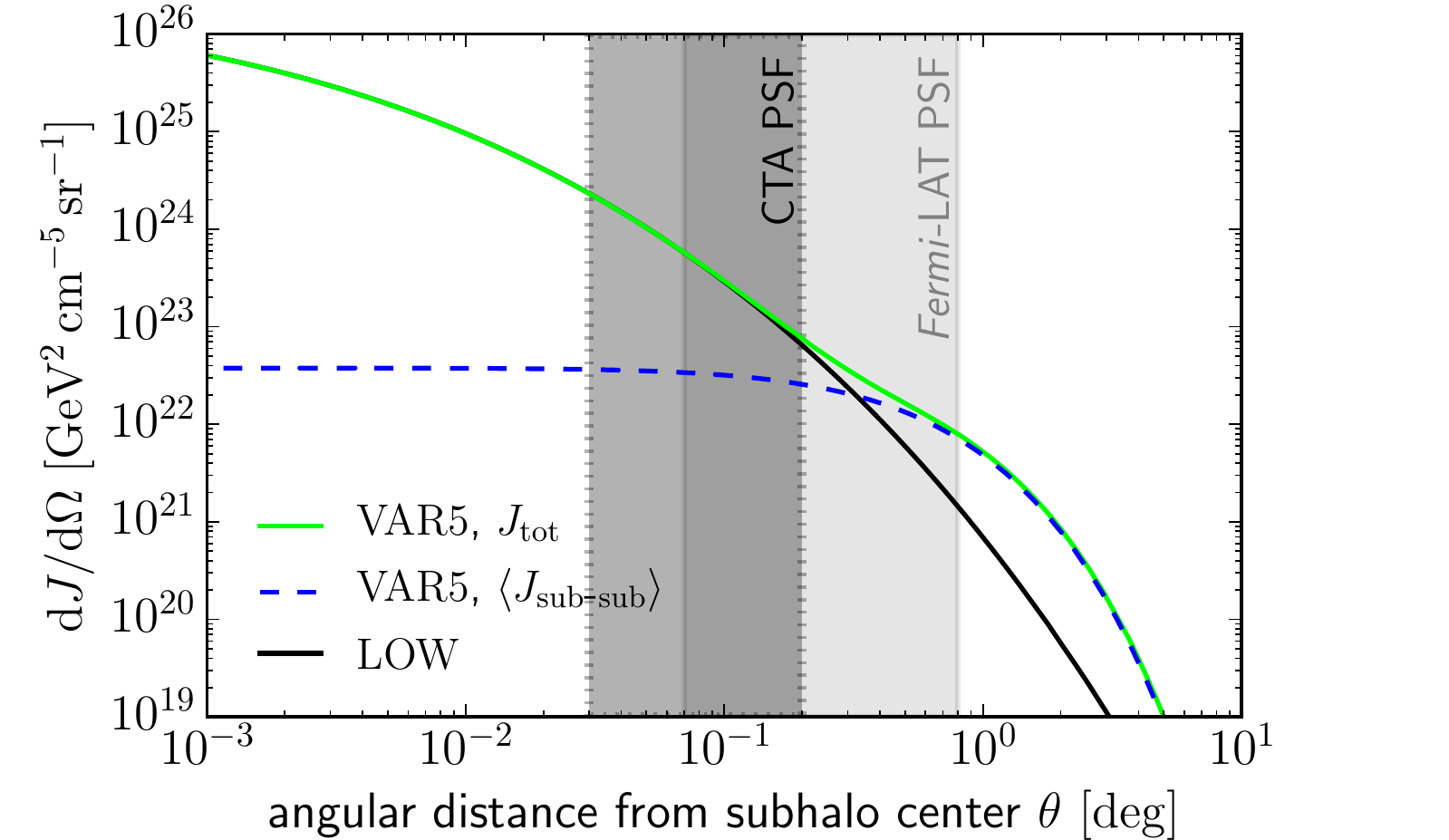}
    \caption{The influence of sub-subhalos as a function of the integration angle. For typical bright detectable DM halos (see \cref{tab:brightest_properties}), the signal with sub-substructures (green solid line) is compared to the signal without them (black solid line). The angular resolutions (68\% containment radius; ``point spread function'', PSF) for CTA \cite{2013APh....43..171B} at $E_{\gamma}\geq\unit[30]{GeV}$ and \fermi{} at $E_{\gamma}\geq\unit[1]{GeV}$  \cite{2009ApJ...697.1071A,2013arXiv1303.3514A} are overlaid (grey shaded areas). For CTA, the sub-substructures contribution is negligible.}
    \label{fig:subsub}
\end{figure}
Including sub-subhalos is only significant in the outskirts of the halos. This is shown in \cref{fig:subsub}, where $\dd J/\dd\Omega$ is plotted as a function from the distance to the centre for the LOW (grey solid line) and VAR5 (dashed blue and green solid lines) models. This finding is in agreement with \cite{2008Natur.456...73S}. For the angular resolution of the background-dominated CTA (grey band), however, they do not play a significant role and we therefore do not include them in the remainder of this study.\footnote{Assuming a spatial dependence of the sub-subhalo concentration---analogous to model VAR5 on the sub-sub level---possibly increases this contribution within the CTA resolution. However, the boost at this second level of substructures is very uncertain and could even be negligible for these objects \cite{2016arXiv160304057M}.}

From the range of substructures properties tested as deviations from the LOW model, we find the distance-dependent concentration parametrisation and boost from sub-subhalos to have the larger effect in terms of the number of halos with the largest $J$-factors. For CTA-like angular resolutions, we conclude that the mass-concentration relation is the most important substructure property to pin down in order to make reliable detectability studies.

 We use the distance-dependent concentration P-VLII by Pieri et al. in the HIGH model and, unless stated otherwise, the remaining of the paper will use the HIGH model as an optimistic template, while LOW remains default.
\begin{figure}[t!]
  \begin{center}
  \renewcommand{\subfigcapskip}{-2.5em}
    \subfigure[Model LOW]{\includegraphics[width=\textwidth]{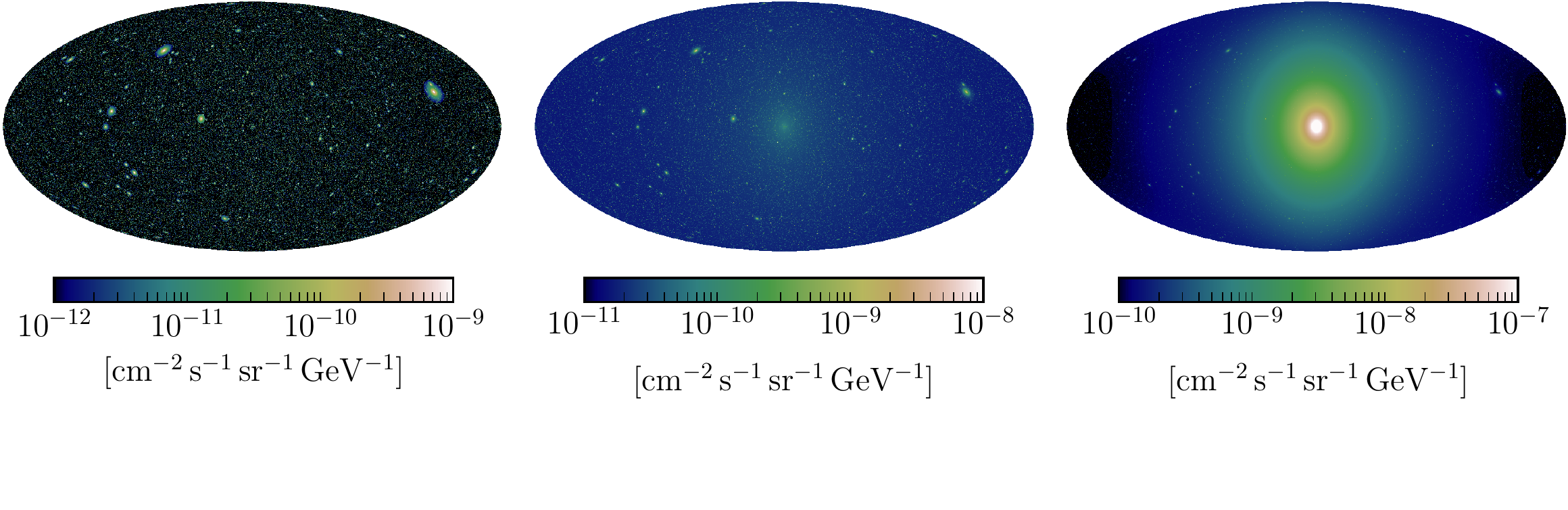}}
    \subfigure[Model HIGH]{\includegraphics[width=\textwidth]{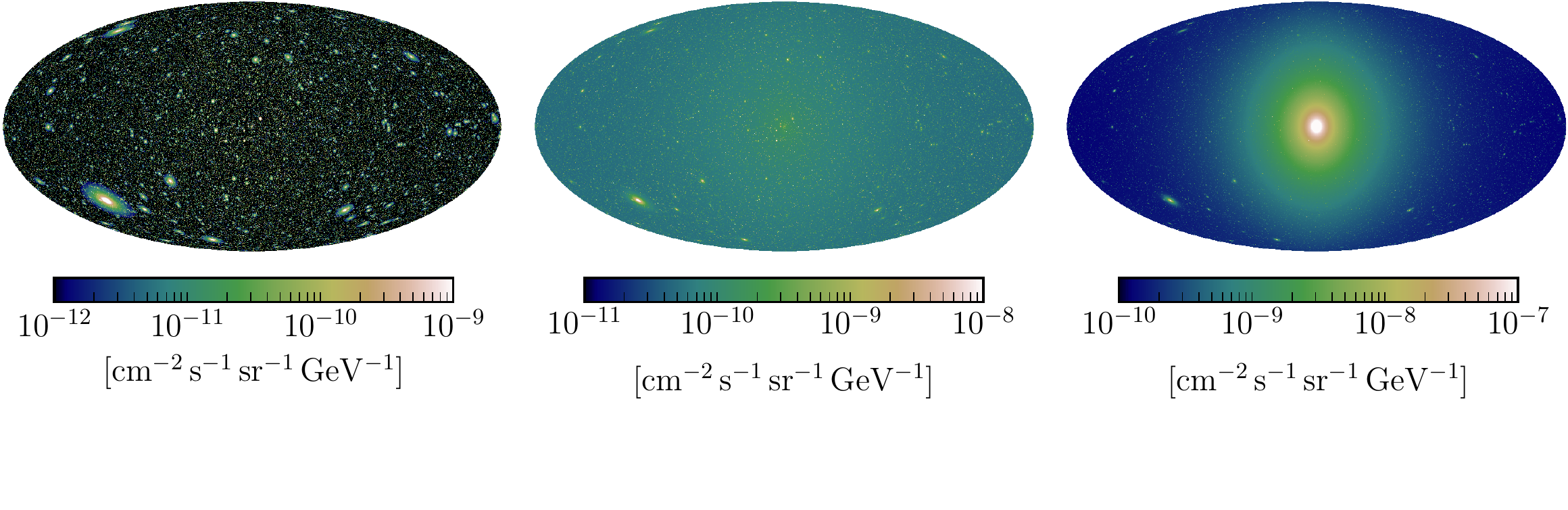}}
  \captionof{figure}{
One statistical realisation of the Galactic differential intensity at $\unit[4]{GeV}$, for the models LOW and HIGH. In the left column, only the flux from the resolved substructures is shown ($\jdrawn$). In the middle column, the flux from all substructures, resolved and unresolved, is shown ($\jdrawn+\jsubs$). In the right column, the total Galactic emission is shown ($\jsmooth$ additionally included). Note the different colour scales between the  columns. The particle physics term is computed from a thermal relic cross-section, $\mchi = \unit[200]{GeV}$, and $\chi\chi\rightarrow b\bar{b}$. 
 The maps are drawn with a \healpix{} resolution $\nside=512$.
}
  \label{fig:Skymaps_compare}
  \end{center}
\end{figure}
For illustration purpose, we display in \cref{fig:Skymaps_compare} the two corresponding differential flux skymaps computed at 4~GeV. The flux is obtained assuming a 200~GeV DM candidate which annihilates exclusively in the $b\bar b$ channel. The left column shows maps of the substructures drawn by \clumpy{}, while the middle column displays the total (resolved+unresolved) substructure contribution. As discussed above, more subhalos are resolved in model HIGH, and the flux of the unresolved component is also higher. The right column displays the total flux in both cases, i.e. including the smooth Galactic halo component, which is the dominant component towards the GC.

\subsection{Comparison of the DM subhalo models to the known Milky Way satellites}
\label{subsec:compare_dSph}

More than twenty dSph galaxies are known to orbit the Milky Way. Formed from the most massive DM subhalos, these objects are prime targets for indirect detection as their DM content, and therefore $J$-factors can been inferred from stellar kinematics, e.g. \cite{2011MNRAS.418.1526C,2015MNRAS.453..849B}. The mass and $J$-factors of these objects are discussed in \cref{app:mvirJ}, \cref{tab:M_and_J} summarises their main properties. These values are used below for a sanity check of our models.

First, the grey shaded area in the top panel of \cref{fig:Histogram1D_Jfactor-Nsub-fullsky} corresponds to the cumulative histogram of $J$-factors built from the known dSph galaxies, plus the SMC, and the LMC. The $J$-factors are reported within an integration angle of $\thetaint=0.5\degs$, as benchmark angular resolution of \fermi{} \cite{2009ApJ...697.1071A,2013arXiv1303.3514A}.  For display purposes, we have used the median values of the $J$-factors only, but we remind the reader that these values may be very uncertain for ultra-faint dSph galaxies. For lower $J$ values, the number of detected dSph galaxies becomes much lower than the number of subhalos measured from the models. This is what one would expect given that the most numerous low-mass halo would not have retained gas and formed stars to become identified as dSph galaxies \cite{2016MNRAS.457.1931S}. The high-$J$ end of the histogram ($J\gtrsim 10^{19}$\;GeV$^2$\;cm$^{-5}$) lies between the LOW and HIGH models but for the very last bin. The latter corresponds to the recently-discovered Triangulum~II galaxy, that we have tentatively analysed here. Note that the authors of \cite{2016MNRAS.461.2914H}, using a non-spherical halo hypothesis have reported a median $J(0.5^\circ)=1.6\times10^{20}$\;GeV$^2$\;cm$^{-5}$ for Triangulum~II, which would bring the last bin down to lie between LOW and HIGH.\footnote{When considering the credible intervals, the Triangulum~II value of \cite{2016MNRAS.461.2914H} and ours are nonetheless compatible within 1$\sigma$.} This behaviour of the MW satellites gives confidence to the fact that the LOW and HIGH models do indeed encompass the uncertainties surrounding the subhalo distribution.

\begin{figure}[t!]
  \begin{center}
    \includegraphics[width=\textwidth]{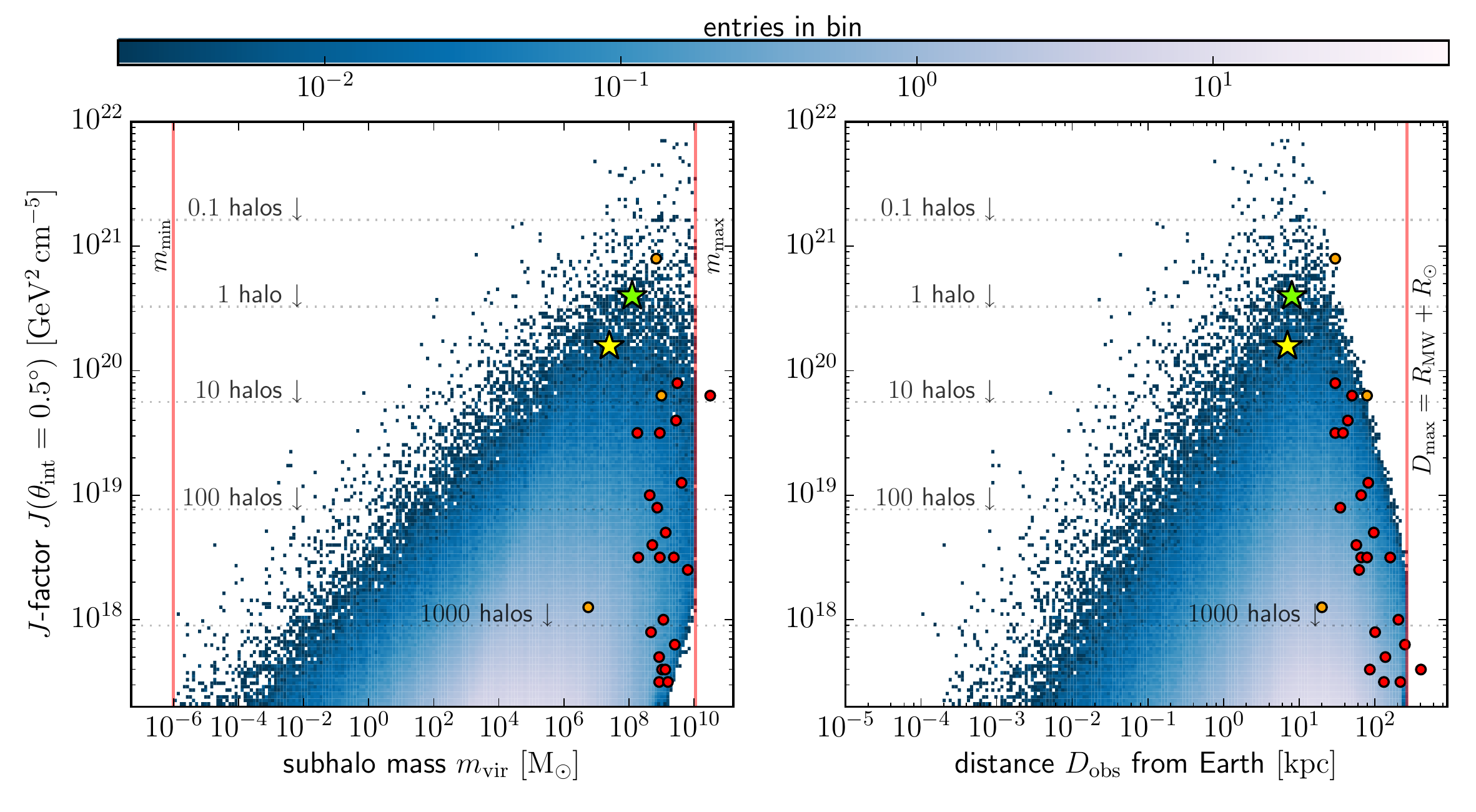}
  \captionof{figure}{
  Relation between the brightness of the subhalos  and their mass (left) or distance to the observer (right). The histograms are shown for the subhalo model HIGH and averaged over 500 simulations. The projection along the vertical axis gives the source count distribution $\langle \dd N/\dd J \rangle$. The dotted lines denote the integrated bins above the respective line, $\Nsubmean_{\rm HIGH}(\geq J)$, identical to what is presented in \cref{fig:Histogram1D_Jfactor-Nsub-fullsky}. The known MW satellites listed in \cref{tab:M_and_J} are displayed as red dots (and orange dots for the dSph discovered most recently). The green asterisk marks the median brightest subhalo expected for {\fermi} and the yellow asterisk for a dark subhalo search with CTA (see \cref{sec:fermi} and \cref{sec:cta}). In this figure, all $J$-values are calculated at $\thetaint=0.5\degs$.}
  \label{fig:Histograms2D_modelRef}
  \end{center}
\end{figure}
Second, the mainly illustrative \cref{fig:Histograms2D_modelRef} gives the location of the detected MW satellites (red and orange dots) in the $J-m_{\rm vir}$ (left) and $J-D_{\rm obs}$ planes (right) on top of the distribution of subhalos of model HIGH. It shows that MW satellite galaxies probe, as expected, the high-mass and high-distance ends of the subhalo population. The horizontal dotted lines indicate how many subhalos are expected in a given realisation, and as above, model HIGH is in excess compared to the known objects. They could be yet-to-discover dSph galaxies or dark halos, and the distribution of entries (shaded blue scale) shows a preference for halos slightly less massive than the known dSph galaxies. Finally, the brightest subhalo expected for \fermi{} and CTA are given by the green and yellow stars, respectively. The \fermi{} subhalo is brighter because of the larger accessible survey area (see \cref{sec:fermi} and \cref{sec:cta} for details).

\subsection{Subhalo angular power spectrum} 
\label{subsec:powspec}

The angular power spectrum (APS) of the subhalo {\gr} sky maps is a powerful tool for DM analyses and provides another cross-check for our analysis. The APS $C_\ell$ of an intensity map $I(\vartheta, \varphi)$ is defined as
\beq
C_\ell = \frac{1}{2\ell + 1} \sum\limits_m |a_{\ell m}|^2 \;,
\label{eq:aps}
\eeq
with $a_{\ell m}$ the coefficients of the intensity map decomposed into spherical harmonics $Y_{\ell m}$,
\beq
I(\vartheta, \varphi) = \sum\limits_{\ell= 0}^{\ell_{\rm max}} \sum\limits_{m= -\ell}^{m= +\ell} a_{\ell m}\, Y_{\ell m}(\vartheta, \varphi).
\label{eq:multipoledecomp}
\eeq

\clumpy{}'s APS calculation relies on the \healpix{}\footnote{\url{http://healpix.sourceforge.net/} package \cite{2005ApJ...622..759G}.} The median and variance of $C_\ell$ are calculated for each of the 500 \jdrawn{} maps produced for all models, with  \healpix{} resolution $\nside=4096$. The power $C_\ell$ caused by $J_{\rm drawn}$ strongly depends on the number of simulated objects, and we show in  the convergence study of \cref{app:APSconvergence} that most of the power at all multipoles is generated by the $\mathcal{O}(100)$ brightest subhalos.\footnote{We showed in \cite{2016CoPhC.200..336B} that the contribution of the unresolved objects, $\jsubs$, to the APS is negligible for $\ell\gtrsim 4$, so this is not discussed further.}
\begin{figure}[t!]
  \begin{center}
    \subfigure{\includegraphics[width=.485\textwidth]{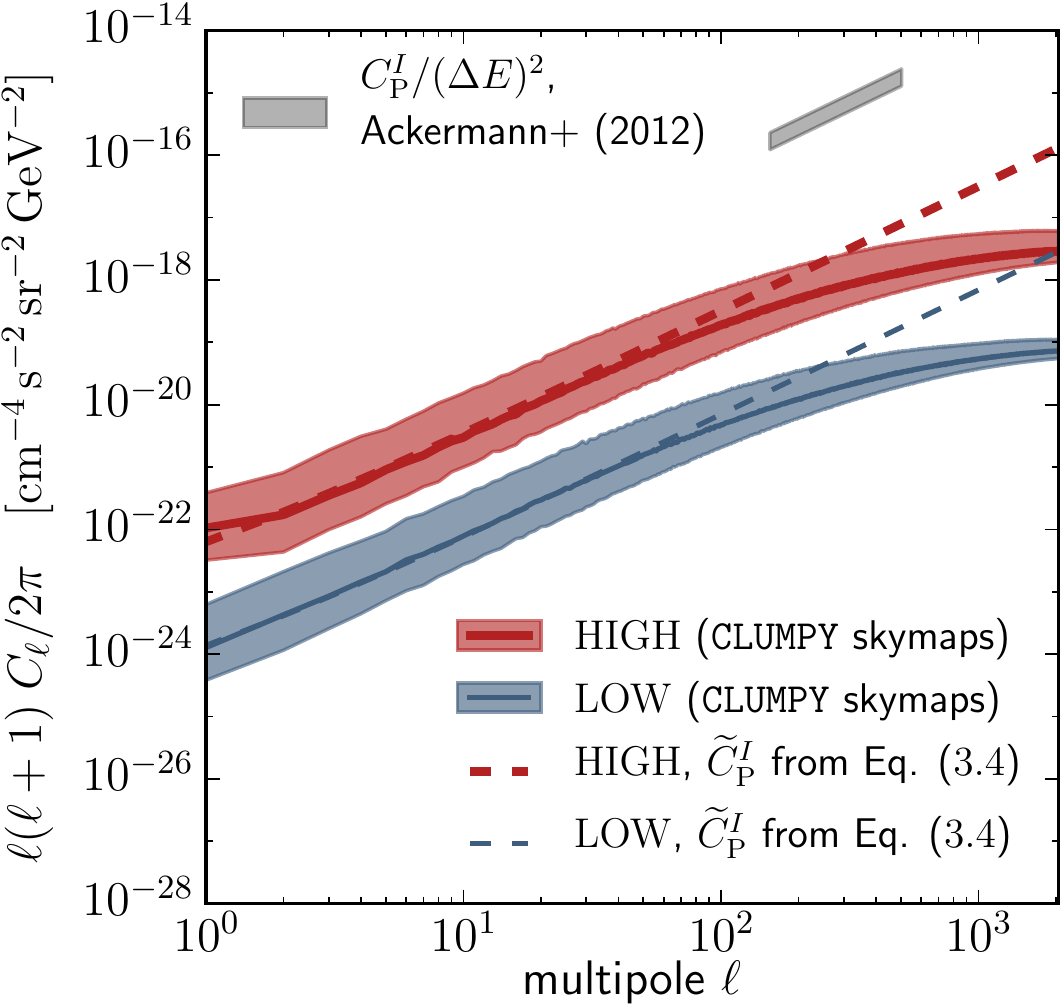}}\hspace{0.022\textwidth}
    \subfigure{\includegraphics[width=.485\textwidth]{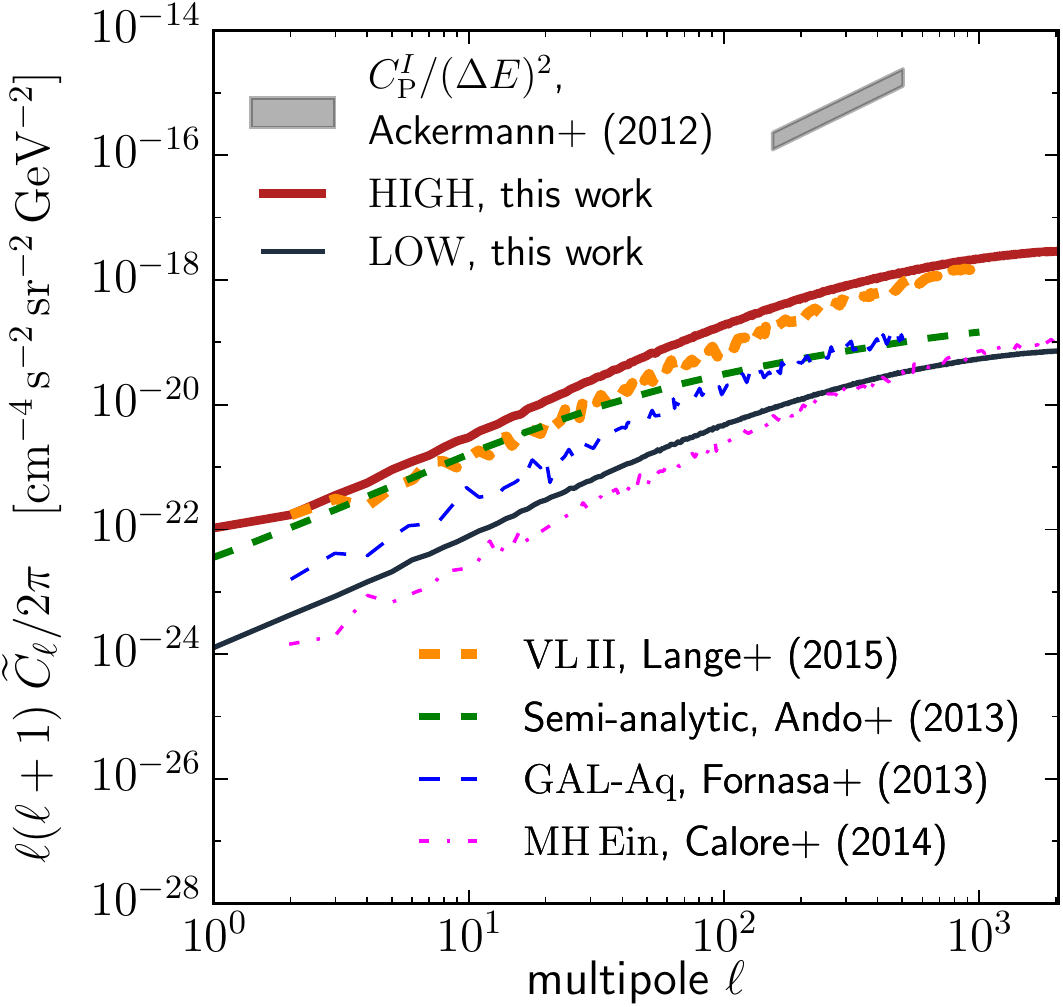}}
  \captionof{figure}{
   {\it Left:}  Differential subhalo intensity APS at $\unit[4]{GeV}$ and its variance on the full sky. The bands show the $68\%$ CI around the median power at each $\ell$ for the models LOW (pale blue) and HIGH (red), obtained from 500 realisations. The dashed lines give the constant $C_{\ell}=\widetilde{C}_P^I$, according to \cref{eq:C1sh_flux_Ando}. At high-$\ell$, the power spectra deviate from the point-source approximation due to the angular extension of the subhalos. The anisotropy level measured in the DGRB, in the multipole range $155\leq \ell \leq 504$, from 22 months of observation with \fermi{}  \cite{2012PhRvD..85h3007A} ($1\sigma$ error band) is shown for comparison (see text for discussion). 
 {\it Right:} Median values for the models LOW (dark grey) and HIGH (red), as on the right figure, compared to other works (dashed colours, see text for details). For the particle physics term, $\mchi = \unit[200]{GeV}$, $\chi\chi\rightarrow b\bar{b}$, and a thermal relic cross-section has been chosen. 
}
  \label{fig:APSi_compare}
  \end{center}
\end{figure}
As already underlined in previous analyses \cite{2009PhRvD..80b3520A,2015MNRAS.447..939L}, the subhalo power spectrum is Poisson-like in the range $\ell \lesssim 500$, and flattens at higher multipoles due to the extended size of the brightest (high-mass) subhalos (see \cref{fig:APSi_compare}). It is possible to connect the subhalo intensity APS to the source count distributions in \cref{fig:Histogram1D_Jfactor-Nsub-fullsky}. For point-like sources and at small angular scales, \cite{2009PhRvD..80b3520A} showed that the angular power is approximately constant (Poisson-like) at all multipoles, and can be calculated by the one-subhalo term:
\beq
C_{\ell}\approx C^{\rm 1 \,sh} \approx \frac{1}{16\pi^2}\;\int\limits_{L_{\rm min}}^{L_{\rm max}}\int\limits_0^{l_{\rm max}}\frac{L^2}{l^2}\,\frac{\dd \overline{n}(l,\,L)}{dL}\,\dd l\, \dd L \equiv C_{\rm P}^I,
\label{eq:C1sh_lum_Ando}
\eeq
where $\overline{n}(l,\,L)$ denotes the number density of the subhalos, averaged over the solid angle $\Omega$. The coordinates $(l,\,\Omega)$ denote spherical coordinates with the observer at $l=0$, and $L$ is the total luminosity of a subhalo. For a point-like DM halo, $L=\frac{\sigmav}{2\,m_{\chi}^2}\frac{\dd \Nsub}{\dd E}\,\lum$, with the luminosity $\lum$ as defined in \cite{2012CoPhC.183..656C}. It is straightforward to show that
\beq
C_{\rm P}^I =  \frac{1}{4\pi} \int_{F_{\rm min}}^{F_{\rm max}}F^2 \frac{\dd \Nsub}{\dd F}\,\dd F,
\label{eq:C1sh_flux_Ando}
\eeq
where $\Nsub$ is the total number of subhalos, and $F$ their flux at the observer. 

\paragraph{APS properties and model comparison.}
To facilitate the comparison to previous works \cite{2015MNRAS.447..939L,2013MNRAS.429.1529F,2014MNRAS.442.1151C}, the APS below is given as differential intensity power at $E=\unit[4]{GeV}$ for a $\mchi = \unit[200]{GeV}$, $\chi\chi\rightarrow b\bar{b}$ annihilation channel with thermal relic cross-section $\sigmav=\unit[3\cdot 10^{-26}]{cm^3\,s^{-1}}$ (same as in \cref{fig:Skymaps_compare}).

The left panel of \cref{fig:APSi_compare} shows the median APS of the LOW and HIGH models and their $68\%$ confidence intervals, CI, along with the median Poisson-like $C_P^I$ given in \cref{eq:C1sh_flux_Ando}, calculated for each model and 500 simulations. Here, the flux $F_{\rm min}$ in \cref{eq:C1sh_flux_Ando} has been set to the flux from the faintest drawn object  in each realisation. As shown in  \cref{app:APSconvergence}, further lowering $F_{\rm min}\rightarrow 0$ adds negligible contribution to the integral. The flux $F_{\rm max}$ in \cref{eq:C1sh_flux_Ando} is $F_{\rm max}=F^{\star}$, the flux from the brightest object in each skymap. The two approaches agree but at high-$\ell$ where the point-like approximation is not valid anymore. Additionally, the medians also differ at low-$\ell$ for the model HIGH, as the spatially isotropic distribution of the objects is violated here (see next paragraph). A main benefit from our approach is to properly propagate the $\ell$-dependent uncertainties on the APS (see left panel), which is also not limited to a single simulation and its limited mass resolution; the cosmic-variance uncertainty is of almost one order of magnitude.

The right panel in \cref{fig:APSi_compare} shows the median $C_{\ell}$ for the models LOW (black) and HIGH (red), which encompass the results based on either the Via Lactea~II \cite{2015MNRAS.447..939L} or Aquarius simulations \cite{2013MNRAS.429.1529F}. The APS from Galactic subhalos obtained by \cite{2013PhRvD..87l3539A} (green dashed line in \cref{fig:APSi_compare}) is also based on the Aquarius simulations.\footnote{The curve for the Galactic APS from \cite{2013PhRvD..87l3539A} has been rescaled for comparison to the other works. We adopted $\dd\Phi^{\rm PP}/\dd E(\unit[4]{GeV},\,\mchi=\unit[200]{GeV},\,\chi\chi\rightarrow b\bar{b})= \unit[5.63\cdot 10^{-32}]{GeV^{-3}\,cm^{3}\,s^{-1}}$, compared to $\Phi^{\rm PP}(E_{\rm min} = \unit[5]{GeV},\,E_{\rm max} = \unit[10]{GeV},\,\mchi=\unit[100]{GeV},\,\chi\chi\rightarrow b\bar{b})= \unit[2.07\cdot 10^{-31}]{GeV^{-2}\,cm^{3}\,s^{-1}}$, thus rescaling by a factor $(0.56/2.07)^2\,\unit{GeV^{-2}}$. However note that the spectral shape for the $\chi\chi\rightarrow b\bar{b}$ channel assumed by \cite{2013PhRvD..87l3539A} slightly differs from ours.} However, they semi-analytically computed the spectrum  from a generalized version of \cref{eq:C1sh_lum_Ando}, additionally taking into account a suppression factor $|\tilde{u}(\ell)|^2$ due to the spatial extension of the subhalos. In the g15784 simulation by \cite{2014MNRAS.442.1151C} (magenta dot-dashed line in \cref{fig:APSi_compare}), only subhalos with masses $>10^{8.6}\,\Msol$ are resolved, which explains the lower overall power. Beyond the validity check, the subhalo APS is an interesting tool for the observational search strategy.  \Cref{fig:APSi_compare} shows that at low multipoles, a small dipole excess ($\ell=1$) is visible for the model HIGH (and none for the model LOW). Therefore, we checked if a spatial bias exists for the location of the brightest clumps. In \cref{fig:Abundance_dir}, we show the probability distribution to find the brightest object at angular distance $\theta$ from the direction of the GC for LOW (pale blue) and HIGH (red). For model LOW, only a marginal spatial preference exists to find the brightest halo in the direction of the GC. This is not the case for model HIGH, where the direction of the GC is clearly preferred.\footnote{Note that we only consider the subhalo emission, not the DM emission from the smooth Galactic halo, which is highly peaked towards the GC. Therefore, this discussion does not apply for the brightness in terms of signal to background ratio.} Cross-checking with model VAR6, we find this change to be mostly due to the distance-dependent concentration of model HIGH: the closer subhalos are to the GC, the more concentrated, hence brighter, they become. We come back later on the fact that the brightest objects are close-by with $D_{\rm obs}\approx \Rsol \approx \mathcal{O}(\unit[10]{kpc})$ from both the observer and the GC.

\paragraph{APS and DM sensitivity.}
The APS can be compared to existing data on the anisotropy in the diffuse \gr{} background (DGRB, see, e.g., \cite{2015PhR...598....1F}). After 22 months of data-taking, \fermi{} reported a significant excess ($>3\sigma$) of \gr{} anisotropy over the photon noise background in the diffuse emission at latitudes $|b|> 30\degs$, in each of four different energy intervals between $\unit[1]{GeV}\leq E \leq \unit[50]{GeV}$, and in the multipole range $155\leq \ell \leq 504$ \cite{2012PhRvD..85h3007A}. We take their {\sc data:cleaned} result, given as appropriately rescaled full-sky equivalent power, in differential form at $\unit[4]{GeV}$. It is shown in \cref{fig:APSi_compare} as a grey-shaded segment vertically extending over the $\pm 1\sigma$ uncertainty range. Accounting for the dominant contribution of distant unresolved blazars \cite{2012PhRvD..86f3004C} and misaligned active galactic nuclei \cite{2014JCAP...11..021D} would lower the differential intensity angular power by a factor of a few, leading to a residual anisotropy that could be attributed to DM. For instance, several authors have used the measurement from \cite{2012PhRvD..85h3007A} and the blazar contribution from \cite{2012PhRvD..86f3004C,2014JCAP...11..021D} to derive upper limits on the  relic cross-section $\sigmav$ \cite{2013PhRvD..87l3539A,2014NIMPA.742..149G,2015MNRAS.447..939L,2015PNAS..11212272S}.

A full analysis of the constraints set by the APS data is beyond the scope of this analysis. Nevertheless, thanks to our self-consistent derivation of the APS and source count distribution of Galactic DM subhalos, we may comment on the sensitivity of the former, and the relative merit of both approaches. Using the result of \cite{2012PhRvD..85h3007A}, based on 22 months of \fermi{} data with 1FGL point sources removed, Ando and Komatsu \cite{2013PhRvD..87l3539A} typically find upper limits of $\sigmav\sim 10^{-24}\,\mathrm{cm^3\; s^{-1}}$ for $\mchi = \unit[200]{GeV}$ and $b\bar{b}$ (see their figure~16). An order-of-magnitude calculation performed from \cref{fig:APSi_compare} (we recall that $C_{\ell}\propto \sigmav^2$) for models LOW and HIGH encompasses this value, with model HIGH yielding the lowest limit. This work also shows that APS-derived limits are affected by a cosmic-variance uncertainty of almost one order of magnitude. Alternatively, looking for dark clumps in the 3FGL catalogue (48 months of data), Schoonenberg et al. \cite{2016JCAP...05..028S} find $\sigmav\sim 10^{-25}\,\mathrm{cm^3\; s^{-1}}$ (see their figure 7). Their underlying subhalo model is in line with our model HIGH prediction (see \cref{fig:Histogram1D_Jfactor-Nsub-Fermi}), and also suffers from a similar variance uncertainty. From these crude considerations, the APS approach appears to be somewhat less favourable than dark clump searches. Considering many effects such as data homogeneity, masking effects, etc., the comprehensive study of \cite{2015MNRAS.447..939L} finds similar results. In the context of a large-sky survey with the CTA instrument, we present the CTA sensitivity to dark clumps in \cref{sec:cta}. Comparing those results to anisotropy searches with CTA, as suggested by \cite{2014JCAP...01..049R}, is left for future work.


\section{Revisiting dark halo searches with \fermi{}}
\label{sec:fermi}

\begin{figure}[t!]
\captionsetup{format=plain}
\begin{minipage}[t]{.485\textwidth}
  \begin{flushleft}
    \includegraphics[width=\textwidth]{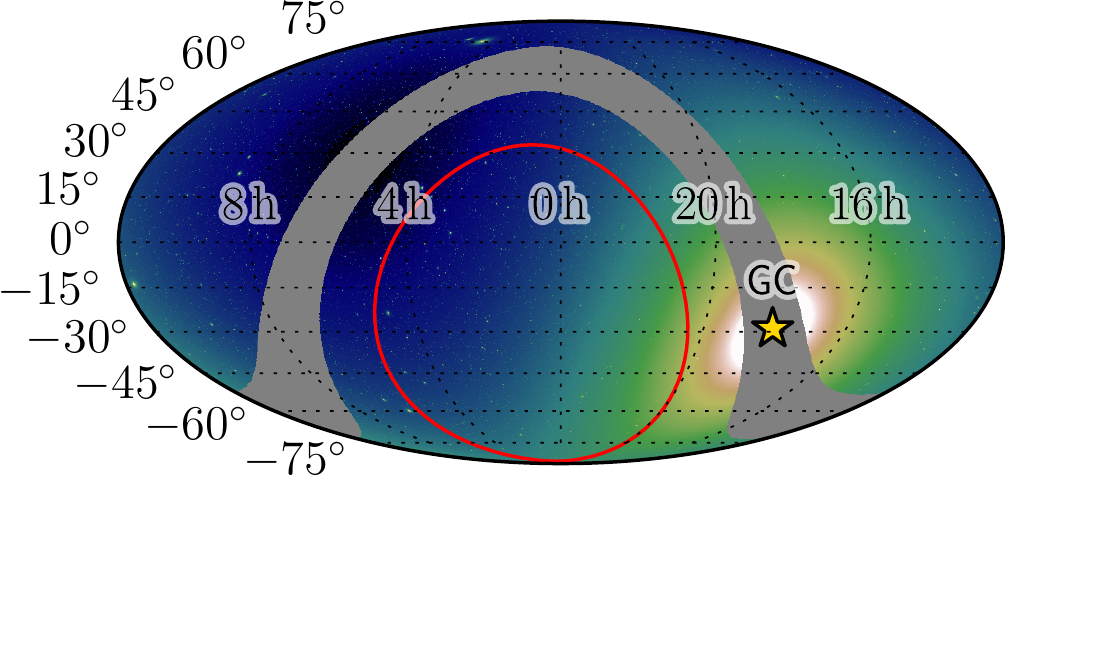}
  \captionof{figure}{ 
Masked skymap (equatorial coordinates)  for our {\fermi} setup (grey band around the Galactic plane) and the assumed CTA survey field in this study (within the red line). The colormap in the back is the same as in \cref{fig:Skymaps_compare} (model LOW, total emission).
}
  \label{fig:ROIskymap}
  \end{flushleft}
\end{minipage} 
\hspace{.02\textwidth}
\begin{minipage}[t]{.485\textwidth}
  \begin{flushright}
     \includegraphics[width=.98\textwidth]{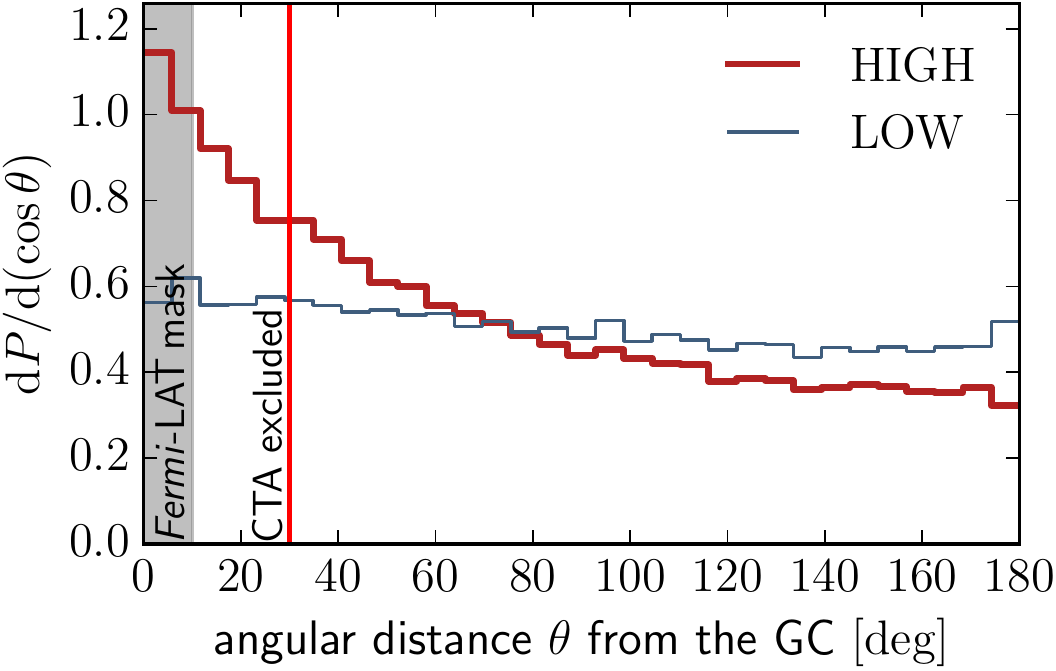}
  \captionof{figure}{
Probability to find the brightest subhalo at the angular distance $\theta$ from the GC. The vertical lines indicate the sky cuts of each instrument (shown in \cref{fig:ROIskymap}). 
}
  \label{fig:Abundance_dir}
\end{flushright}
\end{minipage}
\end{figure}
The subhalo detection prospects for \fermi{} have been investigated several times \cite{2003MNRAS.345.1313S,2008ApJ...686..262K,2009Sci...325..970K,2010ApJ...718..899A,2011PhRvD..83b3518P,2014PhRvD..89a6014B,2015JCAP...12..035B,2016JCAP...05..028S}. This short section aims at comparing our work to the recently published results of \cite{2015JCAP...12..035B,2016JCAP...05..028S} and  at commenting on some differences compared to a CTA-like survey (presented in \cref{sec:cta}). To do so, we now move from the full sky approach of the previous section to a setup tailored to the \fermi{} experiment. Following \cite{2016JCAP...05..028S}, we (i) investigate the subhalo population outside the Galactic plane at $|b|>10\degs$, and we adopt the same region of interest, as shown in \cref{fig:ROIskymap}; (ii) limit the \jfactor{} integration angle to $\thetaint=0.8\degs$, as done by \cite{2016JCAP...05..028S}, describing \fermi{}'s $68\%$ containment radius at $\unit[1]{GeV}$.

\begin{figure}[t!]
\captionsetup{format=plain}
\centering
    \includegraphics[width=0.8\textwidth]{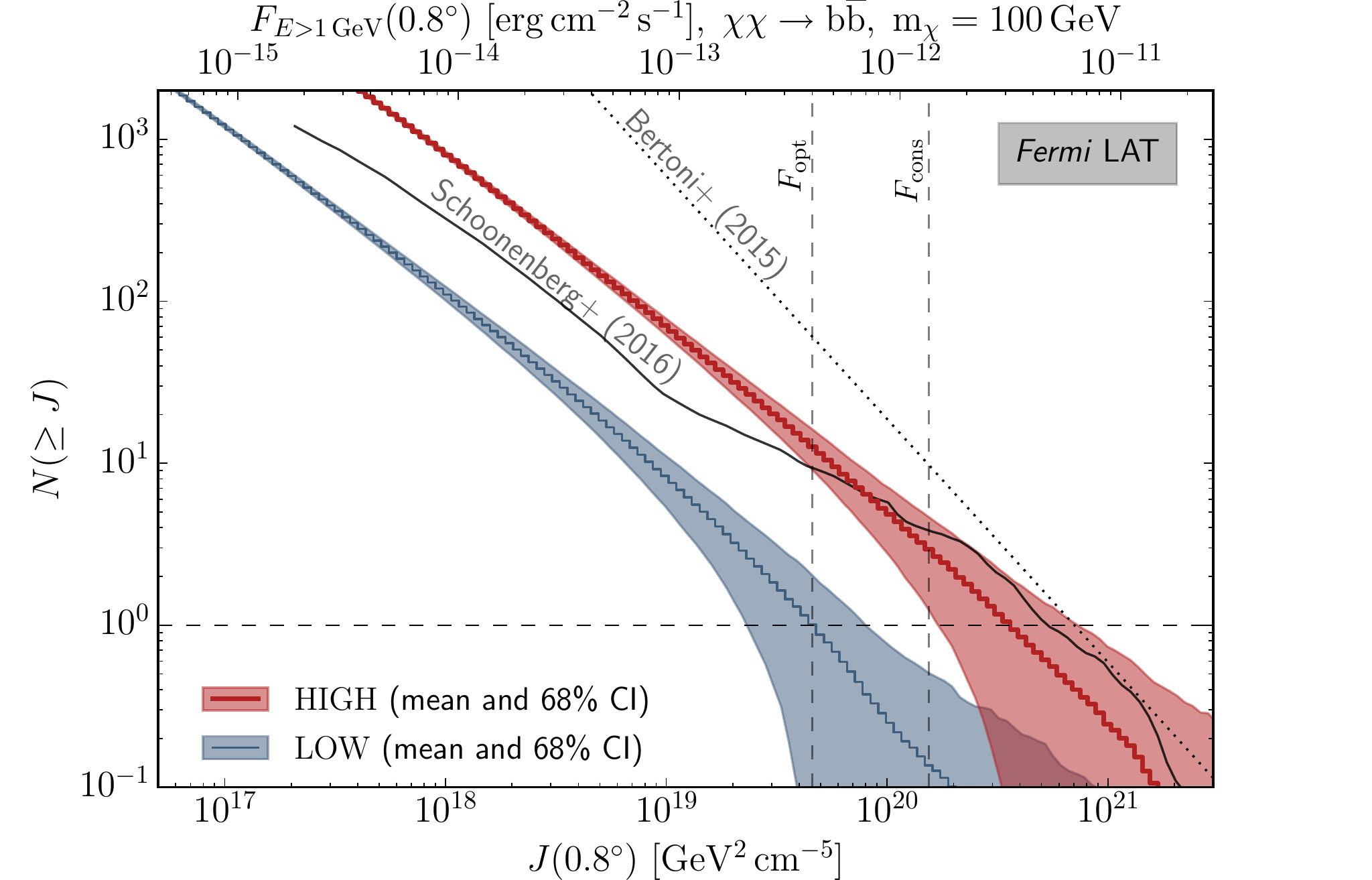}
  \caption{
Cumulative source count distribution of DM subhalos for the {\fermi} setup. The coloured bands denote the $1\sigma$ standard deviation around the mean $\Nsubmean$ from the 500 simulations. The lower $x-$axis gives {\jfactors} and the upper $x-$axis the corresponding flux for a given particle physics model, using $\sigmav = 3\cdot10^{-26}\,\mathrm{cm^3\; s^{-1}}$. The Bertoni et al.\!\! \cite{2015JCAP...12..035B} (dotted line) and Schoonenberg et al.\!\! \cite{2016JCAP...05..028S} results (solid line) are also displayed. The vertical dashed lines show the conservative and optimistic detection thresholds chosen by \cite{2016JCAP...05..028S}. Taking into account an up-to-date LAT sensitivity (see text), these thresholds would move by an approximate factor 2 to the left.
}
  \label{fig:Histogram1D_Jfactor-Nsub-Fermi}
\end{figure}

\Cref{fig:Histogram1D_Jfactor-Nsub-Fermi} shows the cumulative source count distributions for this setup and the subhalo models LOW (pale blue) and HIGH (red). The lower $x$-axis presents the subhalo source count distribution in terms of the particle-physics independent \jfactor{}, while the upper $x$-axis gives the corresponding integrated energy flux distribution above $\unit[1]{GeV}$ (for $m_{\chi}=\unit[100]{GeV}$, thermal annihilation cross-section, and pure annihilation into bottom quarks). This allows us to directly compare our modelling to the findings of Schoonenberg et al.\!\!  \cite{2016JCAP...05..028S} (black solid line)\footnote{We added their distributions of point-like and extended halos which they discuss separately. Note that they use a simplified approach to calculate the {\jfactors}. Performing the full line-of-sight integration in \clumpy{}, we do not find a strict difference between point-like and extended objects.} and, with some limitations, Bertoni et al.\!\! \cite{2015JCAP...12..035B} (black dotted line).
Ref.~\cite{2015JCAP...12..035B} considers only subhalos at $|b|>20\degs$, while we use $|b|>10\degs$, meaning that, compared to our setup, the dotted line should be even higher than that shown in \cref{fig:Histogram1D_Jfactor-Nsub-Fermi}.\footnote{The integration angle used in \cite{2015JCAP...12..035B} is not specified and the comparison is only valid provided that their calculation of the DM spectra is consistent with \cite{2011JCAP...03..051C}.} For low $J$ values, HIGH and LOW nicely encompass the Schoonenberg et al.\!\! results. For the largest {\jfactors}, the HIGH model is consistent with \cite{2016JCAP...05..028S} within uncertainties,\footnote{Ref.~\cite{2016JCAP...05..028S} finds a similar sample variance of $\Nsub$ (68\% CI), which is not shown in \cref{fig:Histogram1D_Jfactor-Nsub-Fermi}.} and in some tension with \cite{2015JCAP...12..035B}.

The authors of \cite{2016JCAP...05..028S} estimate the number of detectable subhalos for a specific DM particle physics model from the \fermi{} detection threshold in the 3FGL, for sources with a similar, relatively hard spectral shape as expected from DM annihilations. In particular considering the $\chi\chi\rightarrow b \bar{b}$ channel,  they assume a conservative detection threshold $F_{\rm cons,\,b\overline{b}}=\unit[1.35\cdot 10^{-12}]{erg\,cm^{-2}\,s^{-1}}$, and an optimistic detection threshold $F_{\rm opt,\,b\overline{b}}=\unit[4.0\cdot 10^{-13}]{erg\,cm^{-2}\,s^{-1}}$ (both fluxes integrated above $\unit[1]{GeV}$). For $F_{\rm opt,\,b\overline{b}}$ and the particle physics model chosen in \cref{fig:Histogram1D_Jfactor-Nsub-Fermi}, model HIGH predicts $13\pm 4$ detectable subhalos to be present in the 3FGL, while still $1\pm 1$ halo could be found for the conservative model LOW; for $F_{\rm cons,\,b\overline{b}}$, model HIGH (LOW) yields $3\pm 2$ ($0.1\pm 0.4$) halos. We adopted the above flux thresholds  and quote the number of detectable clumps based on the 3FGL to ease comparison with the results found by \cite{2016JCAP...05..028S}. However, after eight years of operation, and with the improved Pass8 event reconstruction \cite{2013arXiv1303.3514A}, the LAT sensitivity to faint sources has significantly improved since the 3FGL release. For the background-dominated regime between $\unit[1]{GeV}$ and $\unit[10]{GeV}$, the double exposure yields a factor $\sqrt{2}$ in improved sensitivity, and the Pass8 reconstruction additionally improves the sensitivity by about $30\%$
compared to the Pass7 analysis chain.\footnote{\url{http://www.slac.stanford.edu/exp/glast/groups/canda/lat_Performance.htm}} With this, we approximate the available flux thresholds above $\unit[1]{GeV}$ after eight years observations with Pass8 reconstruction to be $F_{\rm 8 years}  \approx 0.5\, F_{\rm 3FGL}$. Correspondingly, the number detectable clumps increases by factor $\sim 0.5^{1-\alpha}\sim 2$, with $1-\alpha\approx -1$ the slope of the source count distribution in \cref{fig:Histogram1D_Jfactor-Nsub-Fermi}.
 Therefore, assuming $\sigmav\approx\unit[3\cdot10^{-26}]{cm^3\,s^{-1}}$ and for $m_{\chi}\lesssim\unit[100]{GeV}$, DM subhalos might already have  been detected by \fermi{}, even under conservative assumptions about the subhalo model and the detection threshold. Conversely, we remark that an exclusion of the DM hypothesis for most of the unidentified 3FGL objects is consistent with \fermi{}'s finding from stacked dSph galaxy observations,  $\sigmav<\unit[3\cdot10^{-26}]{cm^3\,s^{-1}}$ for $m_{\chi}<\unit[100]{GeV}$ \cite{2015PhRvL.115w1301A}, and these limits are now more stringent after four more years data taking.

We remark that the \fermi{} angular resolution strongly improves at energies above $\unit[1]{GeV}$, reaching a $68\%$ containment radius of $0.2\degs$ 
at $\sim\unit[10]{GeV}$, and $0.1\degs$ above $\sim\unit[100]{GeV}$ \cite{2009ApJ...697.1071A,2013arXiv1303.3514A}. Therefore, we stress that the choice of $\thetaint = 0.8\degs$ in this paragraph is not a description of the LAT performance, but serves for comparison of the source count distribution from \cite{2016JCAP...05..028S}.

For the plausibility of  dark subhalo searches, it is of interest to investigate the physical properties (mass, distance, angular extension) of the brightest subhalo. The median properties of the brightest object within the masked \fermi{} sky and for the model HIGH are presented in \cref{tab:brightest_properties}. The properties of the brightest subhalo  may depend on the angular resolution of the instrument and we find that the object with the largest overall \jfactor{}, $J(\theta_{\rm vir})$, is not necessarily the same as the object with the largest  \jfactor{} within some $\thetaint < \theta_{\rm vir}$. This is accounted for in  \cref{tab:brightest_properties}, and the brightest object is listed separately within the integration angles $\theta_{\rm vir}$, $0.8\degs$ and $0.1\degs$. However, this differentiation results into barely different objects within the resolutions of \fermi{} (in contrast to CTA, which is discussed later). For \fermi{}, the subhalo with the median largest \jfactor{} is typically $\unit[8^{+11}_{-6}]{kpc}$ away, and has a mass between  $\sim\unit[10^7]{\Msol}-\unit[10^9]{\Msol}$. 

We finally remark that in case of  potential DM subhalo candidates, alternative origins for the signal (VHE blazars, milli-second pulsars)  have to be ruled out. Various approaches for disentangling DM and astrophysical sources are presented in \cite{2012PhRvD..86d3504B,2012JCAP...11..050Z,2016JCAP...05..028S,2016ApJ...825...69M}. Especially CTA will be a suitable instrument to perform dedicated follow-up observations of source candidates, and to resolve different origins by the temporal, spectral and spatial morphology of a candidate. Besides, CTA will be able to perform a large-area survey on its own, entering an energy regime beyond the accessibility of \fermi{}. In the following, we will now investigate whether a CTA survey can be used to search for DM subhalos.


\section{CTA sensitivity to dark clumps}
\label{sec:cta}

This section presents the CTA sensitivity to dark clumps (for the sake of readability, all the technical details are postponed to \cref{app:CTAanalysis_details}). We first introduce the salient features of CTA and CTA's extragalactic survey  (\cref{sec:nominal_sensitivity}), present the instrument background (\cref{sec:CTAbackground}), the characteristics of the brightest clump for CTA (\cref{sec:brightest_CTA}), the likelihood and statistics used to draw our CIs (\cref{sec:logL_TS}), and then the ensuing sensitivity for CTA to these objects (\cref{sec:CTA_results}). We underline that the knowledge of the probability distribution of the brightest clumps is mandatory to set robust constraints on DM detection. As discussed and shown in \cref{sec:fat_tails}, this distribution has a long tail, so that the construction of credible intervals (CI) must rely on large samples to be correctly defined. For this reason, all the results presented in this section are based on $10^4$ \clumpy{} runs of the models LOW and HIGH.

\subsection{Observation setup and nominal sensitivity}
\label{sec:nominal_sensitivity}
CTA is the next-generation ground-based {\gr} observatory, using the technique of imaging atmospheric Cherenkov radiation (`Imaging Atmospheric Cerenkov Telescope', IACT). It will feature an unprecedented resolution in energy and angular separation for {\grs} in the range between $30\,\mathrm{GeV}\lesssim E_{\gamma}\lesssim 200\,\mathrm{TeV}$, and, over the whole energy range, an effective collection area of about an order of magnitude larger  than current IACT \cite{2013APh....43....3A}. Having a large field-of-view, CTA will be the first IACT to efficiently perform large-area surveys in VHE (very high energy, $\gtrsim \unit[100]{GeV}$) {\grs}. In particular, it plans to perform an extragalactic survey with an approximately uniform exposure over $25\%$
of the sky \cite{2013APh....43..317D}. This survey aims at an unbiased population study of extragalactic sources, primarily to search for `dark particle accelerators'  without any counterparts at other wavelengths. Analogously, it can be used for a search for Galactic DM annihilation in dark subhalos.

CTA will consist of two arrays, one in the northern and one in the southern hemisphere. We assume, for simplicity, that most of the survey will be performed by the southern array, in circular region around the Galactic south pole ($b<-30\degs$).\footnote{Depending on the final location/latitude of CTA (southern site), a substantial part of the  extragalactic survey area has to be covered by CTA North. Moreover, some of the area we selected may be unreachable for CTA (only observable at elevations $<40\degs$), or the survey may reach smaller Galactic latitudes $|b|\lesssim 10\degs$.} This choice covers the projected $f_{\rm sky}=25\%$, as illustrated in \cref{fig:ROIskymap}, and excludes the area close to the GC (see \cref{fig:Abundance_dir}), where, for the model HIGH,  bright subhalos are found with the highest probability; therefore, scanning lower Galactic latitudes or the combination with a Galactic plane survey could in principle increase the sensitivity to dark halos. Around 400 to 600 hours will be available for the extragalactic survey with CTA \cite{2013APh....43..317D}. With an uniform distribution of this observing time over the whole survey area, a sensitivity to fluxes of about $\unit[2.5\cdot10^{-12}]{cm^{-2}\,s^{-1}}$ to $\unit[1~\cdot~10^{-11}]{cm^{-2}\,s^{-1}}$ above $100\,\mathrm{GeV}$ (approximately corresponding to $0.5\%-2\%$ the flux of the Crab nebula \cite{2015arXiv150806442K}) can be reached for a Crab-nebula-like energy spectrum \cite{2013APh....43..317D}. We discuss further the uniform observation strategy in the context of dark subhalos in \cref{app:obs_streg}.

\subsection{Diffuse {\grs} and residual background}
\label{sec:CTAbackground}

\begin{figure}[t]
\centering
\includegraphics[width=0.62\textwidth]{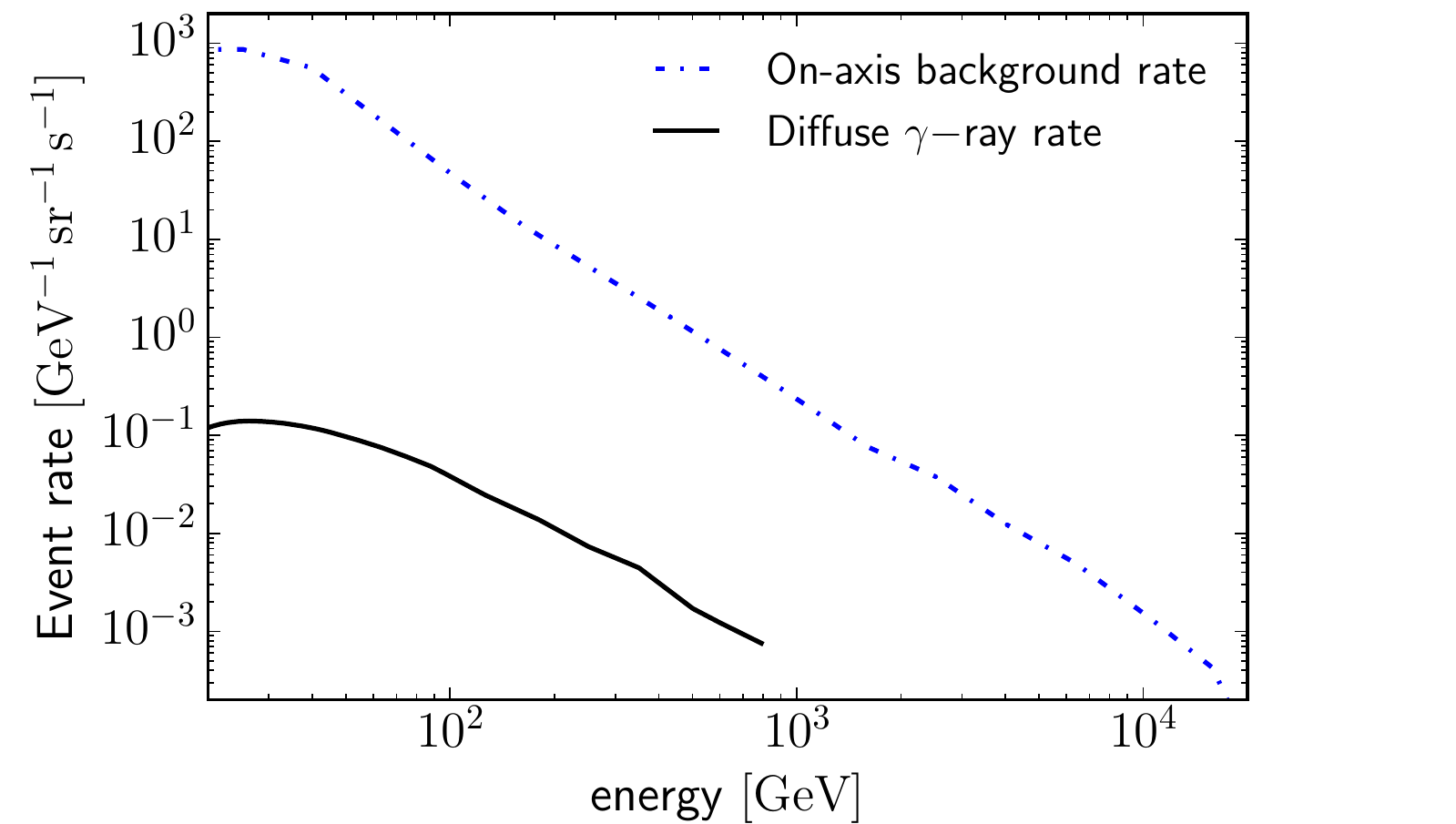}
\caption{{
Diffuse {\gr} and background rates after gamma-hadron separation cuts for the CTA on-axis performance. The diffuse {\grs} (black solid line) are taken from \cite{2015ApJ...799...86A} and comprise the intensity of all {\grs} above $|b|>20\degs$, measured by {\fermi} up to $820\,\mathrm{GeV}$, including the resolved sources.  The background rate (blue dashed-dotted line) is given on-axis, optimised for $30$ minutes of observation (as used throughout this paper).
}\label{fig:plot_rates}}
\end{figure}
\begin{table}[t]
 \centering
  \begin{tabular}{|c|ccc|} \hline
Energy  & Diffuse {\gr} rate & Background rate & {\grs}/background \\
threshold & $\mathrm{[Hz\;\deg^{-2}]}$&  $\mathrm{[Hz\;\deg^{-2}]}$& ratio  \\\hline
$30\,\mathrm{GeV}$  & $2.8\cdot 10^{-3}$ & 6.1   & $0.5\cdot 10^{-3}$ \\
$100\,\mathrm{GeV}$ & $1.1\cdot 10^{-3}$ & 1.1   & $1.1\cdot 10^{-3}$ \\
$300\,\mathrm{GeV}$ & $3.0\cdot 10^{-4}$ & 0.27  & $1.2\cdot 10^{-3}$ \\\hline
  \end{tabular}
 \caption{
Diffuse {\gr} and background rates after gamma-hadron separation cuts for the CTA on-axis performance, integrated over the energy above different energy thresholds.
  }
\label{tab:cta-diffuse-background}
\end{table}

Earth-based {\gr} observatories suffer from a large residual background from cosmic rays. While sophisticated analysis techniques are able to efficiently discriminate the {\gr} signal from the hadronic background, it is almost impossible to separate it from cosmic electrons. In \cref{fig:plot_rates}, we show the residual background rate estimation for CTA (blue dashed-dotted line), after applying background rejection cuts optimised for an observation over $30\,\mathrm{min}$, the typical observing time per field.
This background rate is computed by MC simulations of the detector response to the cosmic-ray intensity, and represents the events passing the analysis cuts. In black, we show the event rate of the diffuse {\grs}, resulting from integrating the {\gr} intensity over the CTA on-axis effective area. We take the  total {\gr} intensity above $|b|>20\degs$, measured by {\fermi} up to $820\,\mathrm{GeV}$ \cite{2015ApJ...799...86A}, to compare it with the CTA residual background. This intensity contains all {\grs} outside the Galactic plane, from resolved and unresolved sources. Thus, this spectrum gives a rather conservatively large value for the total diffuse {\gr} emission outside the Galactic plane. 
From \cref{fig:plot_rates} (differential rates) and \cref{tab:cta-diffuse-background} (integrated rates), it can be seen that the residual background outweighs the diffuse \gr{} emission by a factor of 1000. Therefore, we can safely ignore all diffuse {\gr} backgrounds on top of the residual cosmic-ray background, including those from Galactic and extragalactic DM.

\subsection{Characteristics of the brightest halos}
\label{sec:brightest_CTA}

\begin{figure}[t]
\centering
\includegraphics[width=0.8\textwidth]{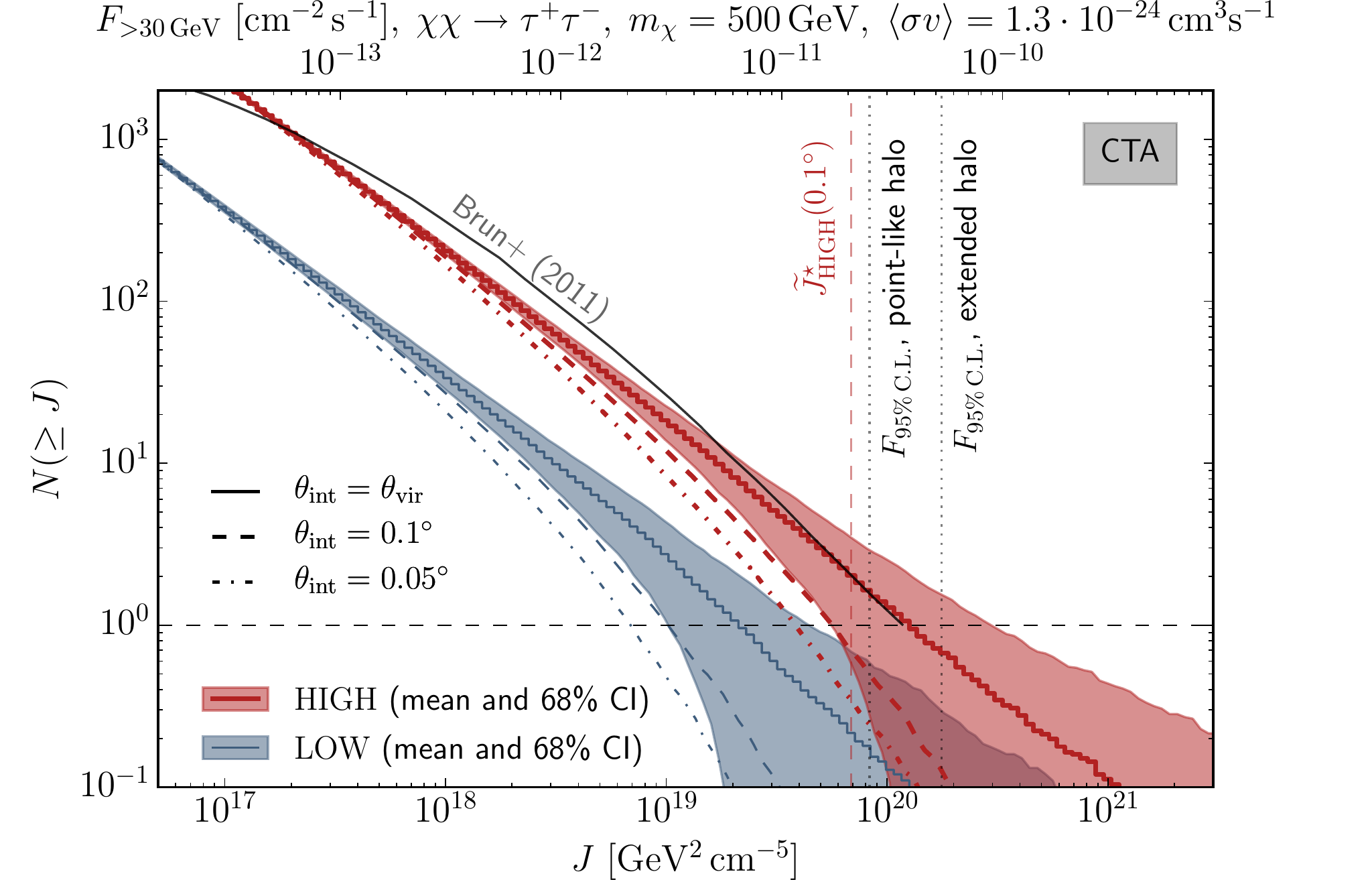}
  \caption{
Cumulative source count distribution of DM subhalos for the  CTA setup. The upper $x-$axis shows the flux level  for the DM particle model to which CTA is most sensitive  (adopting the subhalo model HIGH; see \cref{fig:IndividualSourceSensitivity-CTA}). The annihilation cross-section is chosen so that CTA would observe one subhalo above the flux sensitivity threshold corresponding to the chosen DM annihilation spectrum. Increasing (decreasing) the annihilation cross-section would result in a shift of the upper $x-$axis and the vertical dashed lines to the left (right). We also display the result of Brun et al.\!\! \cite{2011PhRvD..83a5003B}, who used $\thetaint\approx\theta_{\rm vir}$.
}
\label{fig:Histogram1D_Jfactor-Nsub-CTA}
\end{figure}

Similarly to the discussion in \cref{sec:fermi} for \fermi{},  we investigate the source count distribution for a \jfactor{} integration angle tailored to the angular resolution of the instrument, and display the result in  \cref{fig:Histogram1D_Jfactor-Nsub-CTA} for models LOW and HIGH. We present the distributions for the integration angles $\thetaint=0.05\degs$ (angular resolution of CTA at $\gtrsim \unit[1]{TeV}$),  $\thetaint=0.1\degs$ (angular resolution of CTA at $\lesssim \unit[1]{TeV}$), and the full emission, $\thetaint=\theta_{\rm vir}$. Comparing \cref{fig:Histogram1D_Jfactor-Nsub-CTA} (CTA  scenario) to \cref{fig:Histogram1D_Jfactor-Nsub-Fermi} (\fermi{}  scenario) shows that the CTA source count distributions shift to lower \jfactors{}, due to a factor $\sim 3$ smaller survey field, and smaller integration angles. We also show in  \cref{fig:Histogram1D_Jfactor-Nsub-CTA} the subhalo distribution assumed by \cite{2011PhRvD..83a5003B} for a survey field characterised by $f_{\rm sky}=25\%$, based on the VL~II subhalo catalogue. These authors consider an integration over the entire extent of the subhalos (i.e. $\thetaint=\theta_{\rm vir}$), dismissing only highly extended subhalos, and their distribution is in fair agreement with our  model HIGH. This is consistent with the fact that  our model HIGH approximately matches the distribution derived from VL~II also for the \fermi{} setup (see \cref{fig:Histogram1D_Jfactor-Nsub-Fermi}). However, we emphasize that assuming the whole subhalo flux,  $J(\theta_{\rm vir})$, originating from a point source heavily overestimates the actual CTA performance. An exact treatment of the sensitivity must account for the energy dependent angular resolution and the extension of the source, and is done in the next subsections.  In \cref{fig:Histogram1D_Jfactor-Nsub-CTA}, we show the result from this rigorous treatment. The dashed lines show that for the considered DM particle, the sensitivity to the full emission from the brightest, extended halo (model HIGH) roughly corresponds to the sensitivity to a point source with the smaller flux within $0.1\degs$, $J(0.1\degs)$.

\Cref{tab:brightest_properties} lists the median properties of the brightest object  for \fermi{} and CTA at various integration angles. The median $\widetilde{J}^{\star}(\theta_{\rm int})$ is obtained from $10^4$ MC simulations (see also \cref{sec:fat_tails}). In terms of halo properties, the population of brightest sources resembles the one for \fermi{}, and consists of close and rather massive halos; we remind the reader that we have identified the distance-dependence subhalo concentration as the main driver of this behaviour. The high angular resolution of CTA implies that changing the integration angle more drastically affects the brightest source properties than for \fermi{}: the smaller the integration angle, the lighter (and closer) the average brightest object becomes ($\Mvir\gtrsim\unit[10^7]{\Msol}$ for $\thetaint\lesssim0.05\degs$).

The subhalo source count distributions estimated so far include the most massive DM clumps, which would have formed stars and and host the dSph galaxies of the MW. Therefore, VHE \grs{} from DM annihilation in these objects will most likely be discovered by dedicated pointed VHE observations. Including dSph objects into the estimation of the CTA survey sensitivity to dark clumps might therefore bias the number of potentially detectable sources, even more so when deep-sky and large-area optical surveys might discover additional faint dSph galaxies. To study the impact of this issue, we computed the subhalo abundance in the CTA scenario (25\% sky coverage),
neglecting clumps heavier than $\unit[10^6]{\Msol}$ or $\unit[10^7]{\Msol}$ (\Cref{tab:M_and_J} in the appendix shows that all the known dSph galaxies have a median mass larger than $\unit[10^6]{\Msol}$, and most objects are likely to have masses larger than $\unit[10^7]{\Msol}$).
We find that when excluding all objects above $\unit[10^7]{\Msol}$, the median \jfactor{} within $\thetaint = 0.05\degs$ of the brightest subhalo is not affected, with $\log_{10}(\widetilde{J}^{\star}(0.05\degs)/\junits) = 19.7^{+0.3}_{-0.2}\,$. Looking over the full extent of the DM subhalo, we find $\log_{10}(\widetilde{J}^{\star}(\theta_{\rm vir})/\junits) = 20.0^{+0.4}_{-0.3}$, i.e. a factor $\sim 2$ decrease compared to \cref{tab:brightest_properties}. This behaviour is understood as a lighter, but more concentrated halo is then selected as brightest object by the exclusion criterion.
The situation changes when rejecting all objects above $\unit[10^6]{\Msol}$, where we obtain $\log_{10}(\widetilde{J}^{\star}(0.05\degs)/\junits) = 19.5^{+0.3}_{-0.2}\,$ and $\log_{10}(\widetilde{J}^{\star}(\theta_{\rm vir})/\junits) = 19.8^{+0.4}_{-0.3}$. The median brightest dark subhalo is then expected to be a factor 2 dimmer within the central $\thetaint=0.05\degs$. The prospects of survey discoveries are therefore only marginally affected by the distinction between dark and bright DM halos. This point is not considered further in the remainder of this paper, where we use the values in \cref{tab:brightest_properties} (right) to characterize the brightest halo properties for a CTA extragalactic survey.

\begin{table}[t!]
  \begin{center}
\resizebox{\textwidth}{!}{%
  \begin{tabular}{|c|ccc|ccc|} \hline
         	& \multicolumn{3}{c|}{{\fermi} scenario} & \multicolumn{3}{c|}{CTA scenario}	\\
 Median properties of 		& \multicolumn{3}{c|}{($f_{\rm sky}=82.6\%$)}					& \multicolumn{3}{c|}{($f_{\rm sky}=25\%$)}	\\
 brightest subhalo within 					&  $\thetaint=0.1^{\circ}$	&  $\thetaint=0.8^{\circ}$	&  $\thetaint=\theta_{\rm vir}$	&  $\thetaint=0.05^{\circ}$	&  $\thetaint=0.1^{\circ}$	&  $\thetaint=\theta_{\rm vir}$\\\hline
 &&&&&&\\[-0.1cm]

\multirow{-2}{*}{$\widetilde{D}_{\rm obs}^{\star}$ $[\unit{kpc}]$} &	\multirow{-2}{*}{$7^{+10}_{-5}$} & \multirow{-2}{*}{$8^{+11}_{-6}$} & \multirow{-2}{*}{$8^{+12}_{-6}$} & \multirow{-2}{*}{$7^{+10}_{-5}$} & \multirow{-2}{*}{$8^{+12}_{-6}$} & \multirow{-2}{*}{$10^{+16}_{-8}$} \\[3mm]

\multirow{-2}{*}{$\widetilde{R}^{\star}$ $[\unit{kpc}]$} &	\multirow{-2}{*}{$9^{+9}_{-3}$} & \multirow{-2}{*}{$10^{+10}_{-3}$} & \multirow{-2}{*}{$10^{+11}_{-3}$}  & \multirow{-2}{*}{$10^{+9}_{-2}$} & \multirow{-2}{*}{$10^{+10}_{-3}$} & \multirow{-2}{*}{$12^{+15}_{-4}$} \\[3mm]

\multirow{-2}{*}{$\log_{10}(\widetilde{m}_{\rm vir}^{\star}/\Msol)$} &	\multirow{-2}{*}{$7.7^{+1.3}_{-1.5}$} & \multirow{-2}{*}{$8.1^{+1.2}_{-1.6}$} & \multirow{-2}{*}{$8.1^{+1.3}_{-1.5}$} & \multirow{-2}{*}{$7.4^{+1.4}_{-1.4}$} & \multirow{-2}{*}{$7.6^{+1.4}_{-1.5}$} & \multirow{-2}{*}{$8.0^{+1.3}_{-1.6}$} \\[3mm]
 
\multirow{-2}{*}{$\widetilde{r}_{\rm vir}^{\star}$ $[\unit{kpc}]$} &		\multirow{-2}{*}{$6.7^{+12}_{-4.6}$} & \multirow{-2}{*}{$8.8^{+14}_{-6.1}$} & \multirow{-2}{*}{$9.2^{+15}_{-6.3}$} & \multirow{-2}{*}{$5.4^{+9.5}_{-3.5}$} & \multirow{-2}{*}{$5.9^{+11}_{-4.0}$} & \multirow{-2}{*}{$8.1^{+14}_{-5.8}$} \\[3mm]
 
\multirow{-2}{*}{$\widetilde{r}_{\rm s}^{\star}$ $[\unit{kpc}]$} &	\multirow{-2}{*}{$0.13^{+0.42}_{-0.10}$} & \multirow{-2}{*}{$0.19^{+0.55}_{-0.15}$} &  \multirow{-2}{*}{$0.21^{+0.62}_{-0.17}$} & \multirow{-2}{*}{$0.12^{+0.36}_{-0.08}$} & \multirow{-2}{*}{$0.14^{+0.43}_{-0.10}$} & \multirow{-2}{*}{$0.22^{+0.69}_{-0.17}$} \\[3mm]
 
\multirow{-2}{*}{$\widetilde{c}_{\rm vir}^{\star}$} &		\multirow{-2}{*}{$50^{+23}_{-16}$} & \multirow{-2}{*}{$44^{+22}_{-15}$} & \multirow{-2}{*}{$43^{+22}_{-15}$} & \multirow{-2}{*}{$45^{+16}_{-14}$} & \multirow{-2}{*}{$43^{+17}_{-14}$} & \multirow{-2}{*}{$37^{+17}_{-13}$} \\[3mm]
 
\multirow{-2}{*}{$\widetilde{\theta}_{\rm vir}^{\star}$ $[\unit{deg}]$} &	\multirow{-2}{*}{$45^{+16}_{-12}$} & \multirow{-2}{*}{$48^{+15}_{-12}$} & \multirow{-2}{*}{$49^{+14}_{-12}$} & \multirow{-2}{*}{$37^{+16}_{-11}$} & \multirow{-2}{*}{$38^{+15}_{-11}$} & \multirow{-2}{*}{$39^{+15}_{-10}$} \\[3mm]
 
\multirow{-2}{*}{$\widetilde{\theta}_{\rm s}^{\star}$ $[\unit{deg}]$} &	\multirow{-2}{*}{$1.2^{+1.4}_{-0.6}$} & \multirow{-2}{*}{$1.5^{+1.6}_{-0.8}$} & \multirow{-2}{*}{$1.6^{+1.6}_{-0.8}$} & \multirow{-2}{*}{$1.0^{+1.1}_{-0.5}$} & \multirow{-2}{*}{$1.1^{+1.1}_{-0.5}$} & \multirow{-2}{*}{$1.3^{+1.1}_{-0.6}$} \\[3mm]
 
\multirow{-2}{*}{$\widetilde{\theta}_{\rm h}^{\star}$ $[\unit{deg}]$} &	\multirow{-2}{*}{$0.16^{+0.20}_{-0.08}$} &  \multirow{-2}{*}{$0.20^{+0.20}_{-0.10}$} &\multirow{-2}{*}{$0.22^{+0.22}_{-0.11}$} & \multirow{-2}{*}{$0.13^{+0.16}_{-0.05}$} & \multirow{-2}{*}{$0.14^{+0.14}_{-0.07}$} & \multirow{-2}{*}{$0.18^{+0.14}_{-0.08}$} \\[3mm]
 
\multirow{-2}{*}{$\log_{10}\left(\widetilde{J}^{\star}/\junits\right)$} & \multirow{-2}{*}{$20.3^{+0.4}_{-0.3}$} & \multirow{-2}{*}{$20.7^{+0.4}_{-0.3}$}  & \multirow{-2}{*}{$20.8^{+0.5}_{-0.4}$} & \multirow{-2}{*}{$19.7^{+0.3}_{-0.3}$} & \multirow{-2}{*}{$19.9^{+0.4}_{-0.3}$} & \multirow{-2}{*}{$20.3^{+0.5}_{-0.4}$} \\\hline
  \end{tabular}
  }
\end{center}
  \caption{Median properties of the brightest subhalo for the survey setups tailored to the {\fermi} and CTA instruments, and for the subhalo model HIGH. The uncertainties denote the $68\%$ CI around the median. For both instruments, the results for different angular resolutions are given. 
 $D_{\rm obs}$ is the distance from the observer, and $R$ the distance from the GC. $\mvir$ is the subhalo mass. $\rvir$ and $\rscale$ denote its virial and scale radius, ${c}_{\rm vir} = \rvir/\rscale$, and $\theta_{\rm vir,\,s}=\arctan(r_{\rm vir,\,s}/D_{\rm obs})$. $\thetahalf$ is the radius enclosing half of the total emission, $J(\thetahalf) = 0.5\,J(\theta_{\rm vir})$. For reliable medians, the values are obtained from a sample of $10^4$ simulations.
  }
\label{tab:brightest_properties}
\end{table}

\subsection{Likelihood ratio and test statistic (\TS{})}
\label{sec:logL_TS}
We use the open-source CTA analysis software \texttt{ctools}\footnote{\url{http://cta.irap.omp.eu/ctools/}}, based on the \texttt{gammalib} library\footnote{\url{http://cta.irap.omp.eu/gammalib/}} \cite{2016arXiv160600393K} to compute the CTA sensitivity to the median brightest dark subhalo. The \texttt{ctools} framework allows the use of a maximum-likelihood inference of hypotheses $\mathcal{M}$ from  event data, considering all available spatial and spectral information from the data (see \cref{app:likelihood} for the likelihood $\mathscr{L}$). The \texttt{cssens} tool is used to simulate events and subsequently calculate the maximum log-likelihood ratio, with the likelihood ratio $\lambda$ given by
\begin{align}
\lambda= \frac{\max\mathscr{L}(\mathcal{M}_{\rm bkg}(\bm{\Theta}_{\rm bkg})\,|\, \bm{X})}{\max  \mathscr{L}(\mathcal{M}_{\rm sig}(\bm{\Theta}_{\rm sig})+ \mathcal{M}_{\rm bkg}(\bm{\Theta}_{\rm bkg})\,|\, \bm{X})}= \frac{\mathscr{L}(\mathcal{M}_{\rm bkg}(\widehat{\widehat{\bm{\Theta}}}_{\rm bkg})\,|\, \bm{X})}{\mathscr{L}(\mathcal{M}_{\rm sig}(\widehat{\bm{\Theta}}_{\rm sig})+ \mathcal{M}_{\rm bkg}(\widehat{\bm{\Theta}}_{\rm bkg})\,|\, \bm{X})}.
\label{eq:lambda}
\end{align}
with $\bm{X}=  (N_{\rm obs},\,E_{\rm{obs},\,1\ldots \mathit{N}_{\rm obs}},\vec{k}_{\rm{obs},\,1\ldots \mathit{N}_{\rm obs}})$ the mock data\footnote{$N$, $E$, and $\vec{k}$ are the number of photons, energy and direction respectively.}, $\bm{\Theta}$ the adjustable parameters in the models maximising the likelihood, and $\widehat{\bm{\Theta}}$ the corresponding maximum likelihood estimators ($\widehat{\widehat{\bm{\Theta}}}_{\rm bkg}$ under the constraint $\mathcal{M}_{\rm sig}=0$).

For the background fit, we allow the normalisation of the rate to vary. The signal model $\mathcal{M}_{\rm sig}$ for DM consists, according to \cref{eq:flux-general},  of the spatial part of our fixed \jfactor{} map (see next subsection), and a spectral part depending on the particle mass $m_{\chi}$ and the annihilation cross-section $\sigmav$. We scan 24 DM particle masses in $\unit[50]{GeV}\leq m_{\chi} \leq \unit[100]{TeV}$, computing for each $m_\chi$ the \gr{} spectrum $\dd N_{\gamma}^f/\dd E$ from \cite{2011JCAP...03..051C}. For each spectrum, we set the flux level to be the only free parameter, such that $\bm{\Theta}_{\rm sig}=\sigmav$.

We use the logarithm of the likelihood ratio \cref{eq:lambda} as the test statistic TS to exclude the signal hypothesis $\mathcal{M}_{\rm bkg}+\mathcal{M}_{\rm sig}$ (at the confidence level $1-p_{\rm pre}$), namely
\beq
\TS = -2\log \lambda.
\label{eq:ts_def}
\eeq
The $\TS(p_{\rm pre})$ values were calculated directly from a set of MC simulations, and we refer the interested reader to \cref{app:ts_statistics} for the technical details and used \TS{} values. More importantly, we cross-checked our analysis method by calculating the sensitivity to the Sculptor dSph galaxy, and found our results consistent with Carr et al.\!\! (2015) \cite{2015arXiv150806128C}.

\subsection{Sensitivity to dark clumps and comparison to other targets}
\label{sec:CTA_results}

To calculate the CTA sensitivity to the brightest subhalo in the  survey field, we build a template of the median brightest object described in \cref{sec:brightest_CTA}. We choose the \jfactor{} profile to be that of the object found to be the brightest within $\thetaint=0.05\degs$ (see \cref{tab:brightest_properties}). We emphasize that the latter choice only determines the shape of the template halo, and the Likelihood-based sensitivity analysis is run over the full spatial extent of that halo. The instrument response, required in \cref{eq:lambda} and in the likelihood function \cref{eq:likelihood-function}, is taken from \cite{2013APh....43..171B}, using the publicly available results from the `Production 2' simulations.\footnote{\url{http://portal.cta-observatory.org/Pages/CTA-Performance.aspx}} The available instrument response data corresponds to on-axis observations of Crab-nebula-like point sources, which can be safely adopted for the highly cuspy DM template halo (with a half emission radius of $\thetahalf=0.13\degs$, only slightly above the CTA angular resolution). We use the response tables with background rejection cuts optimised for a $30\,\mathrm{min.}$ observation at a constant elevation of $70\degs$ with CTA (southern site),\footnote{We cross-checked the analysis with the 'Production 2' rejection cuts optimized for a $5\,\mathrm{h}$ observation and obtained the same sensitivities.} assuming 4 large-size telescopes, 24 medium-size telescopes and 72 small-size telescopes. Different observation strategies have been proposed to raster the CTA survey field with overlapping observations to obtain a preferably homogeneous exposure \cite{2013APh....43..317D}. We mimic the survey coverage by an on-axis observation of the template halo with a one hour exposure. With this choice, we calibrate our template observation setup to the optimistic benchmark performance projected for the CTA extragalactic survey, and obtain a similar sensitivity to a Crab-like point source as in \cite{2013APh....43..317D}.\footnote{We obtain project a survey sensitivity to fluxes  $\gtrsim\unit[4\cdot 10^{-12}]{cm^{-2}\,s^{-1}}$ above $\unit[100]{GeV}$
(approximately $0.7\%$
 of the Crab nebula flux) for a point source with a Crab-nebula-like spectrum, requiring a test statistic of $\TS = 25$ and without applying a trials correction. This is in an optimistic agreement with \cite{2013APh....43..317D}. For comparison, a point source with a spectrum $\chi\chi\rightarrow \tau^+\tau^-$, $m_{\chi}=\unit[500]{GeV}$ (as discussed in \cref{fig:Histogram1D_Jfactor-Nsub-CTA}), would be detected at $F_{\rm TS=25}=\unit[7.8\cdot 10^{-12}]{cm^{-2}\,s^{-1}}$ above $\unit[100]{GeV}$.}

\begin{figure}[t!]
  \begin{center}
    \includegraphics[width=\textwidth]{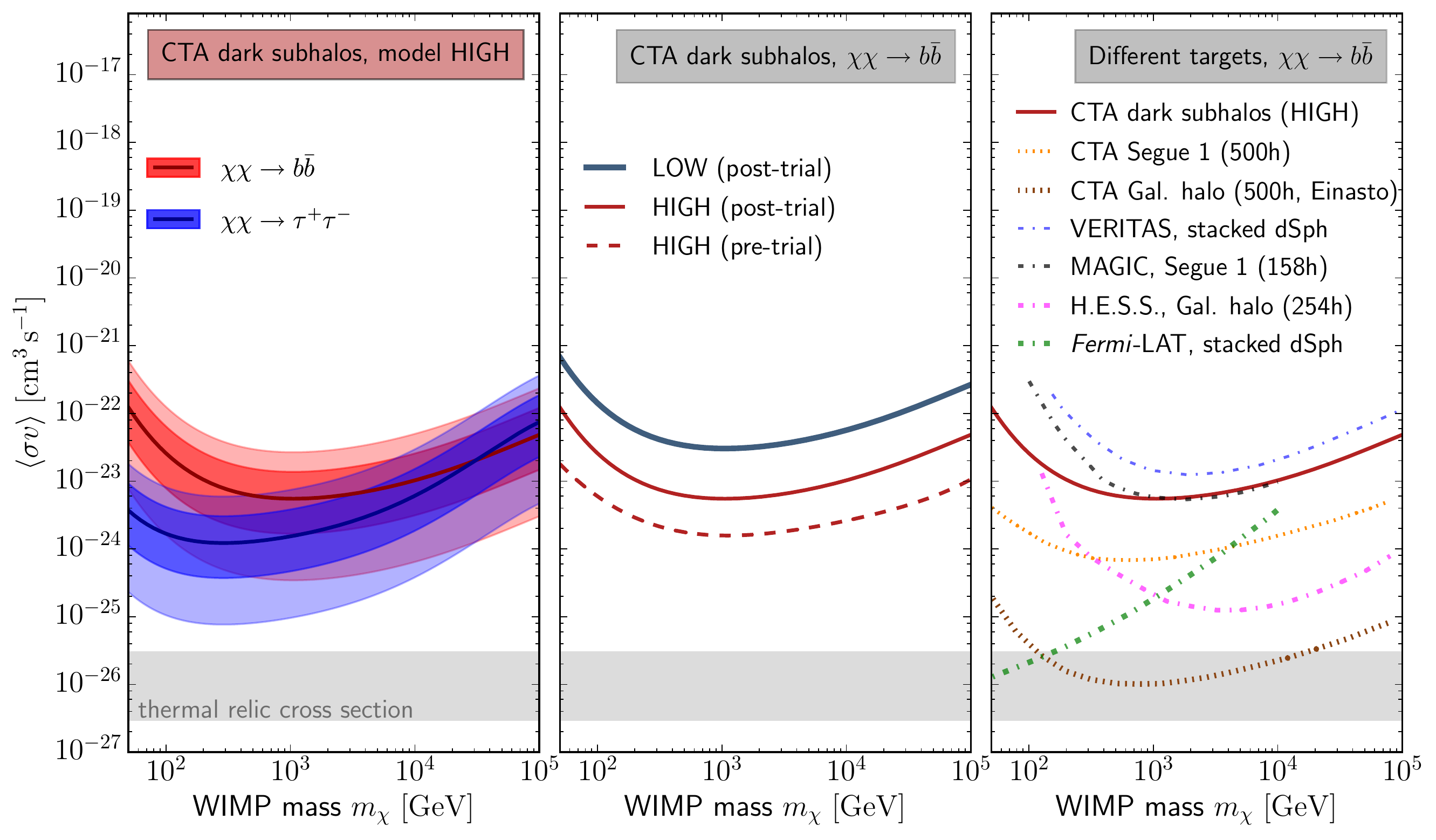}
  \captionof{figure}{ 
Sensitivity of the CTA extragalactic survey to find the brightest Galactic subhalo in the survey FOV. All sensitivities are given at the $95\%$ CL. {\em Left:} median (solid lines) and $68\%$ ($95\%$) {\jfactor} (model HIGH) uncertainty around the median (coloured areas) for $b\bar{b}$ and $\tau^+\tau^-$ annihilation channels. {\em Centre:} comparison of LOW, HIGH (pre- and post-trial). {\em Right:} Comparison of our analysis to the CTA sensitivity for other targets (Segue I and the GC from \cite{2015arXiv150806128C}) and to the limits from running experiments (VERITAS \cite{2015arXiv150901105Z}, MAGIC \cite{2014JCAP...02..008A}, H.E.S.S. \cite{2015arXiv150904123L}, and \fermi{} \cite{2015PhRvL.115w1301A}).
}
  \label{fig:IndividualSourceSensitivity-CTA}
  \end{center}
\end{figure}

In \cref{fig:IndividualSourceSensitivity-CTA}, we show the projected sensitivity of CTA to the WIMP DM annihilation cross-section by searching for Galactic dark subhalos in the planned extragalactic survey:
\begin{itemize}
   \item In the left panel, we present the sensitivity for two benchmark annihilation channels with $100\%$ branching ratio of $\chi\chi\rightarrow b\bar{b}$ and $\chi\chi\rightarrow \tau^+\tau^-$, respectively. The solid lines represent the sensitivity to the template halo with median \jfactor{}. The shaded bands denote the $68\%$ ($95\%$) statistical uncertainty around this median, originating from the {\jfactor} variance. It can be seen that the sample-to-sample variation of the sensitivity scatters over almost one order of magnitude within the 68\% CI, and two orders of magnitude within the 95\% CI. Also, the sample variance has a long tail towards low values of \sigmav{}, such that despite a relative modest median sensitivity, the 95\% CI reaches relatively small annihilation cross-sections. The shape of the \jfactor{} sample variance, which underlies the sensitivity variance, is discussed in detail in \cref{app:meanmedian}.
   \item In the central panel, we show the impact of the model uncertainty onto the sensitivity. The one order of magnitude difference in the predicted fluxes between the model LOW and HIGH translates into the equivalent difference in sensitivity.  We also calculated the sensitivity at the $1-p_{\rm post}\equiv 1-p_{\rm pre}$ confidence level (see \cref{app:prepost_trial}), neglecting the trials penalty (dashed line), that accounts for a putative improvement of the sensitivity by more than a factor of 5. 
   \item In the right panel, we put the sensitivity obtained from this study into broader context. A CTA survey search for dark subhalos provides a less powerful probe for DM annihilation than CTA pointed observations of the Galactic halo, and to a lesser extent, than the MW satellite galaxies. However, these other targets also suffer from systematic errors. \Cref{fig:IndividualSourceSensitivity-CTA} also shows the limits from current experiments for comparison. It is visible that the GC provides the best limit at high energy, whereas the \fermi{} experiment already reaches the CTA parameter space below TeV energies.
   \end{itemize}
Before concluding, we briefly comment on the results of Ref.~\cite{2011PhRvD..83a5003B}, who previously discussed the CTA sensitivity to dark clumps. Based on the Via Lactea~II simulation and a similar survey in area (1/4 of the sky, though towards the GC), they find a more favourable sensitivity than the one we obtain. For instance, in the $b\bar{b}$ channel, our calculation reaches a minimum (pre-trial) $\sigmav\gtrsim 2\times10^{-24}$ to compare to $\sigmav\gtrsim 4\times10^{-26}$ in \cite{2011PhRvD..83a5003B}. Several reasons may be at the origin of this difference. First, our limit is based on the 95\% CL whereas theirs is based on 90\% CL. With this choice, we find a factor of 2 improvement on the pre-trial sensitivity (the post-trial sensitivity is not affected). Second, Ref.~\cite{2011PhRvD..83a5003B} models the CTA instrument characteristics starting from the H.E.S.S. instrumental response, assuming a factor 10 larger effective area and a factor 2 better background rejection. However,  the  improvement of the CTA performance is energy-dependent, and according to the `Production 2' simulations, the largest improvement in differential sensitivity compared to current instruments will be reached at energies above $\sim \unit[1]{TeV}$. Because CTA will be most sensitive to WIMP masses of $m_{\chi}\approx \unit[1]{TeV}$ (annihilation products below $\unit[1]{TeV}$), the applied H.E.S.S. extrapolation most likely overestimated the CTA sensitivity to DM. Third,  Ref.~\cite{2011PhRvD..83a5003B} assumes that the total subhalo \jfactors{}, $J(\theta_{\rm vir})$, is enclosed in the instrumental resolution. This overestimates the flux of the brightest subhalo by another factor $\sim 2$. For these reasons, we are confident our analysis provides a more realistic estimate of the CTA sensitivity to dark subhalos.

\section{Summary and discussion}
\label{sec:discussion}

In this paper, we have revisited the detectability of dark clumps for present and future \gr{} instruments. Using the \clumpy{} code, we have simulated several distributions of the subhaloe in the Galaxy, in order to critically assess the range of potential number of detectable clumps and to identify the most important parameters of such prediction. For each model, several hundreds of skymaps have been generated to obtain the statistical properties (mass, distance, $J$-factor, etc.) for these configurations. The model dubbed HIGH (LOW) provides a realistic (conservative) benchmark model for the number of expected dark clumps. These benchmark models have been used to discuss the prospects of dark clumps detection for the \fermi{} and CTA instruments, and we compared our results and exclusion limits on annihilating DM to several previous calculations. We underline that we have carefully validated our analysis at each step of the calculation (validation and cross-checks for the distribution of $J$ and the APS from \clumpy{}, and for the sensitivity calculations for CTA). Our findings are summarised below. First, for the substructure modelling, we find that:
\begin{itemize}

  \item  The concentration parametrisation is the main uncertainty in the determination of the number of halos with the largest $J$-factors. Indeed, the brightest detectable dark clumps are found to be close by, deep in the potential well of the Galaxy, where the impact of tidal stripping on the concentration and brightness is the most critical. As such, our LOW model serves as a baseline for a conservative estimate, in which no tidal stripping is accounted for (subhalos assumed to be as `field' halos'). Our HIGH model is based on \cite{2011PhRvD..83b3518P}, which agrees well, in the regime that matters for detecting dark clumps, with the very recently estimated distance-dependent concentration parametrisation of Ref.~\cite{2016arXiv160304057M}. This makes HIGH a likely realistic benchmark model, though further improvements and understanding of the stripping effect is necessary to strengthen this conclusion. 
  
  \item The number of calibration subhalos $N_{\rm calib}$ (between $10^8$ and $\unit[10^{10}]{\Msol}$) and the boost from sub-subhalos in the external parts of the subhalos are the next impacting, though subdominant factors. First, moving from $N_{\rm calib}=300$ in model HIGH to half this number in model VAR6, as hinted at by hydrodynamical simulations \cite{2015MNRAS.447.1353M,2016MNRAS.457.1931S}, translates into a similar decrease for $J$ of the brightest object and the corresponding sensitivity. It is thus desirable to better constrain this number. Second, the boost factor is expected to be at most of a few for $10^7-10^8\Msol$ subhalos \cite{2016arXiv160304057M,2014MNRAS.442.2271S}. Compared to our conservative analysis in which this boost was discarded, this could slightly improve the prospects for \fermi{}, but probably not for CTA: for background-dominated instruments like CTA, the best sensitivity is achieved at the angular resolution (to decrease the background), for which the `boosted' outskirts of the subhalos are not encompassed.
   
  \item The uncertainties in the other parameters (inner profile, index of the clump mass distribution, spatial distribution, width of the mass-concentration distribution) impact the overall level of the diffuse DM emission in the Galaxy, hence changes the contrast of the subhalos. However, it leaves mostly unchanged the number and signal strength of the brightest clumps.

\end{itemize}

A second aspect of our analysis, made possible by the reasonable running time of \clumpy{}, is the determination of the statistical properties of the detectable clumps. We have shown that:

\begin{itemize}

  \item The brightest clumps are typically located at $\sim10-20$~kpc from the GC and from us, with concentration values $\sim 40-50$. \fermi{} and CTA probe slightly different populations, with larger and more massive subhalos for \fermi{} ($M\lesssim\unit[10^8]{\Msol}$, $\thetahalf\lesssim0.2^\degs$, and $J\sim\unit[3\cdot10^{20}]{\junits}$) than for CTA ($M\gtrsim\unit[10^7]{\Msol}$, $\thetahalf\gtrsim0.1^\degs$, and $J\sim\unit[10^{19}]{\junits}$). These halo masses and $J$-factors are close to the values obtained for dSph galaxies, and this raises the question whether the calculated sensitivity applies for truly dark halos or objects that could be discovered as nearby and very faint dSph galaxies in the future. However, we have shown that the calculated sensitivities are only mildly degraded (less than a factor 2) when rejecting masses above $M>\unit[10^6]{\Msol}$ in a CTA large-sky survey scenario. In any case, CTA is sensitive to smaller mass ranges than \fermi{}. This difference is related to the angular resolution of the instruments ($\sim 0.05^\circ$ for CTA compared to $\gtrsim 0.1^\circ$ for \fermi{}), illustrating that instruments with better angular resolution are sensitive to lower mass subhalos, which are more likely to be dark. CTA will therefore provide a complementary view to the observation of MW satellite galaxy observations.
 
 \item The properties of the brightest clumps derived from the statistical assessment underlie a large sample variance. The \jfactor{} of the brightest clump varies by almost a factor 10 at 68\% CI (100 at 95\% CI). In particular, the distribution has a long tail towards large $J$-values. On the one hand, this means that a dark DM subhalo might be detected even for a relatively small annihilation cross-section. On the other hand, the large sample variance introduces a large systematic uncertainty for the limits derived from non-observation of DM candidates in the surveys. 
   
  \item Tidal effects of subhalos, modelled here via the distance-dependent concentration, leads to an anisotropy in the distribution of the directions of the brightest clumps, as seen in the small dipole power excess (compared to models using field halos for the concentration leading to an isotropic distribution). The impact of this effect on detectability and sensitivity is not straightforward to assess: (i) the contrast of the dark clumps w.r.t. the DM diffuse emission decreases towards the GC, which should mostly affect the detectability when pointing towards $\theta\lesssim 10^\circ$; (ii) however, this should be balanced with the existence of a preferential direction to search for these dark halos (4 times more likely to lie towards the GC direction than towards the anticentre), for which the analysis could be optimised. In any case, the number of bright dark halos searched for in an analysis is small (by definition) and variance dominated at low$-\ell$. A dedicated study is necessary to assess the potential benefit of an optimised search.
  
\end{itemize}
Finally, a last aspect of our analysis deals with the sensitivity of CTA, and to some extent of \fermi{}, to dark clumps, compared to previous calculations and other targets:
\begin{itemize}
  \item For an observation setup such as that of \fermi{}, a comparison of the expected number of clumps to previous calculations confirms that Ref.~\cite{2015JCAP...12..035B} obtain a subhalo abundance slightly larger  than our model HIGH. Our results are more in line with those of Ref.~\cite{2016JCAP...05..028S}, which discusses possible explanations for the difference to the results by \cite{2015JCAP...12..035B}. However, compared to \cite{2016JCAP...05..028S}, we do not find two different populations and behaviours `point-like' vs `extended' source in our analysis (we perform the full integration along the line of sight, whereas \cite{2016JCAP...05..028S} do not). As a result the shape of the number of expected dark clumps as a function of $J$ is not the same. The imprint of the population of DM galactic subhalos also shows on the APS, and our models LOW and HIGH encompass previously published studies.
  
  \item For a CTA large-sky survey scenario, we have based our sensitivity calculation on the projected extragalactic survey \cite{2013APh....43..317D}. We provide both a pre-trial and post-trial sensitivity, as scanning a quarter of the sky results in a million independent trials that must be accounted for in setting the limits. For pure annihilations into $b\bar{b}$, we find the best post-trial median sensitivity reach of $\sigmav\sim 6\times10^{-24}~\unit{cm^3~s^{-1}}$ at $\mchi\sim 1$~TeV,  and for the $\tau^+\tau^-$ annihilation channel, the best sensitivity of $\sigmav\sim 1\times10^{-24}~\unit{cm^3~s^{-1}}$ at $\mchi\sim 500$~GeV. The 68\% (95\%) CI changes these values by a factor $\sim5$ ($\sim10$). In any case, taking the post-trial limit into account worsens the sensitivity by a factor $\sim 5$.
  
  \item The sensitivity reach of CTA (in model HIGH) for dark clumps in the extragalactic survey is quite competitive with dedicated dSph galaxies observations. Looking at the sensitivity based on the most-likely values favours the $500$~hrs single bright dSph observation over that of the extragalactic survey (factor $\sim3$), but we recall that both suffer from large uncertainties. In particular, it has been shown that Segue~1's $J$-factor (used in our comparison plots) could be significantly lower than previously estimated \cite{2015arXiv150608209B,2016MNRAS.461.2914H}. In any case, the best target remains the GC region.
\end{itemize}

To conclude, we have shown that a CTA dark halo search provides a complementary view on a different population of subhalos compared to observing dSph galaxies. The planned CTA astronomical surveys will therefore nicely complement the dedicated DM searches programme (a similar analysis as done here could be performed for decaying dark matter). Given the unprecedented angular resolution of CTA, the search for small-scale anisotropies in its data, e.g., as proposed by \cite{2014JCAP...01..049R}, may provide another complementary constraint on \gr{} emission from Galactic dark clumps to that we have presented in this article.

\acknowledgments

This article has gone through internal review by the CTA Consortium, and we thank Hannes Zechlin and Josep Mart\'i for their careful reading and suggestions that helped to improve the manuscript.  We also thank the anonymous referee for useful suggestions that have improved the quality of the paper.  We warmly thank V.~Bonnivard for helping out with the Jeans analysis of the recently discovered dSph galaxies, and L. G\'erard for valuable help with the CTA analysis.  We also thank I. Sadeh for useful discussions  and comments to the manuscript, and R. Kieokaew for performing useful studies preceding this work. This work has been supported by the Research Training Group 1504, ``Mass, Spectrum, Symmetry'', of the German Research Foundation (DFG), by the ``Investissements d'avenir, Labex ENIGMASS", and by the French ANR, Project DMAstro-LHC, ANR-12-BS05-0006.  Some of the results in this article have been derived using the \healpix{} package \cite{2005ApJ...622..759G}.


\appendix

\section{Halo overdensity definition $\Delta$ and cosmology}
\label{app:definitions}

At a given redshift $z$, the enclosed mass $M_\Delta$ in a dark matter halo is generically defined as the mass contained within a radius $R_\Delta$ inside of which the mean interior density is $\Delta$ times the critical density $\rho_c$:
\begin{equation}
   M_\Delta = \frac{4\pi}{3}R_\Delta^3\times \Delta\times\rho_c.
\end{equation}
The spherical top-hat collapse model provides the virial overdensity $\Delta=\Dvir$ \cite{1980lssu.book.....P} which, for the family of flat cosmologies, $\Omega_m+\Omega_\Lambda=1$ (i.e., $\Omega_K=0$), can be approximated by  $\Dvir{} \simeq (18\pi^2 + 82x -39 x^2)/\Omega_m(z)$, with $x=\Omega_m(z)-1$, $\Omega_m(z)=\Omega_m(0)(1+z)^3/E^2(z)$, and $E^2(z)=\Omega_m(0)(1+z)^3+ \Omega_\Lambda$ \cite{1998ApJ...495...80B}. In this paper we use $\Omega_m(0)=0.308$ and $\Omega_\Lambda=0.692$ \cite{2014A&A...571A..16P}, leading to $\Dvir\simeq 332$.

In principle, the virial radius \Rvir{} can be used as a criterion to identify bound objects. In the spherical collapse model, this radius can be interpreted as a threshold separating a region within which the material is virialized and an external region where mass is still collapsing onto the object. Because the details of the collapse and virialization are not-well understood, several choices have been made in the literature to identify halos in simulations, such as $\Delta=\Dvir,\,200,\,500,\,\dots$ (see, e.g., \cite{2001A&A...367...27W}). We refer the reader to \cite{2015ApJ...799..108D} for a study on whether there exists a more `universal' choice.

\clumpy{} works with $\Delta=\Dvir$, and it uses conversion formulae from \cite{2010MNRAS.404..502G} to convert `$200$' to `${\rm vir}$' quantities, whenever necessary. This choice can in principle impact the estimation of the halo and subhalo mass shown in several plots of this study. However, in \clumpy{}, the mass is mostly a proxy to calculate structural parameters of subhalos, via the $c_\Delta-M_\Delta$ relationship: the exact definition for $\Delta$ does not matter as long as the calculation and conversions are done consistently---the uncertainties on the signal are dominated by our ignorance on the dark matter subhalo population, not by the conversion factors. Then, for comparisons to the mass or $J$-factor of real data, as discussed in \cref{app:mvirJ}, the uncertainties are larger than the difference one would obtain by using different choices for $\Delta$.
Finally, we could also ask how sensitive is the $c_\Delta-M_\Delta$ relationship to cosmological parameters. For instance, in \cref{sec:sets_of_models}, the results based on the Via Lactea~II \cite{2010ApJ...718..899A}, Aquarius \cite{2008MNRAS.391.1685S}, and ELVIS \cite{2014MNRAS.438.2578G} simulations all start with different cosmologies (from WMAP 3 to WMAP 7 \cite{2011ApJS..192...18K}). However, \cite{2016arXiv160304057M} finds a very weak dependence of $c_\Delta-M_\Delta$ on the cosmological parameters, within their statistical dispersion. We refer the reader to \cite{2010arXiv1005.0411C} for a pedagogical introduction to the link between various choices of $\Delta$, the $c_\Delta-M_\Delta$ relationship, and the cosmological parameters.

\section{Mean and variance from analytical approximations}
\label{app:meanmedian}

Analytical formulae are always useful to cross-check numerical calculations and to better identify the underlying important parameters. Under the approximation of power-law source count distributions, we focus in this appendix on (i) the mean and median of the number of dark clumps above a given $J$-factor, which is ultimately related to the sensitivity of an instrument to dark clumps and on (ii) the mean and variance of the angular power spectrum of the $\gamma$-ray signal. The analytical results are compared to the direct calculations from our simulations in several places in the paper.

\subsection{Power-law source count distribution}

As seen in \cref{fig:Histogram1D_Jfactor-Nsub-fullsky}, the mean number of halos $\Nsubmean(\,>J)$ whose annihilation factor is larger than $J$ can be approximated by a power-law distribution over a large range of $J$-factors,
\beq
\Nsubmean(\,>J) \equiv \left\langle \; \int\limits_{J}^{\infty} \frac{\dd \Nsub}{\dd J'}\,\dd J'\,\right\rangle \approx \left(\frac{J}{J_{\rm lim}}\right)^{1-\alpha}\;,
\label{eq:PL_scd}
\eeq
where $J_{\rm lim}$ is defined by $\Nsubmean(\,>J_{\rm lim})=1$, and provided that $\alpha> 1$.\footnote{\Cref{eq:PL_scd} can be easily generalized including an exponential cut-off at $J_{\rm c}$, $\Nsubmean(\,>J)=\left({J}/{J_{\rm lim}}\right)^{1-\alpha}\,\exp[-(J-J_{\rm lim})/J_{\rm c}]$, obtaining the form of a Schechter luminosity function \cite{1976ApJ...203..297S}.}

\subsection{Mean and median of the brightest DM halo}\label{sec:fat_tails}

We define the probability $P_{\,\geq 1}$ to obtain at least one object brighter than a given flux $J$ as
\beq
P_{\,\geq 1}(J)= \sum\limits_{\Nsub=1}^{\infty} p[\Nsub(\,>J)\,|\,\Nsubmean(\,>J)],
\eeq
with $p(\Nsub\,|\,\Nsubmean)$ the probability density to obtain exactly $\Nsub$ objects brighter than $J$ for an expectation value of $\Nsubmean$. If $p(\Nsub\,|\,\Nsubmean)$ follows a Poisson distribution, then the cumulative density function $P_{\,\geq 1}$ is given by
\beq
P_{\,\geq 1}(J) = 1- \exp[- \Nsubmean(\,>J)]\,.
\label{eq:PL_scd_cdf}
\eeq
For example, \cref{eq:PL_scd_cdf} implies that one obtains at least one subhalo brighter than $J_{\rm lim}$ with a chance of $1-e^{-1}=63\%$. For $\Nsubmean(>J)$ a power-law distribution,  \cref{eq:PL_scd}, plugged into \cref{eq:PL_scd_cdf}, gives the probability density function
\beq
\frac{\dd P_{\,\geq 1}}{\dd J}(J) =\frac{\alpha-1}{J_{\rm lim}}\, \exp\left[-\left(\frac{J}{J_{\rm lim}}\right)^{1-\alpha}\right]\,\left(\frac{J}{J_{\rm lim}}\right)^{-\alpha}.
\label{eq:PL_scd_pdf}
\eeq
The expectation value $\overline{J}_{\geq 1}$ is calculated from \cref{eq:PL_scd_pdf} and the median  $\widetilde{J}_{\,\geq 1}$ from \cref{eq:PL_scd_cdf}. If the subhalo distribution can   be approximated by a power-law distribution,  \cref{eq:PL_scd}, then $\overline{J}_{\geq 1}$ and $\widetilde{J}_{\,\geq 1}$ are suitable approximations for the mean and median of the brightest halo:
\begin{align}
\overline{J}^{\star} &\approx \overline{J}_{\,\geq 1} = \int_0^{\infty} J\,\frac{\dd P_{\,\geq 1}}{\dd J} \,\dd J = \Gamma\left(\frac{1}{1-\alpha}+1\right)\times J_{\rm lim},
\label{eq:J_mean-J_lim}\\
\widetilde{J}^{\star} &\approx \widetilde{J}_{\geq 1} = \log(2)^{1/(1-\alpha)}\times J_{\rm lim}.
\label{eq:J_median-J_lim}
\end{align}
The probability distribution~\cref{eq:PL_scd_pdf} is defined for $\alpha>1$ and always positively skewed, even w.r.t. to a log-normal distribution. The long tail follows a power-law proportional to $J^{-\alpha}$, and thus the mean is only defined for $\alpha>2$ ($\overline{J}_{\geq 1}\rightarrow\infty$ for $\alpha\leq 2$). The mean is also always larger than the median value. For all our models in \cref{tab:model-parameters} we obtain $\alpha\gtrsim 2$, with a steepening of the $\Nsubmean(>J)$ distribution at the brightest halos (see \cref{fig:Histogram1D_Jfactor-Nsub-fullsky}). Even if a mean brightest halo could be computed in all our models, the median can always be defined (even for $\alpha\rightarrow 2$) and offers a much better characterisation of the brightest object.

\begin{figure}[t!]
\begin{center}
\includegraphics[width=0.62\textwidth]{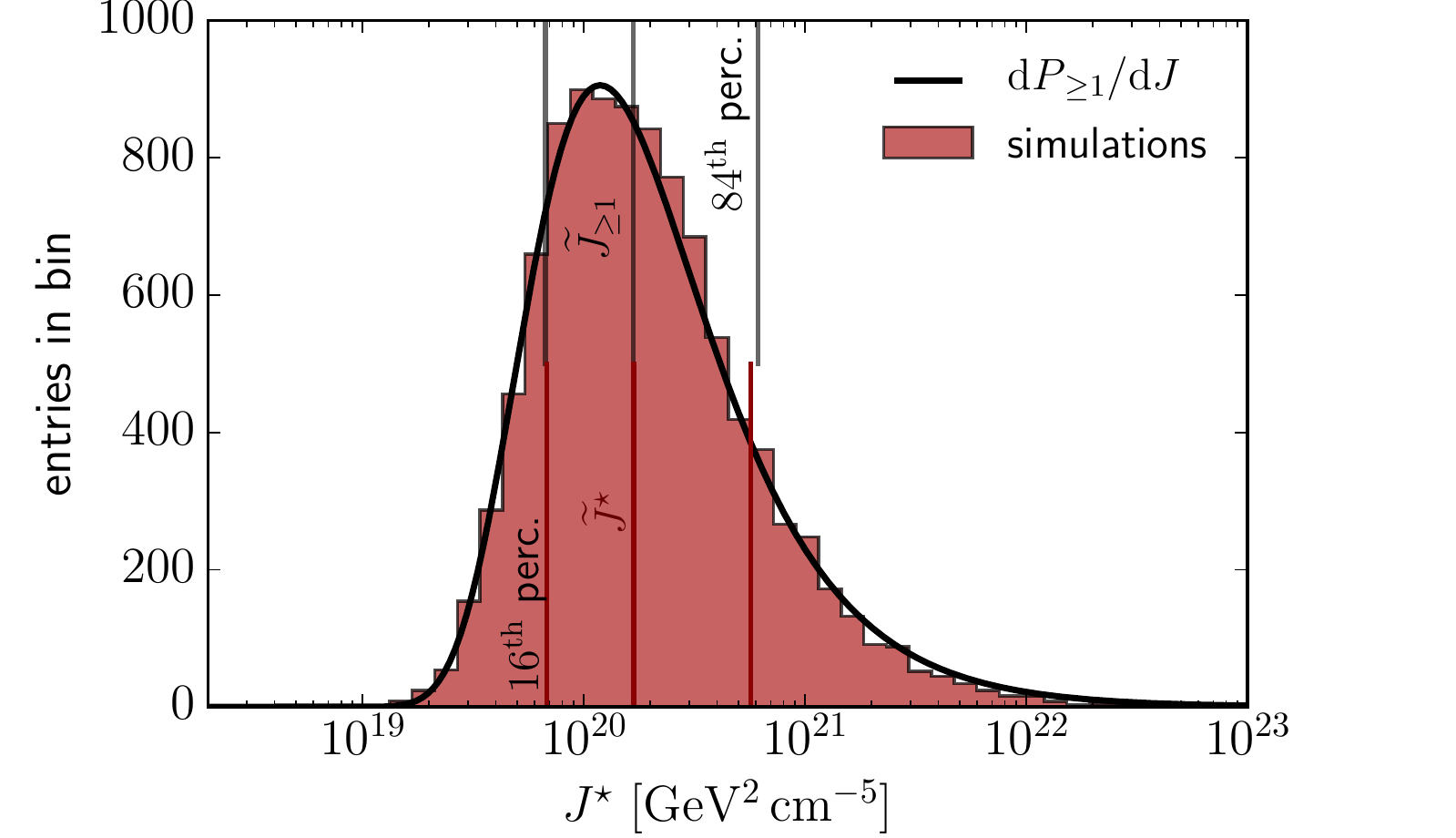}
\caption{
 Probability distribution  of the brightest subhalo $J^{\star}$ within the CTA survey FOV for model HIGH; $J^{\star}$ is chosen here w.r.t. to an integration angle $\thetaint=0.05\degs$, but integrated over $\theta_{\rm vir}$ (total emission). The histogram (red) is based on $10^{4}$ skymap realisations. The line relies on the power-law approximation $\dd P_{\,\geq 1}/\dd J$ from \cref{eq:PL_scd_pdf}, with $J_{\rm lim}=1.2\cdot 10^{20}\,\junits$ (fixed by the \nth{37} percentile from the skymap realisations) and $\alpha= 2.06$ (obtained from a least-square fit to the histogrammed skymap realisations). The vertical lines give the \nth{16}, \nth{50}, and \nth{84} percentiles---from the skymap realisations (bottom half) or from the distribution $\dd P_{\,\geq 1}/\dd J$ (top half)---encompassing the $68\%$ CI.
}
\label{fig:brightestClump-dpdF}
\end{center}
\end{figure}

The success of this analytic approximation is illustrated in \cref{fig:brightestClump-dpdF}, where the distributions from the power-law approximation (black line) and from the direct calculation over $10^4$ samples (red histogram) are compared for model HIGH. The two results are in excellent agreement. As stated in the main text, the skewness means that the limits on DM clump detection are extremely sensitive to the long tail of the distribution. The quantiles defining the 68\% CIs are also shown: they are used for the sensitivity calculations in \cref{sec:cta}. We remark that similar calculations as presented in this section recently have  been carried out in the  context of star cluster luminosities by \cite{2014MNRAS.438.2355D}.

\subsection{APS mean, median, and variance}
\label{app:APSapprox}
For a particular realisation of a point-like source count distribution, the intensity power $C_{\rm P}^I$ defined in  \cref{eq:C1sh_flux_Ando} scatters. The mean number of halos $\Nsubmean(\,>F)$ brighter than a flux $F$ behaves like $\Nsubmean(\,>L)$, i.e. it can be approximated by a power-law. Similarly to \cref{eq:PL_scd}, we define $F_{\rm lim}$ such as $\Nsubmean(\,>F_{\rm lim})=1$, and obtain
\beq
\Nsub(F_{\rm min},\,F^{\star}) =  \int\limits_{F_{\rm min}}^{F_{\rm max}=F^{\star}} \left\langle \frac{\dd  \Nsub}{\dd F'} \right\rangle \,\dd F'  \approx \left(\frac{F_{\rm min}}{F_{\rm lim}}\right)^{1-\alpha} - \left(\frac{F^{\star}}{F_{\rm lim}}\right)^{1-\alpha},
\label{eq:nsub_explicitfmax}
\eeq
by integrating $\langle \dd \Nsub/\dd F \rangle$ up to the flux of the brightest halo $F^{\star}$. Analogously, by using the power-law $\langle \dd \Nsub/\dd F \rangle = (\alpha -1)/F_{\rm lim} \cdot \left(F/F_{\rm lim}\right)^{-\alpha}$, $\alpha>1$, and replacing $F_{\rm min}$ by \Nsub{} from \cref{eq:nsub_explicitfmax}, one can rewrite \cref{eq:C1sh_flux_Ando} as
\beq
\label{eq:APS_PL_approx}
C_{\rm P}^I(F^{\star},\,\Nsub) \approx \frac{1}{4\pi\beta}\;F_{\rm lim}^{\;2}\;\left[\left(\frac{F^{\star}}{F_{\rm lim}}\right)^{3-\alpha}-\left\{\left(\frac{F^{\star}}{F_{\rm lim}}\right)^{1-\alpha}+ \Nsub\right\}^{\;-\beta} \right],
\eeq
with $\beta \equiv {(3-\alpha)}/{(\alpha - 1)}$ and $\alpha\neq 3$. Thanks to this approximation, $C_{\rm P}^I$ now became only a function of the brightest object $F^{\star}$ and of the total number of objects $\Nsub$. A further simplification arises assuming $\Nsub\!\rightarrow\!\infty$, in which case $F_{\rm min}\!\rightarrow\!0$ and \cref{eq:APS_PL_approx} simplifies to
\beq
C_{\rm P}^I(F^{\star},\,\Nsub) \;\;\stackrel{\Nsub\rightarrow \infty}{\approx}\;\; \frac{1}{4\pi\beta}\;F_{\rm lim}^{\;2}\;\left(\frac{F^{\star}}{F_{\rm lim}}\right)^{3-\alpha}.
\label{eq:APS_PL_largeNsub}
\eeq
Note that $C_{\rm P}^I$ becomes highly sensitive to a finite $N$ ($F_{\rm min}\neq 0$) in the case $\alpha\rightarrow 3$, such that the latter approximation is only applicable for $\alpha$ sufficiently smaller than $3$, and does not hold for $\alpha \geq 3$. For the case of \cref{eq:APS_PL_largeNsub} being valid, the median $\widetilde{C}_{\rm P}^I$ is directly related to $\widetilde{F}^{\star}$
\beq
\widetilde{C}_{\rm P}^I(F^{\star},\,\Nsub) \;\;\stackrel{\Nsub\rightarrow \infty}{\approx}\;\;  C_{\rm P}^I(\widetilde{F}^{\star}) \approx
 \frac{1}{4\pi\beta}\;F_{\rm lim}^{\;2}\;\log(2)^{-\beta}.
\label{eq:APS_PL_largeNsub_median}
\eeq
It is also useful (see \cref{app:APSconvergence}) to fix $\Nsub\equiv \Nsubmean$ in \cref{eq:APS_PL_approx}, such that $C_{\rm P}^I$ only becomes a function of $F^{\star}$. Then again, $\widetilde{C}_{\rm P}^I(F^{\star})\approx  C_{\rm P}^I(\widetilde{F}^{\star})$ holds, and one can simplify the ratio of the medians to
\beq
\frac{\widetilde{C}_{\rm P}^I(F^{\star};\,\Nsubmean)}{\widetilde{C}_{\rm P}^I(F^{\star};\,\Nsub\rightarrow \infty)} \approx 1 - \left(1 + \frac{\Nsubmean}{\log(2)}\right)^{\;-\beta}\,.
\label{eq:aps_convergence_hardapprox}
\eeq
Finally, one can extract the probability density function $\dd P/\dd C$ ($C\equiv C_{\rm P}^I$),
\beq
\frac{\dd P}{\dd C}(C) =  \frac{\dd P_{\,\geq 1}}{\dd F}\left(F^{\star}(C)\right)\times \left|\frac{\dd F^{\star}}{\dd C}(C)\right|,
\label{eq:rand_transformation}
\eeq
from which one can calculate analytic expressions for the mean $\overline{C}_{\rm P}^I$ and standard deviation $\sigma_{C_{\rm P}}$. The latter expression shows that the PDF of $C_{\rm P}^I$ is proportional to the PDF of $F^{\star}$, as given in \cref{eq:PL_scd_pdf}. This is in fact observed in our simulations (see \cref{fig:APSi_compare}), where the power at each $\ell$ shows a variance skewed w.r.t a log-normal distribution, with a long tail towards high $C_{\ell}$ values.

We stress that the above discussion only involves the variance of the APS from the varying flux of the objects. For randomly distributed objects on the sphere, the APS additionally scatters because of their random positions in space. This variance, which decreases for larger multipoles, can, on the full sky, be estimated by \cite{2006PhRvD..73b3521A}:
\beq
\sigma_{C_{\ell}} = C_{\ell}\,\sqrt{\frac{2}{2\ell +1}}\,.
\eeq

\section{APS convergence}
\label{app:APSconvergence}

As described in \cite{2012CoPhC.183..656C,2016CoPhC.200..336B}, \clumpy{} relies on a combination of the calculation of the mean signal \jsubs{} from subhalos and the calculation of individual drawn clumps $J_{\rm drawn}$ to ensure a quick calculation of skymaps. A critical parameter of a \clumpy{} run is the relative error $RE_{J_{\rm drawn}}$ \cite{2012CoPhC.183..656C}, which ultimately controls the number $N_{\rm sub}$ of clumps to be drawn. In practice, a critical distance is obtained for each mass decade, by requiring the relative error of the signal integrated from $l_{\rm crit}$ to remain lower than this user-defined precision requirement (level of fluctuation selected w.r.t. the mean signal). This reduces the number of clumps to draw in the Galaxy from $\sim10^{15}$ to $\sim10^{4}$ for  angular resolutions $\thetaint\sim 0.1\degs$ and $RE_{J_{\rm drawn}}=5\%$. This appendix shows that this choice is a good compromise between speed and precision, and that it ensures convergence of our results to the expected values up to the highest multipoles the considered $\gamma$-ray instruments are sensitive to. The fair agreement with the approximate analytical further validates the \clumpy{} code.

\begin{figure}[t]
\begin{center}
\includegraphics[width=0.62\textwidth]{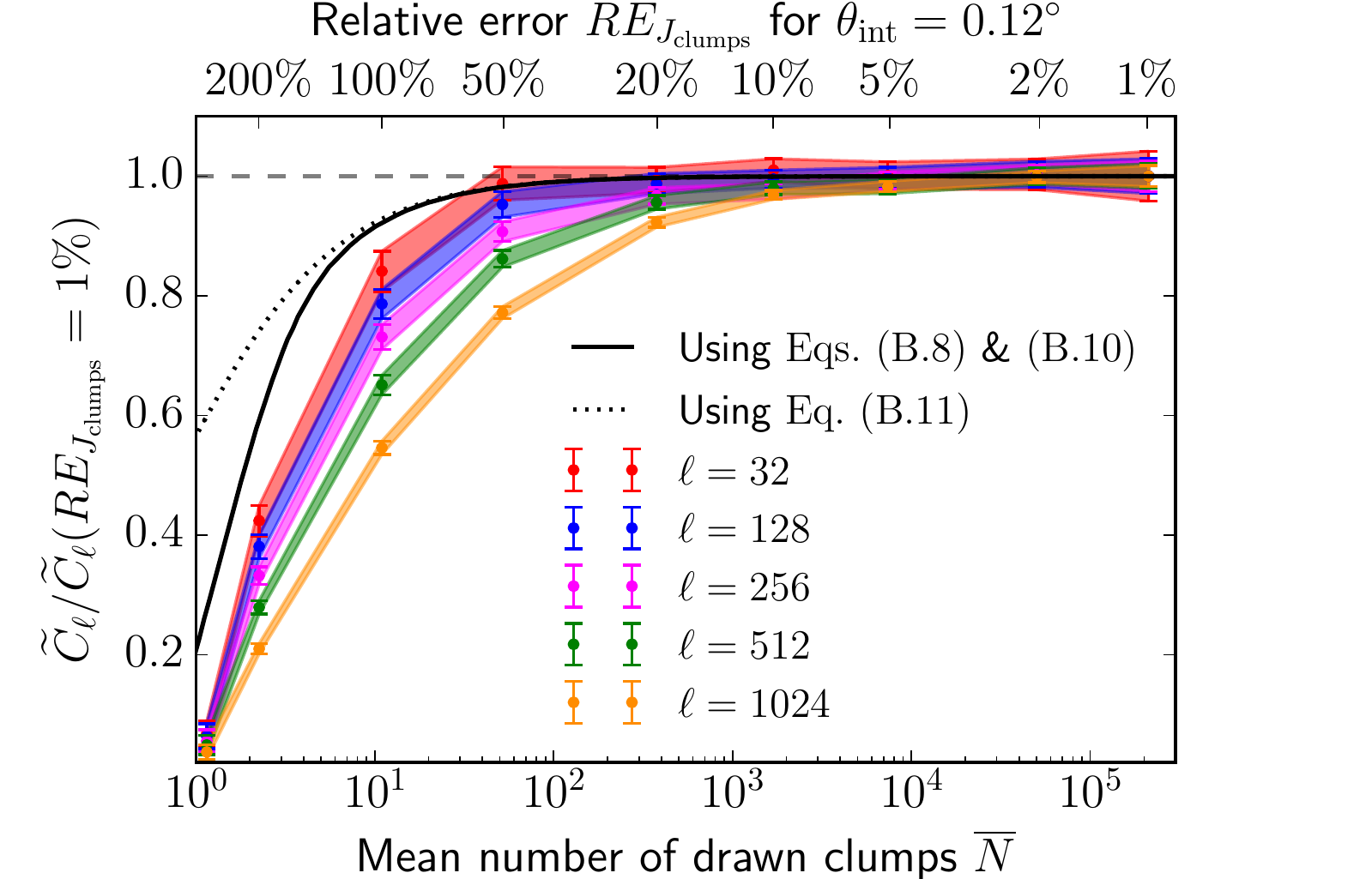}
\caption{
Convergence study  of the median power $\widetilde{C}_{\ell}$ (integration angle $\thetaint=0.12\degs$, i.e. \healpix{} $\nside=512$ of the \jfactor{} maps) as a function of $RE_{J_{\rm drawn}}$ (upper $x$-axis) or, equivalently, the number of drawn clumps $\overline{N}_{\rm sub}$ (lower $x$-axis). The different colours show different multipoles $\ell$ for the model LOW, with error bars from \cref{eq:error_of_median}. For comparison, we also show two analytical calculations (see \cref{app:meanmedian}), based on \cref{eq:APS_PL_approx} (black solid line) or the further approximated \cref{eq:aps_convergence_hardapprox} (dotted line).
}
\label{fig:APSi_rsestudy}
\end{center}
\end{figure}
\Cref{fig:APSi_rsestudy} shows the ratio $\widetilde{C}_{\ell}/\widetilde{C}_{\ell}(N_{\rm sub}\rightarrow \infty)$ for DM subhalos (model LOW) at different multipoles $\ell$, calculated over $N_{\rm sample}=5000$ simulations. The sample median was chosen because of its robustness compared to the mean, which is ill-defined for a power law index close to $\alpha=2$. The sample size $N_{\rm sample}=5000$ was taken in order to reach at least a $5\%$ relative accuracy of the ratio of the medians at the lowest multipoles, as calculated by the sample error of the median (shown by the coloured area)
\beq
\Delta \widetilde{C} = \frac{1}{2\,\frac{\dd P}{\dd C}(\widetilde{C})\, \sqrt{N_{\rm sample}}}.
\label{eq:error_of_median}
\eeq
The APS converges differently at different multipoles, but reaches 95\%
 of the overall power $\widetilde{C}_{\ell}(N_{\rm sub}\rightarrow \infty)$ at all multipoles $\ell \leq 1024$ whenever $N_{\rm sub} \gtrsim 10^{4}$ drawn objects. All power spectra presented in this work meet this requirement. For comparison purpose, the black dotted line shows the expected convergence calculated from the $\ell$-independent power-law ($\alpha=2.03$) approximation \cref{eq:aps_convergence_hardapprox}, which gives a fair description of the sample calculation for fixed $\overline{N}_{\rm sub}$. This approximation reveals the interesting result $\widetilde{C}_{\rm P}^I(\overline{N}_{\rm sub}=1)/\widetilde{C}_{\rm P}^I(N_{\rm sub}\rightarrow \infty)=0.57$, indicating that in median, the brightest point-like object accounts for more than $50\%$ to the overall power. The additional degree of freedom of a Poisson-distributed \Nsub{} from sample to sample shifts down the median power for a low average $\overline{N}_{\rm sub}$.  This is accounted for in a more accurate numerical calculation of the median from \cref{eq:APS_PL_approx}, divided by \cref{eq:APS_PL_largeNsub_median}, which gives an even fairer agreement with the sample median (black solid line).

\section{Virial mass and and $J$-factor for Milky Way satellites}
\label{app:mvirJ}

This appendix presents the derivation of \Mvir{} and $J(0.5^\circ)$ factors for spectroscopically confirmed satellite galaxies in the Milky Way. We recall that these values are used in \cref{fig:Histogram1D_Jfactor-Nsub-fullsky} and \cref{fig:Histograms2D_modelRef} for comparisons purpose with the drawn clumps of our simulations. 

\paragraph{Jeans and Markov Chain Monte Carlo (MCMC) analysis}
Most of our values come from a reprocessing the chains from the recent analyses of classical and ultrafaint dSphs of \cite{2015MNRAS.453..849B} for the pre-2015 dSphs, \cite{2015ApJ...808L..36B} for Ret~II, and  \cite{2016ApJ...819...53W} for Tuc 2. In these papers, a Jeans analysis of the light and velocity data coupled to an MCMC engine was used to recover the DM profiles, and then extract the median values and credible intervals on several quantities deriving from these profiles (e.g., mass, $J$-factor).\footnote{The robustness of the analysis has been validated on mock data~\cite{2015MNRAS.446.3002B}. All the tools to achieve of the steps of the analysis are public, they are described in the second release of the \clumpy{} code \cite{2012CoPhC.183..656C,2016CoPhC.200..336B}.} Moreover, in the last few months, and since the published results mentioned above, kinematic data of several new dSph galaxies have become publicly available. We therefore take the opportunity of this study to apply the same Jeans analysis on these dSphs, allowing for a more complete census for the objects shown in several plots of this paper. For these  ultrafaint objects the number of spectroscopically measured stars is generally $\sim 5-10$. Draco~II data has been taken from \cite{2016MNRAS.458L..59M}, Horologium~I from \cite{2015ApJ...811...62K}, Pisces~II from \cite{2015ApJ...810...56K} and Triangulum~II from \cite{2015ApJ...814L...7K}.\footnote{Thirteen member stars in Triangulum~II have recently been measured by \cite{2016ApJ...818...40M} but this dataset, showing distinct behaviours in the inner and outer part of the dSph galaxy, has not been used here.} We do not consider in this study the faint and/or uncertain Bo\"{o}tes~II \cite{2009ApJ...690..453K,2014ApJ...794...89K} and Hydra~II \cite{2015ApJ...804L...5M,2015ApJ...810...56K} dSph galaxies. We cannot exclude that several of the dSph galaxies in our selection suffer from tidal effects (see, e.g., \cite{2011AJ....142..128W} for Willman 1), but only Sagittarius dSph galaxy is undoubtedly flagged for this effect (see, e.g., \cite{2015MNRAS.446.3110K} and references therein) and was therefore also excluded from this sample.

\paragraph{Tidal mass as a proxy for the virial mass}
As recalled in \cref{sec:sets_of_models}, \clumpy{} works with virial quantities so that comparisons must be made w.r.t. this choice. The quantity \mvir{} depends on the definition of \rvir{} which, strictly speaking, depends on cosmological parameters. However, as discussed in \cite{2001A&A...367...27W}, the values of the quantities \mvir{} and $m_{\Delta=200\,,500\,\dots}$ are tightly correlated. Moreover, the differences between the mass reconstructed from these various definitions is much smaller than the uncertainties we obtain from the MCMC analysis ($\sim 2$ orders of magnitude). We therefore use the mass inside the virial radius \rvir{} as a measure of the total mass inside a DM halo. The authors of \cite{2008MNRAS.391.1685S} have shown the tidal radius to be a good proxy of the virial radius, this property holding even in the presence of baryons \cite{2015MNRAS.447.1353M}. The tidal radius is defined as \cite{2008gady.book.....B}
\begin{equation}
r_\text{tid} = \left[ \frac{m(r_\text{tid})}{[2-\dd \ln
      M_{\text{MW}}/\dd \ln R]\times M_{\text{MW}}(R)}
  \right]^{(1/3)} \times R\;,
\end{equation}
where $M_{\text{MW}}(R)$ is the mass of the MW enclosed within the galactocentric distance $R$ of the dSph, and $m(r_\text{tid})$ is the subhalo mass inside $r_{\rm tid}$. For each model of our MCMC analysis, we compute the enclose mass within $r_{\rm tid}$ and define it to be $m_{\rm vir}$ of that particular model. We underline that integrating up to $r\gg r_{\rm tid}$ instead of $r_\text{tid}$ only changes by a factor of 2 the total mass, which is much smaller than the uncertainties derived from the MCMC analysis.
\begin{alphafootnotes}
\begin{table*}
\begin{center}
\begin{tabular}{|l|cc|cc|cc|} \hline
  Name                      &   $D_{\rm obs}$  & $r_h$ &\multicolumn{2}{c|}{$\log_{10}[m_{\rm vir}/\Msol]$}&\multicolumn{2}{c|}{$\log_{10}[J(0.5^\circ)/({\rm GeV}^{2}~{\rm cm}^{-5})]$} \\
                            & [kpc]&  [kpc]&Median&   68\% CI   &Median&  68\% CI     \\
\hline
$^\star${\it Draco\;II}     &  20  &  0.019& 6.80 & [3.63-9.20] & 18.1 & [14.9-20.5] \\
{\it Segue\;I}              &  23  &  0.03 & 6.71 & [4.43-9.16] & 17.2 & [14.9-19.0] \\
{\it Ursa Major\;II}        &  30  &  0.14 & 9.49 & [8.31-10.9] & 19.9 & [19.3-20.5] \\
{\it Reticulum\;II}         &  30  &  0.032& 8.95 & [7.46-10.7] & 19.5 & [18.8-20.5] \\
$^\star${\it Triangulum\;II}&  30  &  0.034& 8.90 & [7.26-11.1] & 20.9 & [19.7-22.3] \\
{\it Segue\;II}             &  35  &  0.03 & 8.87 & [7.13-10.5] & 18.9 & [17.8-20.0] \\
{\it Willman\;I}            &  38  &  0.02 & 8.26 & [7.08-10.3] & 19.5 & [18.9-20.7] \\
{\it Coma}                  &  44  &  0.08 & 9.45 & [8.05-10.9] & 19.6 & [18.9-20.4] \\
{\it Tucana\;II}            &  57  &  0.165& 8.72 & [7.38-10.3] & 18.6 & [17.7-19.5] \\
Ursa Minor                  &  66  &  0.28 & 8.64 & [8.31-9.24] & 19.0 & [18.9-19.1] \\
{\it Bo\"otes\;I}           &  66  &  0.24 & 8.95 & [8.13-10.0] & 18.5 & [18.1-19.2] \\
Sculptor                    &  79  &  0.26 & 8.29 & [8.06-8.68] & 18.5 & [18.5-18.7] \\
$^\star${\it Horologium\;I} &  79  &  0.03 & 8.84 & [7.06-10.9] & 19.8 & [18.8-21.3] \\
Draco                       &  82  &  0.20 & 9.62 & [8.85-10.5] & 19.1 & [18.8-19.5] \\
Sextans                     &  86  &  0.68 & 9.02 & [8.57-9.68] & 17.6 & [17.4-17.8] \\
{\it Ursa Major\;I}         &  97  &  0.32 & 9.13 & [8.34-10.2] & 18.7 & [18.3-19.3] \\
Carina                      & 101  &  0.24 & 8.68 & [8.16-9.35] & 17.9 & [17.8-18.1] \\
{\it Hercules}              & 132  &  0.33 & 8.93 & [8.24-9.92] & 17.5 & [16.8-18.2] \\
Fornax                      & 138  &  0.67 & 8.93 & [8.67-9.26] & 17.7 & [17.6-17.8] \\
{\it Leo\;IV}               & 160  &  0.11 & 7.87 & [6.61-9.29] & 16.2 & [14.6-17.5] \\
{\it Canes Venatici\;II}    & 160  &  0.07 & 9.38 & [8.10-10.8] & 18.5 & [17.6-19.7] \\
{\it Leo\;V}                & 180  &  0.03 & 8.11 & [6.88-9.38] & 16.1 & [15.0-17.3] \\
$^\star${\it Pisces\;II}    & 182  &  0.058& 7.75 & [6.35-9.72] & 16.9 & [15.2-18.4] \\
Leo\;II                     & 205  &  0.14 & 9.06 & [8.37-10.0] & 18.0 & [17.8-18.6] \\
{\it Canes Venatici\;I}     & 218  &  0.57 & 9.20 & [8.74-9.92] & 17.5 & [17.3-17.9] \\
Leo\;I                      & 250  &  0.24 & 9.41 & [8.63-10.2] & 17.8 & [17.6-18.3] \\
{\it LeoT}                  & 407  &  0.18 & 9.12 & [8.27-10.4] & 17.6 & [17.1-18.6] \\[1mm]
\hline
                            &      & Type  &Central&  Min/Max   &Central&   Min/Max   \\[1mm]
{\bf LMC}                   & 50   &  SBm  & 10.5 & 10.2 / 11.4 & 19.8 & 19.1 / 21.7 \\
{\bf SMC}                   & 62   &  dIrr & 9.81 & 9.43 / 10.6 & 18.4 & 16.3 / 18.6 \\
\hline
\end{tabular}
\caption{The Galaxy satellites are ordered by distance and the columns correspond to (from left to right): name, distance, typical half-light radius (for dSphs) or type (for Magellanic clouds), median and 68\% CI virial mass, median and 68\% CI $J$-factor in an integration angle of 0.5$^\circ$. In our calculations, we do not account for the uncertainties on the distance $D_{\rm obs}$ and the half-light radius $r_h$ of the dSph galaxies, and only quote the values we have actually used in the analysis. Objects in {\em italic} are confirmed ultrafaint dSphs while the ones denoted by the star symbol [$^\star$] are the most recently discovered objects, tentatively analysed here. The {\bf boldface} objects correspond to the large and small Magellanic clouds, for which the central and Min/Max values represent the range of modelling uncertainties taken from the literature (see text).}
\label{tab:M_and_J}
\end{center}
\end{table*}
\end{alphafootnotes}

\paragraph{dSph, SMC and LMC values}
\Cref{tab:M_and_J} gathers the distance and median (and 68\% CI) of the mass $M_{\rm vir}$ and $J(0.5^\circ)$ for all the dSphs mentioned above. Given the very recent status of some of these spectroscopic measurements and the sometime intriguing kinematics they suggest (e.g., Triangulum~II \cite{2016ApJ...818...40M}), we urge caution in interpreting the new $J$-factors (especially the large J-factor of Triangulum~II) of all the objects denoted by the star symbol [$^\star$] in the table.\footnote{Near the completion of this paper, the non-spherical Jeans analysis of dSph galaxies by \cite{2016MNRAS.461.2914H} became available on arXiv and the $J$-factors of all dSph galaxies of \cref{tab:M_and_J} (including the most recently discovered) are compatible at 1$\sigma$ with their values but for Segue~2 and UMa~1. Also using \clumpy{}, \cite{2016arXiv160400838G} reports similar value for Triangulum~II as quoted here.} We also include estimation for the LMC and SMC in boldface. For these irregular objects, a spherical Jeans analysis cannot be applied, and we have taken the values from \cite{2009MNRAS.395..342B,2013Natur.495...76P,2015arXiv151103346B} for the LMC, and from \cite{2009MNRAS.395..342B,2014ApJ...780...59G,2015arXiv151103346B,2016PhRvD..93f2004C} for the SMC.

\section{Details of the CTA analysis}
\label{app:CTAanalysis_details}

To avoid too lengthy a discussion in the core of the text, several aspects of the CTA analysis were postponed to this appendix. The key points developed here are related to the observation strategy (\cref{app:obs_streg}), the post-trial sensitivity (\cref{app:prepost_trial}), the calculation of the maximum log-likelihood ratio (\cref{app:likelihood}), and the robustness and values of the test statistic (\cref{app:ts_statistics}) used in our analysis.

\subsection{Observation strategy}
\label{app:obs_streg}
For isotropically distributed sources on the sky, one may ask if a large field survey with a shallow exposure is the most sensitive strategy to detect these objects. It has been shown that observing a small portion of the sky with the same available total time is not beneficial for most source populations, and the loss of chance to encounter a very bright object within a small field of view outweighs the gain in deep-exposure sensitivity \cite{2013APh....43..317D}. We show below that this also applies to Galactic DM subhalos isotropically distributed on large scales. The isotropy assumption does not strictly hold for model HIGH, but still yields a fair characterisation of the subhalo abundance.

We assume isotropically distributed sources  on the sky obeying a power-law source count distribution, \cref{eq:PL_scd}. Then one obtains for the mean number $ \Nsubmean$ of subhalos with a flux above a given threshold and on a survey area $A$:
 \beq
 \Nsubmean(\,>F,\;A)=\left(\frac{F}{F_{\rm lim,\,fullsky}}\right)^{1-\alpha}\;\frac{A}{\unit[4\pi]{sr}}\,.
 \label{eq:nsub_fovsize}
 \eeq
so that
\beq
 \frac{\Nsubmean(\,>F_1,\;A_1)}{\Nsubmean(\,>F_2,\;A_2)}  =\left(\frac{F_1}{F_2}\right)^{1-\alpha}\;\frac{A_1}{A_2}\,.
 \eeq
For a fixed total observation time $T$ available to uniformly cover an area of the total size $A$, then an area within $A_{\rm FOV}$ can be observed for $t=T\times (A_{\rm FOV}/A)$. As the sensitivity to a flux $F$ goes as $1/\sqrt{t}$, we have $F_1/F_2=\sqrt{t_2/t_1}$, and $t_1/t_2=A_2/A_1$, so that
\beq
\frac{\overline{N}_{\rm detectable}(A_1,\,T)}{\overline{N}_{\rm detectable}(A_2,\,T)} = \left(\frac{A_1}{A_2}\right)^{\frac{3-\alpha}{2}},
\label{eq:best_fovsize}
\eeq
with $A_1$ and $A_2$ the observed area on the sky and $\overline{N}_{\rm detectable}$ the mean number of detectable objects. Thus, for a power-law index $\alpha<3$, the average number of detectable subhalos is $\overline{N}(A_1)>\overline{N}(A_2)$ for $A_1>A_2$ independent of $T$, and the probability of detecting an object from the  population rises for increasing the survey area. All our subhalo models meet this requirement, with $2\lesssim \alpha<3$. It is also useful to extract from \cref{eq:nsub_fovsize} the relation
 \beq
\frac{F_{\rm lim,\,A_1}}{F_{\rm lim,\,A_2}}=\left(\frac{A_1}{A_2}\right)^{\frac{1}{\alpha-1}}.
 \label{eq:flim_fovsize}
 \eeq
For a power-law distribution, according to \cref{eq:J_mean-J_lim,eq:J_median-J_lim}, this ratio also holds for the mean and median brightest halo,
 \beq
\frac{F_{\rm lim,\,A_1}}{F_{\rm lim,\,A_2}}=\frac{\overline{F}_{A_1}^{\star}}{\overline{F}_{A_2}^{\star}}=\frac{\widetilde{F}_{A_1}^{\star}}{\widetilde{F}_{A_2}^{\star}}\,.
 \label{eq:flim_fovsize2}
 \eeq
For a power-law index $\alpha\approx 2$, the ratio of mean/median fluxes (\jfactors{}) of the brightest halo within two survey fields $A_1,\,A_2$ is proportional to the ratio of the field sizes. 
We remark that this result also applies to the APS behaviour discussed in \cref{app:APSapprox}. Adopting \cref{eq:APS_PL_largeNsub_median}, one obtains for $\alpha\approx 2$
 \beq
\widetilde{C}_{\rm P}^I(A_1) \approx \left(\frac{A_1}{A_2}\right)^2\;\widetilde{C}_{\rm P}^I(A_2) \,.
 \label{eq:cp_fovsize}
 \eeq
For $A_2=\unit[4\pi]{sr}$, it is $A_1/A_2 = f_{\rm sky}$, the fraction of the sky. Note the $f_{\rm sky}^2$ scaling of \cref{eq:cp_fovsize} in the latter case.

\subsection{Pre- and post-trial sensitivity}
\label{app:prepost_trial}

To realistically assess a survey sensitivity, one has to account for the trials when searching for a signal from unspecified locations. Assuming an average $68\%$ containment radius of the CTA point spread function of $0.05\degs$, corresponding to a containment area of $\unit[2.4\cdot 10^{-6}]{sr}$, scanning a quarter of the sky results in approximately $\pi/(2.4\cdot 10^{-6}) = 1.3\cdot 10^6$ independent trials.\footnote{By chance, this number comes close to the frequency one expects a $5\sigma$ background up-fluctuation in  $1/p=3.5\cdot 10^6$ repetitions of a random experiment.} In order to reject a background fluctuation in the survey search at a trials corrected confidence level $1-p_{\rm post}$, we calculate the required confidence level, $1-p_{\rm pre}$, for the template observation setup according to
\beq
p_{\rm pre} = 1-(1-p_{\rm post})^{1/N_{\rm trials}}.
\eeq
Presenting our results at the $1-p_{\rm post}=95\%$ confidence level, with $N_{\rm trials}= 1.3\cdot 10^6$, we thus require $p_{\rm pre}=3.9\cdot 10^{-8}$. This corresponds to a Gaussian one-sided confidence level, $p=\alpha/2$, of $5.4\,\sigma$.

\subsection{Likelihood}
\label{app:likelihood}
The full unbinned likelihood function for a specific model $\mathcal{M}$ considered in our study is
\beq
\mathscr{L}(\mathcal{M}\,|\, N_{\rm obs},\,E_{\rm{obs},\,1\ldots \mathit{N}_{\rm obs}},\vec{k}_{\rm{obs},\,1\ldots \mathit{N}_{\rm obs}})=p(N_{\rm obs}\,|\, N_{\rm pred}(\mathcal{M}))\times\! \prod_1^{N_{\rm obs}} p(E_{\rm obs,\,i},\, \vec{k}_{\rm obs,\,i}\,|\,\mathcal{M}).
\label{eq:likelihood-function}
\eeq
Here, $N_{\rm obs}$ is the total number of observed events, $E_{\rm obs,\,i}$ and $\vec{k}_{\rm obs,\,i}$ denote the reconstructed energy and angular direction of each event. For Poisson statistics, 
\beq
p(N_{\rm obs}\,|\, N_{\rm pred})=\frac{N_{\rm pred}^{\quad N_{\rm obs}}\,e^{-N_{\rm pred}}}{N_{\rm obs}!}.
\eeq
 The total number of predicted events, $N_{\rm pred}$, is calculated by
\beq
N_{\rm pred}(\mathcal{M}) = T_{\rm obs}\,\int\limits_{E_{\rm min}=30~{\rm GeV}}^{E_{\rm max}=200~{\rm TeV}}\int\limits_{\Delta \Omega_{\rm obs}}\, p(E_{\rm obs},\, \vec{k}_{\rm obs}\,|\,\mathcal{M})\;\dd E_{\rm obs}\, \dd \Omega_{\rm obs},
\eeq
with $T_{\rm obs}$ the duration of the observation.\footnote{We do not include the time coordinates of the events into the likelihood calculation, as both the event rates from background and  DM annihilation are assumed to be constant in time.} The probability $p(E_{\rm obs},\, \vec{k}_{\rm obs}\,|\,\mathcal{M})$ for each event is given by the differential intensity $\dd \Phi_{\mathcal{M}}/(\dd E\, \dd \Omega)$ predicted by the model $\mathcal{M}$, integrated over the effective area, $A_{\rm eff}$, and convolved with the energy and angular response of the instrument,
\beq
p(E_{\rm obs},\, \vec{k}_{\rm obs}\,|\,\mathcal{M}) = \!\!\!\!\!\!\!\!\!\int\limits_{E,\, \Omega,\, A_{\rm eff}(E)}\!\!\!\!\!\!\!\!\! \,p(E_{\rm obs}\,|\,E,\,\vec{k}) \times p(\vec{k}_{\rm obs}\,|\,E,\,\vec{k}) \times \frac{\dd \Phi_{\mathcal{M}}}{\dd E \,\dd \Omega}(E,\,\vec{k})\;\;\dd A\,\dd E\, \dd \Omega,
\eeq
with $E$ and $\vec{k}$ the true energy and direction of the event.\footnote{In general, the effective area additionally depends on the time-dependent zenith and azimuthal coordinates of the observation, i.e. $A_{\rm eff} = A_{\rm eff}(E,\,\vec{k},\,t)$, which would introduce a time dependence to $p(E_{\rm obs},\, \vec{k}_{\rm obs},\,t\,|\mathcal{M})$ and to the likelihood~\cref{eq:likelihood-function}. However, for simplicity, we perform our study with a constant $A_{\rm eff}$.} For computation reasons, we neglect the energy dispersion of the events, i.e. we set $p(E_{\rm obs}\,|\,E,\,\vec{k})= \delta(E-E_{\rm obs})$, whereas the point spread function, $ p(\vec{k}_{\rm obs}\,|\,E,\,\vec{k})$, is modelled as a two-dimensional Gaussian with energy-dependent width $\sigma(E)$. 

For the signal model $\mathcal{M}_{\rm sig}$, $\dd \Phi_{\mathcal{M}_{\rm sig}}/(\dd E\, \dd \Omega)$ is the {\gr} intensity from DM annihilation, given by \cref{eq:flux-general}. For the background model $\mathcal{M}_{\rm bkg}$, $\dd \Phi_{\mathcal{M}_{\rm bkg}}/(\dd E\, \dd \Omega)$ corresponds to the residual cosmic ray background after cuts, and we directly make use of the background rate model shown in \cref{fig:plot_rates}, so that\footnote{The background rate model is taken for the chosen $A_{\rm eff}$ and thus is also constant in time for our study.}
\beq
p(E_{\rm obs},\, \vec{k}_{\rm obs}\,|\,\mathcal{M}_{\rm bkg}) = \frac{\dd N_{\rm bkg}}{\dd E_{\rm obs}\, \dd \Omega_{\rm obs}\,\dd t}=f(E_{\rm obs},\, \vec{k}_{\rm obs}). 
\label{eq:bck_model}
\eeq

\subsection{Distribution and values of the test statistic (\TS{})}
\label{app:ts_statistics}

For the test statistic $\lambda$ defined in \cref{eq:ts_def}, the hypothesis $\mathcal{M}_{\rm bkg}+\mathcal{M}_{\rm sig}$ has one more degree of freedom than the hypothesis background only. In the limit of a large data sample (event number) and provided the physical bound $\sigmav \geq 0$, \TS{} is expected to be distributed according to \cite{2011EPJC...71.1554C,2015APh....62..165C}
\beq
p(\TS) = \frac{1}{2}\,\delta(\TS)+\frac{1}{2}\, \chi^2_{k=1}(\TS).
\label{eq:ts_pdf}
\eeq
However, we found from $10^5$ MC simulations that for our setup, the \TS{} distribution is poorly described by \cref{eq:ts_pdf}, and depends on the spectral shape of $\mathcal{M}_{\rm sig}$. \Cref{tab:ts_distribution} shows the used test statistic (\TS{}) values for the sensitivity calculation in \cref{sec:cta}. We  performed $N_{\rm sim}=10^5$ MC simulations distributed over 24 spectra. We then merged the spectra into five groups, and  calculated $\TS(p_{\rm pre}=0.05)$ separately for each group. Because performing $10^8$ MC simulations for $p_{\rm pre}=3.9\cdot 10^{-8}$ was not feasible, we fitted an exponential tail to our  distributions, and computed $\TS(p_{\rm pre}=3.9\cdot 10^{-8})$ from this extrapolation. \Cref{tab:ts_distribution} shows that the obtained \TS{} values (left and middle columns) approximately correspond to the assumption of a $\chi^2_{k=1}$ distribution (right columns, constant and independent of the DM spectrum), but deviate for very low and high DM masses. 
\begin{table}[t]
\centering
 \begin{tabular}{|c|ccc|ccc|} \hline
         & \multicolumn{3}{c |}{$p_{\rm pre}=0.05$ ($1.6\,\sigma$)} & \multicolumn{3}{c |}{$p_{\rm pre}=3.9\cdot 10^{-8}$ ($5.4\,\sigma$)}\\[0.1cm]
 \mchi{} range [GeV] & $\chi\chi\rightarrow b\bar{b}$ & $\chi\chi\rightarrow \tau^+\tau^-$ & $0.5\,\chi^2_{k=1}$ & $\chi\chi\rightarrow b\bar{b}$ & $\chi\chi\rightarrow \tau^+\tau^-$ & $0.5\,\chi^2_{k=1}$ \\\hline
~$50-100$    & 1.5 & 1.5 & 2.71 & 24 & 28 & $29.1$ \\
$150-500$   & 2.5 & 2.9 & 2.71 & 31 & 32 & $29.1$ \\
~$600-1000$  & 3.4 & 2.5 & 2.71 & 35 & 30 & $29.1$ \\
$1500-7500$ & 3.0 & 1.9 & 2.71 & 34 & 28 & $29.1$ \\
$10^4-10^5$ & 2.0 & 1.1 & 2.71 & 29 & 26 & $29.1$ \\
\hline
 \end{tabular}
\caption{
Test statistic values used for the CTA sensitivity analysis in this study.  The values given in the left and middle columns ($\chi\chi\rightarrow XX$) are obtained from our MC calculation, the right columns are expected from Wilks' theorem ($0.5\,\chi^2_{k=1}$). The pre-trial p-value from the left block results into a post-trial p-value of 0.05 for $N_{\rm trials}= 1.3\cdot 10^6$. See \cref{app:prepost_trial} for pre- and post-trial definitions.
}
\label{tab:ts_distribution}
\end{table}


\bibliography{hutten_etal}
\end{document}